\documentclass[twocolumn,showpacs,preprintnumbers,amsmath,amssymb,superscriptaddress,prb]{revtex4-1}
\usepackage{epsfig,amsopn}
\usepackage{graphicx}
\usepackage{epstopdf}
\usepackage{sidecap}
\usepackage{color}
\usepackage{caption}
\usepackage{subfig}
\usepackage{amsmath,amssymb}
\usepackage{amsthm}
\usepackage{textcomp}
\usepackage{enumerate}
\usepackage{soul}

\newcommand\beq{\begin{equation}}
  \newcommand\eeq{\end{equation}}
\newcommand\bqa{\begin{eqnarray}}
\newcommand\eqa{\end{eqnarray}}

\newcommand{\kk} {\ensuremath{{\bf k}}}
\newcommand{\qq} {\ensuremath{{\bf q}}}

\newcommand{\moire}{moir$\acute{e}~$}

\begin{document}
\title{Trigonal Warping, Satellite Dirac Points and Multiple Field Tuned Topological Transitions in Twisted Double Bilayer Graphene
}
\author{Priyanka Mohan}
\affiliation{Department of Theoretical Physics, Tata Institute of Fundamental Research,
  Homi Bhabha Road, Mumbai 400005, India }
\author{Unmesh Ghorai}
\affiliation{Department of Theoretical Physics, Tata Institute of Fundamental Research,
  Homi Bhabha Road, Mumbai 400005, India }
\author{Rajdeep Sensarma}
\affiliation{Department of Theoretical Physics, Tata Institute of Fundamental Research,
  Homi Bhabha Road, Mumbai 400005, India }
\date{\today}
\begin{abstract}
We show that the valley Chern number of the low energy band in twisted
double bilayer graphene can be tuned through two successive
topological transitions, where the direct band gap closes, by changing the electric field perpendicular
to the plane of the graphene layers. The two transitions with Chern
number changes of $-3$ and $+1$ can be
explained by the formation of three satellite Dirac points around
the central Dirac cone in the \moire Brillouin zone due to the presence
of trigonal warping. The satellite
cones have opposite chirality to the central Dirac cone. Considering
the overlap of the bands in energy, which lead to metallic states, we
construct the experimentally observable phase
diagram of the system in terms of the indirect band gap and the
anomalous valley Hall conductivity. We show that while most of the
intermediate phase becomes metallic, there is a narrow
parameter regime where the transition through three insulating phases
with different quantized valley Hall conductivity can be seen.
We systematically study the effects of variations in the model parameters on
the phase diagram of the system to reveal the importance of
particle-hole asymmetry and trigonal warping in constructing the phase
diagram. We also study the effect of changes in interlayer tunneling
on this phase diagram.
\end{abstract}

\maketitle
% \tableofcontents
% \newpage
\section{Introduction}

The ability to experimentally control the angle of twist between
layers of two-dimensional materials placed on top of each other has
led to the field of twistronics: the manipulation of electronic
properties of materials by controlling the twist angle~\cite{CFFWTKJ2018,CarrMassattFangCazeauxLuskinKaxiras2017}. The materials
can range from graphene~\cite{BistritzerMacDonald2011} or few layers of graphene~\cite{SCWLWZTLTWTYMSYZ2020,MorellPachecoChicoBrey2013}, to graphene-boron
nitride heterostructures~\cite{Spanton62} to transition metal dichalcogenides~\cite{ZhangWangWatanabeTaniguchiUenoTutucLeRoy2020}, expanding
the class of Van der Waals bonded two-dimensional materials~\cite{Novoselovaac9439} with
interesting and tunable electronic properties. The most celebrated member of this family is the twisted
bilayer graphene, where the system shows an almost flat band around
particular twist angles called magic angles~\cite{BistritzerMacDonald2011,CFFWTKJ2018,CFDFTLSWTKAJ2018}. The presence of
strongly correlated phases like superconductivity~\cite{CFFWTKJ2018,Yankowitz1059}, correlated insulator~\cite{CFDFTLSWTKAJ2018}, magnetism~\cite{Sharpe605,PoZouVishwanathSenthil2018} etc. in
these materials around the magic angle has led to a surge in both
theoretical and experimental activity in this field.

Twisted double bilayer graphene (TDBLG)~\cite{koshino2019,ChebroluChittariJung2019,HaddadiWuKruchkovYazyev2020} consists of two layers of bilayer
graphene (which are themselves stacks of two layers of graphene)
placed on top of each other and rotated with respect to each other. In
these systems, the twist angle where flat bands are formed can be tuned relatively easily by
application of pressure~\cite{LinZhuNi2020}. Correlated insulating states
~\cite{LHKLRYNWTVK2020,BZTWMT2019,AdakSinhaGhoraiSanganiVarmaWatanabeTaniguchiSensarmaDeshmukh2020} with possible
magnetic order have been seen in these systems near magic angle. On
doping, the system has also shown superconductor-like
phases\cite{LHKLRYNWTVK2020}.

% While a simple minimal model with additional
% particle hole symmetry has been proposed to describe this
% system~\cite{Macdonald_minimal}, it has been experimentally shown that
% the metallic and insulating behavi

Electric field applied perpendicular to the plane of the material is
widely used in simple bilayer graphene to modify its electronic
properties, from tunable band gaps~\cite{ZTGHMZCSW2009,CNMPSNGGN2010} to occurrence of Lifshitz
transitions~\cite{VMKBSTWFIE2015,VBSWTIEMF2014} in the Fermi surface
of the system. The electronic properties of TDBLG in the presence of
electric field have been studied previously both theoretically~\cite{ChebroluChittariJung2019,LeeKhalafLiuLiuHaoKimVishwanath2019}
and experimentally~\cite{LHKLRYNWTVK2020,BZTWMT2019,AdakSinhaGhoraiSanganiVarmaWatanabeTaniguchiSensarmaDeshmukh2020}. While
a simple minimal model having particle hole symmetry has been
proposed to describe this system~\cite{koshino2019,ChebroluChittariJung2019,BZTWMT2019}, it has been
experimentally shown that the metallic and insulating behaviour of the
system is actually consistent with a more detailed model which does
not have particle-hole symmetry~\cite{AdakSinhaGhoraiSanganiVarmaWatanabeTaniguchiSensarmaDeshmukh2020}.

In this paper, we study the electronic properties of the TDBLG as a function of the twist angle and the
applied electric field (or equivalently the potential difference
between the layers) and present a topological
phase diagram of the system. We show that as a function of electric field,
the system undergoes two topological transitions, where the direct
band gap between the conduction and the valence band of the system vanishes. The valley
Chern number of the low energy valence band changes by $-3$ and $+1$
at these two transitions respectively. There are two ways to stack layers of AB bilayer graphene and twist them: the AB-AB stacking and the AB-BA stacking. We find that the Chern numbers for the AB-AB stacking and the AB-BA stacking are different; however, the changes in the Chern numbers across the transition are same for both stacking.

While the Chern numbers for TDBLG have been predicted before~\cite{ChebroluChittariJung2019}, in this paper, we provide an explanation for the transitions using the splitting of the single Dirac point into four different Dirac points. Of these, the central one at $K_M$ has opposite chirality to the three satellite Dirac points. This is similar to the Lifshitz transition in a BLG in perpendicular electric field. However, unlike BLG, in presence of tunneling between the twisted layers, the location of these Dirac points can be moved up or down in energy by tuning the electric field, resulting in non-trivial topological transitions. At low and intermediate twist
angles $(\theta < 2.8^\circ)$, these points are separated in energy. As a function of
increasing electric field, the gap first closes at the satellite Dirac
points, leading to a Chern number change of $-3$, while the gap
closing at higher electric field occurs at the central Dirac point,
leading to a Chern number change of $+1$. These two transitions come
closer and almost merge at large twist angles. The formation of the
satellite Dirac points and the separation of the two transitions in
the system are driven by the presence of trigonal warping, which is
known to create satellite Dirac points in simple bilayer graphene as well~\cite{RozhkovSboychakovRakhmanovNori2016}.

While the direct band gap and the Chern number are well defined
theoretical quantities for TDBLG, the overlap of the bands in energy
space implies that they are not directly related to observable
quantities. Focusing on the undoped system, we study the indirect
band gap, which can be directly linked to transport properties, and
the anomalous valley Hall conductivity which can be measured in
non-local transport measurements. The overlapping bands lead to
Fermi surface and metallic behaviour with non-quantized values of the
anomalous valley Hall conductivity in two regimes: (a) low twist
angles and low electric fields and (b) high twist angles and high
electric fields. However, the system remains insulating at high twist angle and low
electric field as well as at low twist angle and high electric fields,
with associated quantized valley Hall conductivity. In between, there is a regime
of intermediate twist angles where the system passes between three
gapped states through two gap closing transitions, revealing the
intermediate phase with non-trivial quantized valley Hall conductivity.

The hopping parameters in a TDBLG are amenable to change under
pressure, which is one of the reasons it is easier to tune this
system's electronic properties. Keeping this in mind, we also do an
extensive survey of how the phase diagram changes as various hopping
parameters in the model are varied. We find that the particle-hole
symmetric minimal model completely misses the phenomenology of gap
closing at finite electric fields. We show that while breaking the
particle hole symmetry results in a single transition, the presence of
the trigonal warping term is crucial to obtain the phenomenology of
multiple transitions with the pattern of Chern number changes seen in
this system. We also study how the phase diagram changes with the
ratio of inter-plane hoppings between twisted layers and find that
decreasing the ratio makes the intermediate gapped phase more (less)
visible in AB-AB (AB-BA) stacked TDBLG.

We now provide a brief road map of the paper. In section \ref{model},
we describe the model for the band structure of TDBLG in detail,
introducing all the relevant parameters in the process. In section
\ref{dirgap}, we present the band structure and the direct bandgap of
the system as a function of the twist angle and the electric field and
find multiple gap-closing transitions. In section \ref{chern}, we
discuss the Chern numbers of the low lying bands in the system. In
section \ref{observable}, we focus on the indirect band gap and the
valley Hall conductivity and present the phase diagram in terms of
these observable quantities. In section \ref{parameter}, we discuss the
variation of the phase diagram with the model parameters before
concluding with a summary of our key results.

% We note that in the two valleys (of the original graphene
% Hamiltonian), the changes are reversed, and the time reversal
% invariant system does not show any net Chern number when both valleys
% are taken into account.

%\vspace{1.5cm}
% \begin{figure}%
% \includegraphics[width=0.35\textwidth]{figures/bilayer_coupling_image.PNG}
% \caption{The AB stacked bilayer graphene lattice: The in-layer couplings in both the layers are 
% denoted by $\gamma_0$ while the  three inter-layer couplings are $\gamma_1$, $\gamma_3$ and $\gamma_4$.}
% \label{Fig:setup}
% \end{figure}

\section{Models of TDBLG \label{model}}

Twisted double bilayer graphene is formed by stacking two layers of
bilayer graphene (BLG) on top of each other, and rotating the layers with
respect to each other. Before we start describing model Hamiltonians
for twisted systems, let us review the Hamiltonian for a single Bernal
(AB) stacked bilayer graphene.
\begin{figure}[t!]%
\centering
 \includegraphics[width=\columnwidth]{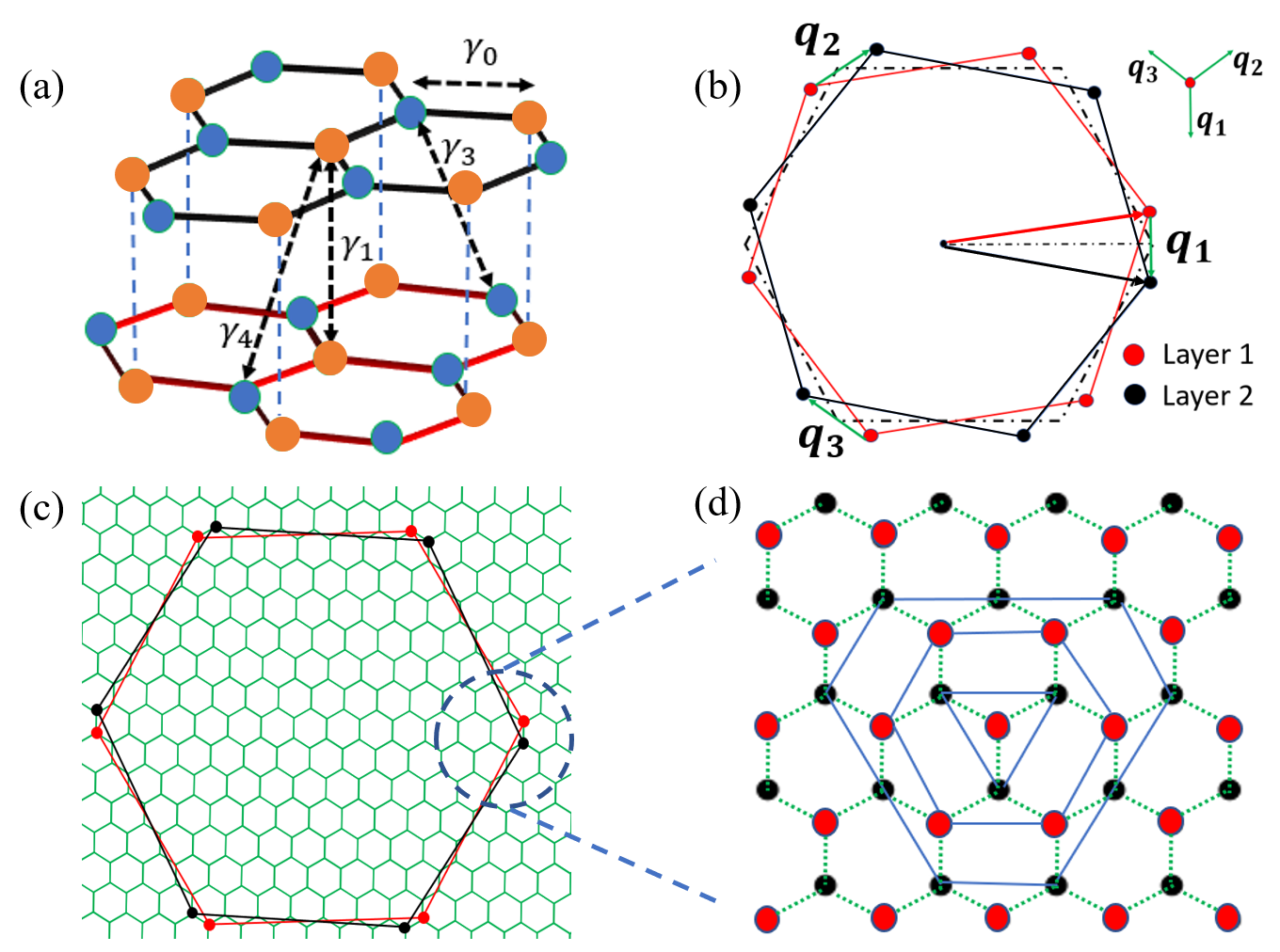}
\caption{(a) A single sheet of AB (Bernal) stacked bilayer graphene. The intralayer hopping $\gamma_0$ and the three inter-layer hoppings within the
bilayer $\gamma_1$, $\gamma_3$ and $\gamma_4$ are clearly marked. (b) The red and black hexagons represent the Brillouin zones of the two twisted layers. The $K$ points of the rotated BZ are 
related to each other by the vectors $\qq_{1,2,3}$. These three
vectors together create the \moire Brillouin zone. The dashed hexagon
is the reference frame with respect to which the layers are rotated by
$\pm \theta/2$. (c) The tiling of the graphene Brillouin zone by the
\moire Brillouin zones (green). The red and black hexagons represent
the tilted Brillouin zones in the two layers. This figure illustrates
the difference in scale between the original graphene Brillouin zone and the
\moire Brillouin zone. (d) The increasing orders of approximation in constructing the Hamiltonian matrix are demonstrated. The innermost blue triangle represents the lowest order, the middle hexagon represents the second order and the outer blue polygon represents the third order. The black and red dots
corresponds to the $K$ points in the Brillouin zones of the two layers in panel (c).}
\label{Fig:setup}
\end{figure}

The AB stacked bilayer graphene
consists of honeycomb lattices of graphene layers, where the $A$
sublattice of the top layer lies directly above the $B$ sublattice of
the bottom layer and the $B$ sublattice of the top layer lies at the center of the bottom layer hexagon, as shown in Fig ~\ref{Fig:setup}(a). One of the main
parameters of the tight-binding Hamiltonian for the bilayer graphene is
the in-plane nearest neighbour hopping along the graphene sheets,
$\gamma_0\sim 2.1354~eV$, leading to a graphene
Fermi velocity of $v_0=10^6 ~m/s$ ~{\cite{GeimNovoselov2007}}. The inter-layer
hopping is dominated by the hopping between the $A$ and $B$ sublattice
sites which lie on top of each other, with a tunneling
amplitude, $\gamma_1\sim 0.4~ eV$ ~{\cite{koshino2019,MalardNilssonEliasBrantPlentzAlvesNetoPimenta2007,ZhangLiBasovFoglerHaoMartin2008,NetoGuineaPeresNovoselovGeim2009}}. Due to this strong
tunnel-coupling, the sites on $A$ sublattice in the top layer and $B$ sublattice in the bottom layer which are just on top of each other are called dimerized sites, while
the $B$ sublattice sites in the top layer and $A$ sublattice sites in
the bottom layer are called non-dimerized sites.

In addition to the main hopping parameters described above, there are
three more parameters in the tight binding Hamiltonian of
a bilayer graphene. The most important of these is the interlayer hopping between
the non-dimerized sites in the top and bottom layer, $\gamma_3$. This trigonal warping term reduces the symmetry of the low energy band dispersion near the Dirac points from an azimuthal symmetry to a $C_3$ symmetry. We will later see that
this plays an important role in determining the topological phases of
TDBLG in electric field. The value of $\gamma_3$ ranges between $100~meV$  and $400~meV$ in the literature
~\cite{MalardNilssonEliasBrantPlentzAlvesNetoPimenta2007,KuzmenkoCrasseeMarelBlakeNovoselov2009,RozhkovSboychakovRakhmanovNori2016}. 
There is also a hopping between the dimerized site in one layer and a non-dimerized
site in another layer, $\gamma_4$, with values ranging between  $40~meV$ and $140~meV$~\cite{KuzmenkoCrasseeMarelBlakeNovoselov2009,NetoGuineaPeresNovoselovGeim2009,mccannkoshino2013}.
The final parameter is a potential difference
between dimerized and non-dimerized sites, $\Delta^{'} \sim 50~meV~{\cite{koshino2019,mccannkoshino2013}}$. These hopping parameters are also
clearly shown in Fig.~\ref{Fig:setup}(a).

Using a four component basis consisting of sublattices $(A_1,B_1,A_2,B_2)$, and
expanding around the Dirac point in a particular valley (say $K$ with valley index $\zeta=+1$), the
AB bilayer graphene Hamiltonian can be written as
\beq	
H_{AB} (\kk)= 
\left( \begin{array}{cc}
H_0(\kk) & g^\dagger(\kk)\\
         g(\kk) & H'_0(\kk) \end{array}\right),
     \label{Eq:H:BLG}
\eeq
where
\begin{widetext}
\beq
H_0(\kk) =
\left( \begin{array}{cc}
	0  & -\hbar v_0 k_- \\
	-\hbar v_0 k_+ & \Delta'
\end{array}\right) ~~,~~
H'_0({\bf k}) 
=
\left( \begin{array}{cc}
	\Delta'  & -\hbar v_0 k_- \\
	-\hbar v_0 k_+ & 0
\end{array}\right) ~~ \text{and}~~
g({\bf k}) 
=\left( \begin{array}{cc}
	\hbar v_4 k_+  & \gamma_1 \\
	\hbar v_3 k_-  & \hbar v_4 k_+
\end{array}\right).
\label{Eq:H:BLG:1}
\eeq
\end{widetext}
Here $k_{\pm} = \zeta k_x\pm \mathbf{i} k_y$, where the momenta are measured
from the Dirac point. $v_0$, $v_3$, $v_4$ are related to the hopping
amplitudes through $v_i =\frac{\sqrt{3}a}{2}\gamma_i$, where $a\approx0.246~nm${~\cite{koshino2019}} is
the graphene lattice constant. We note that both $\gamma_4$ and $\Delta^{'}$ breaks the particle-hole
symmetry of the low energy dispersions in the model, while $\gamma_3$
introduces a trigonal warping to the band structure. The Hamiltonian
for the $K'$ valley can be obtained by putting $\zeta=-1$ , while for
a $BA$ stacking, $H_0$ and $H'_0$ will exchange their positions and $g(\kk)$ will get conjugated to give $H_{BA}(\kk)$.

As mentioned earlier, there are two ways to stack layers of AB bilayer graphene and twist them: the AB-AB stacking and the AB-BA stacking. Here, the ordered list of sublattice indices indicate the dimerized sites in each layer, starting from the top. At zero twist angle, the strongest interlayer tunnelings naturally occur between these sublattices in the respective layers.
In both cases, the twist angle between the bilayer graphene sheets lead to a tiling of the original graphene Brillouin zone by
smaller \moire Brillouin zones(MBZ), as shown in Fig.~\ref{Fig:setup}(c).

To
understand the structure of the \moire Brillouin zone, we consider the
Brillouin zones of the top and bottom bilayer graphene, rotated about
each other by an angle $\theta$, as shown by the red and black
hexagons in Fig.~\ref{Fig:setup}(b). One can also consider a reference
Brillouin zone, shown by the dashed hexagon in Fig.~\ref{Fig:setup}(b), so that the Brillouin
zones of the top and bottom layer are rotated by $\pm \theta/2$ with
respect to this reference. The Dirac points in the two layers are now
separated from each other by wave vectors $\qq_1,\qq_2,\qq_3$, which are
aligned at angles of $2\pi/3$ with respect to each other.
The
difference in momentum between nearby Dirac points in the two layers
sets the scale of the \moire Brillouin zone to be  $k_M =k_D \sin(\theta/2)$ where $k_D = 8\pi/(3a) $ is the graphene Brillouin zone vector
and $\theta$ is the twist angle.  We will use the standard notation
with a subscript $M$  to
denote the high symmetry points of the \moire Brillouin zone, e.g. the \moire Brillouin zone center is denoted by $\Gamma_M$.
\begin{figure*}[t]
     \includegraphics[width=0.76\textwidth]{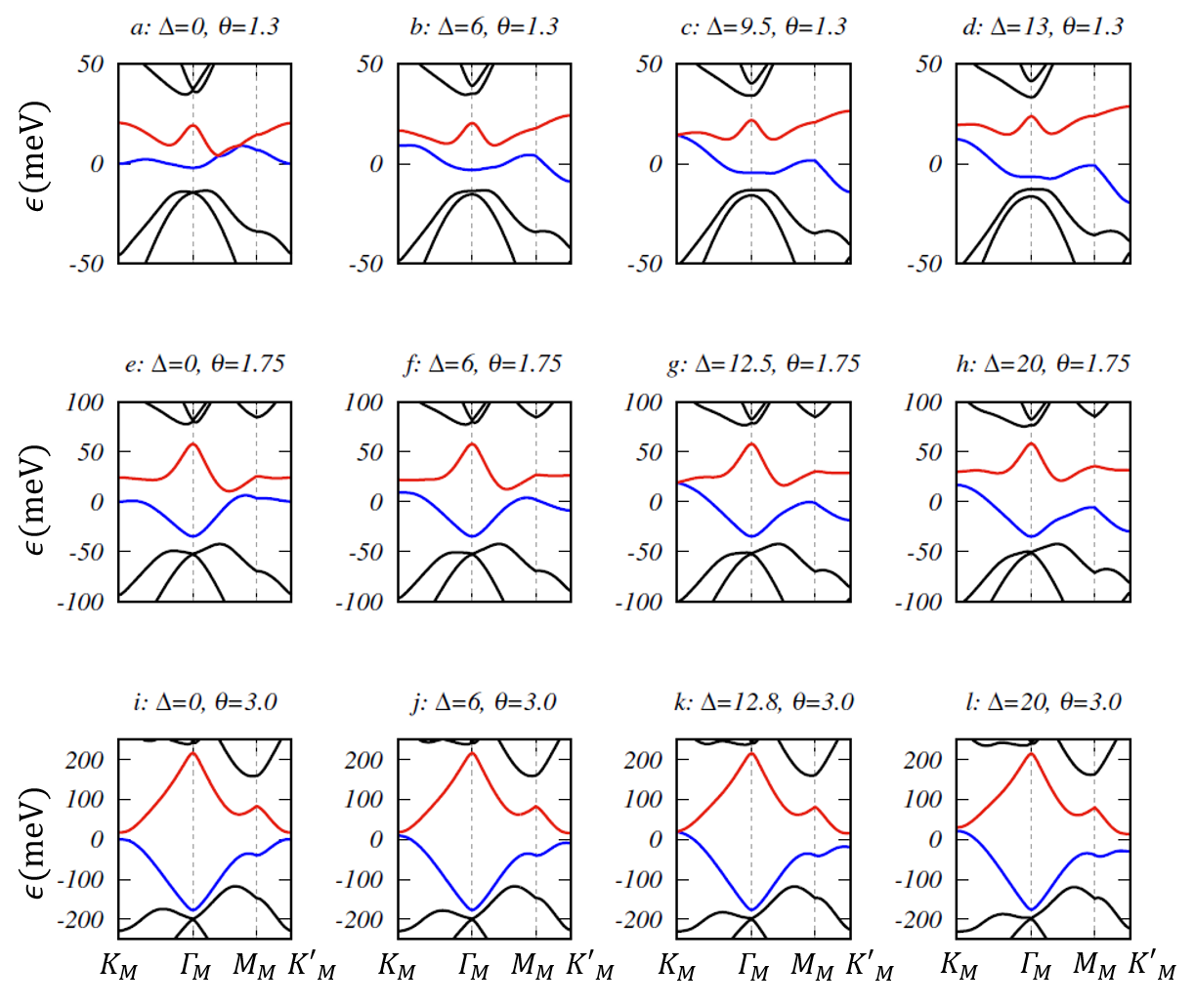}
\caption{Band dispersions in an AB-AB stacked TDBLG along the high symmetry axes for different values of the
  twist angle $\theta$ and the potential difference across
  each layer, $\Delta$. The conduction band is shown in red, while the
  valence band is shown in blue. Other bands are shown in black. (a)
  -(d): Band dispersion at a relatively low angle of $1.3^\circ$. (a)
  corresponds to $\Delta=0$, where the valence and the conduction
  bands touch each other at some $\kk$ points. The dispersions at
  $\Delta=6.0~meV$ (b)  and $\Delta=9.5~meV$ (c) show a decreasing
  gap between the bands at $K_M$ point, with the gap vanishing at
  $\Delta=9.5~meV$ (c). The dispersion at $\Delta=13~meV$ (d) shows
  the gap opening up once again. (e)-(h): Band dispersions at an
  intermediate twist angle of $1.75^\circ$ for $\Delta=0$ (e),
  $\Delta=6~meV$ (f), $\Delta=12.5~meV$ (g) and $\Delta=20~meV$
  (h). At $\Delta=0$, valence and conduction bands do not touch each
  other. The $K_M$ point gap closes at $\Delta=12.5~meV$ (g) before
  reopening, as seen in (h). (i)-(l): Band dispersions at a relatively
  high twist angle of $3.0^\circ$ for $\Delta=0$ (i),
  $\Delta=6~meV$ (j), $\Delta=12.8~meV$ (k) and $\Delta=20~meV$
  (l). The trends are similar to that for $1.75^\circ$, with a $K_M$
  point gap closing at $\Delta=12.8~meV$ (k).}
\label{Fig:BandDispAB-AB}
  \end{figure*}

If we start with electrons in the top BLG with momentum $\kk$ in the
\moire Brillouin zone, the
interlayer hopping between the twisted layers couples this electron to
electrons in the lower BLG in the three nearest \moire Brillouin
zones with momenta $\kk +\qq_{i}$, with $i=1,2,3$. This is shown
clearly in Fig.~\ref{Fig:setup}(b).  The wave vectors
$\qq_{1(2)(3)}$ shown in Fig.~\ref{Fig:setup} are related to
each other by the \moire reciprocal lattice vectors. Thus the minimum
size of the Hamiltonian matrix is $16$ ($4$ BLG Hamiltonians
with $4\times 4$ matrix for each).
% This constitutes the lowest order
% terms in the Hamiltonian. 
% Each of these three points $\kk +\qq_{i}$ would be connected by the
% interlayer tunneling to points at  $\kk +\qq_{i}+\qq_{j}$ in other
% \moire Brillouin zones, forming a Hamiltonian matrix of ever-increasing
% size. In practice, this size is numerically cut-off the higher order Hamiltonian matrix. 
% Under the rotation, the momentum dependent  elements in Eq.\ref{Eq:H:BLG:1} would transform as $\kk\rightarrow {\mathcal R}(\pm \theta/2) \kk$ depending
% on the counter-clockwise or clockwise rotation. The standard procedure in calculations is to symmetrically rotate each bilayer by an angle $\pm\theta/2$ 
% to achieve a total twist angle of $\theta$.
% {\rcol Need more details about defn of $\kk$ rotation  matrix etc. Please fill up this part. }
For AB-AB stacking, this can be written as
\begin{widetext}
\beq
H_{AB-AB}(\kk)=\left(\begin{array}{cccc}%
                      H_{AB}(\kk, \theta/2) & U_1^\dagger & U_2^\dagger &U_3^\dagger\\
                      U_1 & H_{AB}(\kk+\qq_1, -\theta/2) & 0 & 0\\
                      U_2 & 0 & H_{AB}(\kk+\qq_2,-\theta/2)& 0\\
                       U_3 & 0 & 0& H_{AB}(\kk+\qq_3,-\theta/2) 
                    \end{array}\right),
                  \label{Eq:HAB-AB}
\eeq
\end{widetext}
where $H_{AB}(\kk,\theta)$ is the BLG Hamiltonian written in a
co-ordinate system which is rotated by an angle $\theta$, with
$k_{\pm}=ke^{\pm i\phi} \rightarrow  ke^{i(\pm \phi+\theta)}$, where $\phi$ is the azimuthal angle. Note
that we use the symmetric reference frame about which the top and
bottom $BLG$ are rotated by $\pm \theta/2$. Here 
\beq
U_n=\left(\begin{array}{cc}%
            0 & \tilde{U}_n\\
            0 & 0 \end{array}\right) ~,~ \tilde{U}_n=\left(\begin{array}{cc}%
            u & u'e^{-\mathbf{i}\frac{2\pi \zeta}{3}(n-1)} \\
            u'e^{\mathbf{i}\frac{2\pi \zeta}{3}(n-1)}& u \end{array}\right)
\eeq
and $u$ and $u'$ are the interlayer hopping between $AA/BB$ and
AB sites respectively. There are a wide range of values used for $u$
and $u'$ in the literature, with some consensus around $u'=100~meV$
and $u/u'=0.6-0.8$~{\cite{LeeKhalafLiuLiuHaoKimVishwanath2019, KoshinoYuanKoretsuneOchiKurokiFu2018, 
NamKoshino2017, UchidaFuruyaIwataOshiyama2014, MoonKoshino2013, ChebroluChittariJung2019, DaiXiangSrolovitz2016, 
WijkSchuringKatsnelsonFasolino2015}}.

We note that this constitutes the lowest order terms in the Hamiltonian. 
Each of these three points $\kk +\qq_{i}$ would be connected by the
 interlayer tunneling to points at  $\kk +\qq_{i}+\qq_{j}$ in other
 \moire Brillouin zones, forming a Hamiltonian matrix of ever-increasing
 size. In general one should
carry out the tiling of the full graphene Brillouin zone
with the \moire Brillouin zones. The momentum points which get
connected in this process are shown in
Fig.~\ref{Fig:setup}(d), where red dots indicate electrons in top BLG while
black dots indicate electrons in the bottom BLG. In practice, as shown
by Bistritzer et. al ~{\cite{BistritzerMacDonald2011}}, one can truncate this
expansion at reasonably low sizes to get accurate description of low
energy bands. In our calculations, we work with $340 \times 340$ sized
matrices, which correspond to taking seven layers of nearby shells \moire Brillouin zones. This ensure low energy dispersion error below $0.3\% $.
  \begin{figure}[h]%
 \includegraphics[width=\columnwidth]{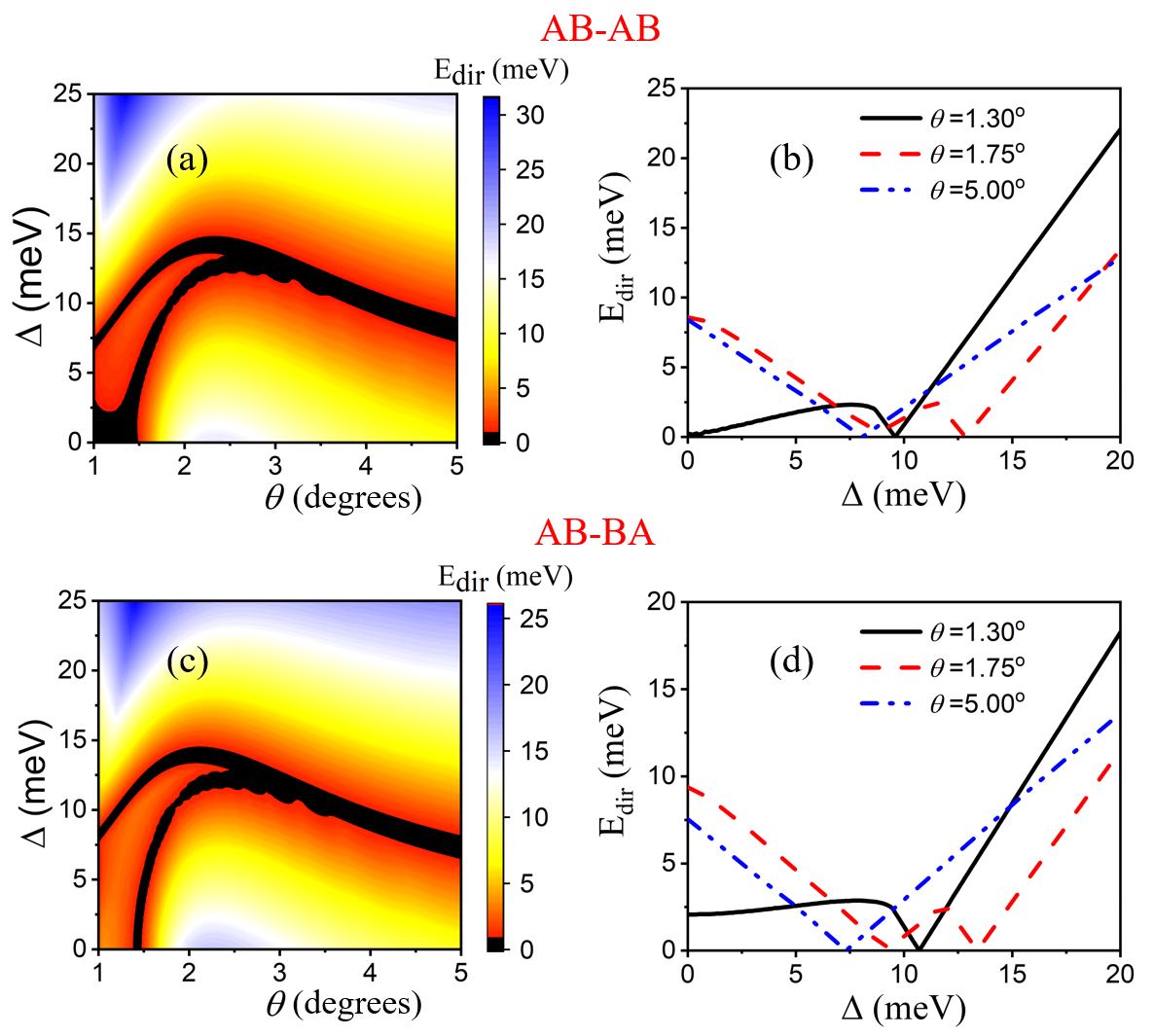}
 \caption{(a): A color plot of the direct band gap between the
   conduction and valence bands in the twist angle
   ($\theta$)-potential difference ($\Delta$) plane for the AB-AB
   stacked TDBLG. There are two visible black lines with some width in the figure where the direct
 band gap practically vanishes. (b) Representative line cuts showing the variation
of the direct band gap $E_{dir}$ with $\Delta$ at fixed $\theta$: At
low $\theta=1.3^\circ$ (solid black line), there is a single gap closing
around $\Delta=9.5 meV$. At
intermediate $\theta=1.75^\circ$ (dashed red line), there are two gap
closings at $\Delta=9 meV$ and $\Delta=13 meV$. At
high $\theta=5^\circ$ (dot-dashed blue line), the two gap closings have
merged  around $\Delta=8 meV$. (c) and (d) are same plots as (a) and (b) respectively for AB-BA stacked TDBLG.} 
 \label{Fig:DirBandGap} 
\end{figure}

We note that the Hamiltonian for the AB-BA stacking can be obtained
from Eq.~\ref{Eq:HAB-AB} by replacing the lower three $H_{AB}$
s by $H_{BA}$. The addition of a perpendicular electric field is
modeled as a uniform potential difference across the layers, with the
topmost layer having a potential $3\Delta/2$ and the lowermost layer
having a potential $-3\Delta/2$. In the Hamiltonian, this corresponds
to augmenting $H_{AB}$ ($H_{AB}$ or $H_{BA}$) of the top (bottom) BLG layer by
a potential term $V_{T(B)}$, where 
\beq
V_T=\left(\begin{array}{cc}
            \frac{3}{2}\Delta \mathbb{I}_2 & 0\\
        0 & \frac{1}{2}\Delta \mathbb{I}_2 \end{array}\right),  ~~V_B=\left(\begin{array}{cc}
            -\frac{1}{2}\Delta \mathbb{I}_2 & 0\\
        0 & -\frac{3}{2}\Delta \mathbb{I}_2 \end{array}\right).
            \eeq
The detailed model which we use can be replaced by a
simpler ``minimal'' model to obtain some of the qualitative properties
of TDBLG ~\cite{koshino2019,ChebroluChittariJung2019}. In our notation, this corresponds to setting $\gamma_3$,
$\gamma_4$ and $\Delta'$ to zero. The explicit dependence of the
Hamiltonian on the twist angle is also
neglected in the minimal model; i.e. $H_{AB}(\kk,\pm\theta/2) $is set
to $H_{AB}(\kk,0) $. We would however like to note that the
minimal model imposes additional particle-hole symmetry~\cite{song2019} and gets some
of the qualitative features of experiments wrong, as shown in
Ref.~{\onlinecite{AdakSinhaGhoraiSanganiVarmaWatanabeTaniguchiSensarmaDeshmukh2020,LHKLRYNWTVK2020}}. In this paper, we will see that the
minimal model fails to capture the full complexity of the topological
phase diagram of the system.

\section{Electronic Structure and Band Gap in TDBLG}
\label{dirgap}
 \begin{figure*}[t]
 \includegraphics[width=\textwidth]{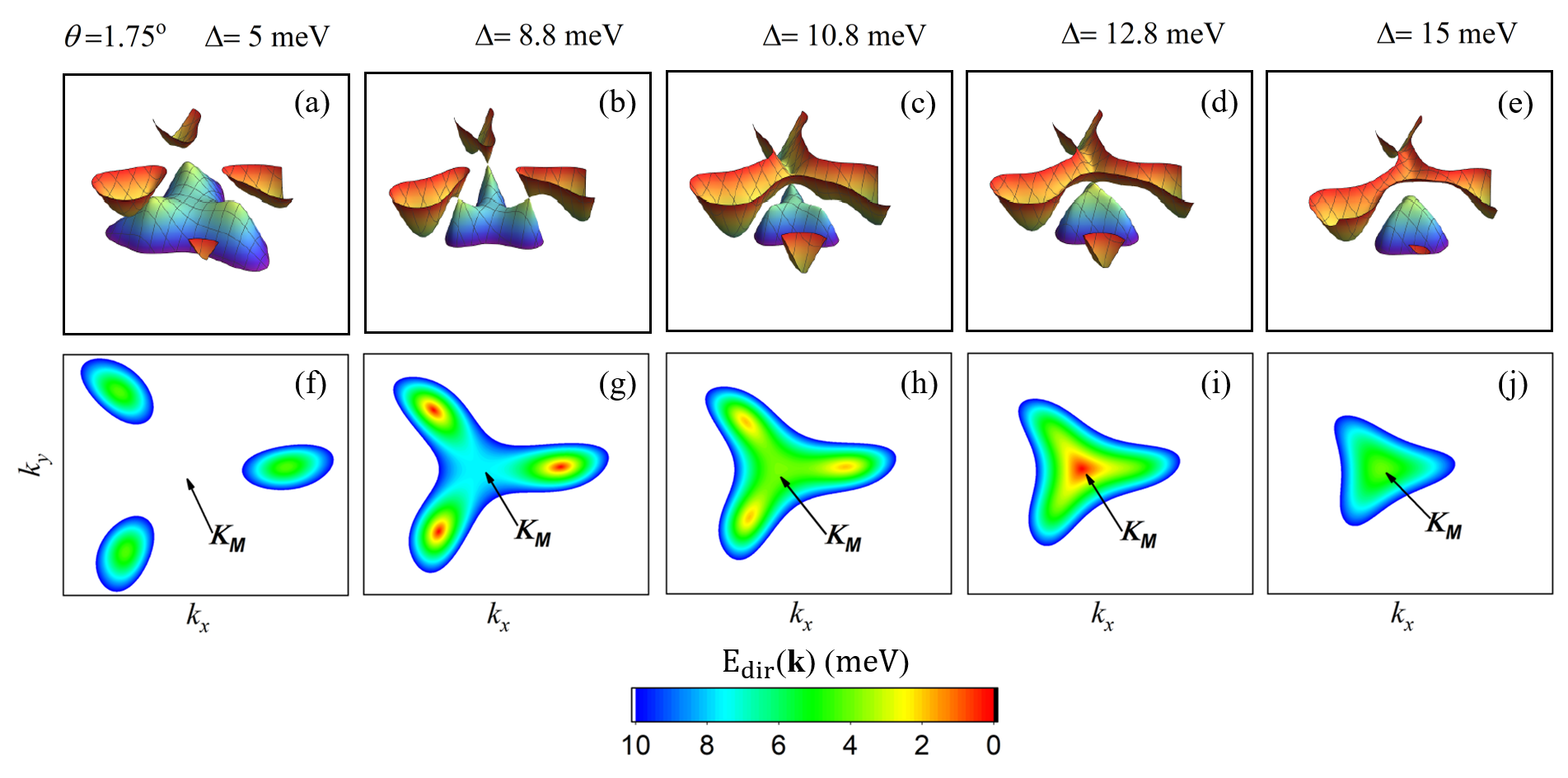}
 \caption{(a)-(e): A $3$ dimensional representation of the valence and the
   conduction bands near the $K_M$ point of the MBZ
   for an intermediate twist angle of $1.75^\circ$ for AB-AB stacked TDBLG. (f)-(j): Color
   plot of the momentum dependent gap $E_{dir}(\kk)$ near the $K_M$
   point for the band structures shown in (a) -(e) respectively.  (a)
   and (f) corresponds to
   $\Delta=5~meV$, where the system is gapped. (b) and (g) correspond
   to the first gap closing transition at $\Delta=8.8~meV$. The gap
   closes at the three satellite Dirac points, as clearly seen in
   (b). In (g) this corresponds to the three red spots in the
   plot. (c) and (h) correspond to $\Delta=10.8~meV$ where the gap has
   opened up again in the intermediate phase. (d) and (i) correspond
   to the second gap closing transition at $\Delta=12.8~meV$. It is
   clear both from(d) and (i) that the gap closes at the $K_M$
   point. (e) and (j) correspond to $\Delta =15~meV$ in the high field
   phase, where the gap has opened up once again.}
 \label{Fig:Satellite_Dirac} 
 \end{figure*}
We first consider the electronic structure of twisted double bilayer
graphene, focusing on the two low energy bands close to zero energy.
While we will later look at how the
  electronic properties change with system parameters, in this section
  we will use the parameters in Ref.~\onlinecite{koshino2019} to get a picture
  of the electronic structure in this system; i.e. we will use
  $\gamma_0= 2.135~ eV$, $\gamma_1=400~meV$, $\gamma_3=320~ meV$, $\gamma_4=44~
  meV$, $\Delta^{'}=50~meV$ for the bilayer graphene parameters and
  $u'=97.5~meV$ and $u=79.7~meV$ for the interlayer tunneling
  parameters of the TDBLG. 
%   The \moire super-lattice connects points on
%   the \moire Brillouin zone with points in nearby Brillouin zones,
%   leading to a very large matrix in general. We will numerically cut off this
%   matrix at a size of $340$ corresponding to taking $7$ nearby
%   layers
%   of \moire Brillouin zones to obtain the band structure.
   \subsection{AB-AB stacked TDBLG}
  
  Let us first focus our attention on AB-AB  stacked TDBLG. While the experimental focus on the twisted systems has
  been concentrated around $1.1^\circ$, in this
  paper we will expand our search to explore the system at twist
  angles from $1^\circ$ to $5^\circ$.

  The band dispersions of the system along the high symmetry axes
  at a low twist angle
    of $1.3^\circ$ (close to magic angle) is shown for increasing
    electric fields corresponding to potential differences of $\Delta =0,~6,~9.5$ and $13~meV$
  in Fig.~\ref{Fig:BandDispAB-AB}(a) -(d) respectively. The conduction
  band dispersion $\epsilon_{1}(\kk)$ and the valence band dispersion
  $\epsilon_{-1}(\kk)$  are plotted in red and blue colors
  respectively. At zero external electric field,
  Fig.~\ref{Fig:BandDispAB-AB}(a) shows that the bands
  touch each other at two points between the $\Gamma_M$ and $M_M$ points{~\cite{koshino2019}}.
  To understand the variation of the band structure with electric fields, it is useful to focus on the gap between the conduction and valence band dispersions at the $K_M$ point in
  the \moire Brillouin zone. This is large at $\Delta=0$. With increasing electric field, the
  gap decreases at $\Delta=6~meV$, and is very small at $\Delta
  \sim 9.5~meV$, as seen in Fig.~\ref{Fig:BandDispAB-AB}(b) and (c)
  . At larger electric field, the gap increases again, as seen from
  the dispersion at $\Delta=13 ~meV$ in
  Fig.~\ref{Fig:BandDispAB-AB}(d). Thus, there is at least one gap
  closing transition at a finite electric field at low twist
  angles. Note that the gap closes only at the $K_M$ point and not at
  the $K_M'$ point; since the gap vanishes at odd number of points, one can expect a change in the topology of the
  bands at the gap closing point.

  In Fig.~\ref{Fig:BandDispAB-AB}(e)-(h), we plot the band dispersion
  of the system at an intermediate twist angle of $1.75^\circ$ for
  increasing $\Delta =0,~6,~12.5$ and $20~meV$
  respectively. The key difference from the low twist angle of
  $1.3^\circ$ is that in this case, there are no gap closings at zero
  electric field, as seen in Fig.~\ref{Fig:BandDispAB-AB}(e). From
  Fig.~\ref{Fig:BandDispAB-AB}(f) and Fig.~\ref{Fig:BandDispAB-AB}(g),
  we see that the gap at $K_M$ point
 continuously  decreases at $\Delta=6~meV$, and closes around $\Delta\sim
  12.5~meV$. The gap increases at higher electric fields, as seen from the
  dispersion at $\Delta=20~meV$ in Fig.~\ref{Fig:BandDispAB-AB}(h). We
  plot the band dispersion at a relatively higher twist angle of
  $3^\circ$ for
  increasing $\Delta =0,~6,~12.8$ and $20~meV$ in Fig.~\ref{Fig:BandDispAB-AB}(i)-(l)
  respectively. The pattern here is very similar to the case of
  $1.75^\circ$, with the gap closing around $\Delta \sim 12.8~meV$, before
  increasing once again.
\begin{figure*}[t]
     \includegraphics[width=0.76\textwidth]{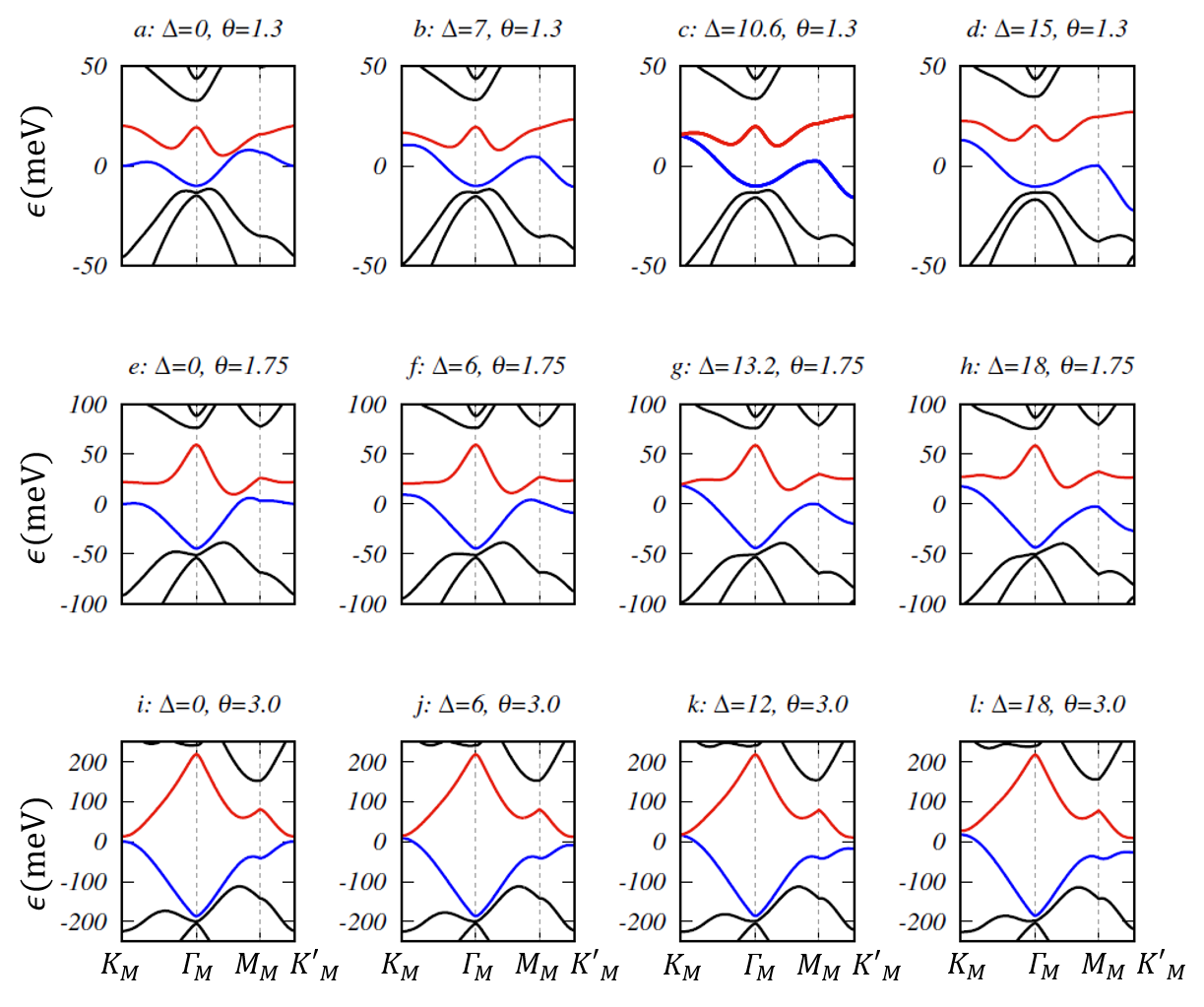}
\caption{Band dispersions in an AB-BA stacked TDBLG along the high symmetry axes for different values of the
  twist angle $\theta$ and the potential difference across
  each layer, $\Delta$. The conduction band is shown in red, while the
  valence band is shown in blue. Other bands are shown in black. (a)
  -(d): Band dispersion at a relatively low angle of $1.3^\circ$. (a)
  corresponds to $\Delta=0$, where the valence and the conduction
  bands touch each other at some $\kk$ points. The dispersions at $\Delta=7.0 ~meV$ (b)  and $\Delta=10.6 ~meV$ (c) show a decreasing
  gap between the bands at $K_M$ point, with the gap vanishing at $\Delta=10.6 ~meV$ (c). The dispersion at $\Delta=15.0 ~meV$ (d) shows
  the gap opening up once again. (e)-(h): Band dispersions at an
  intermediate twist angle of $1.75^\circ$ for $\Delta=0$ (e),
  $\Delta=6~meV$ (f), $\Delta=13.2 ~meV$ (g) and $\Delta=18.0 ~meV$
  (h). At $\Delta=0$, valence and conduction bands do not touch each
  other. The $K_M$ point gap closes at $\Delta=13.2 ~meV$ (g) before
  reopening, as seen in (h). (i)-(l): Band dispersions at a relatively
  high twist angle of $3.0^\circ$ for $\Delta=0$ (i),
  $\Delta=6~meV$ (j), $\Delta=12.0 ~meV$ (k) and $\Delta=18.0 ~meV$
  (l). The trends are similar to that for $1.75^\circ$, with a $K_M$
  point gap closing at $\Delta=12.0 ~meV$ (k).}
\label{Fig:BandDispAB-BA}
\end{figure*}

The gap closings motivate us to look at the direct band gap between
the valence and conduction bands as a function of electrostatic
potential difference and twist angle. We define the direct band gap between the bands as
%
% \beq
% E_{dir} =min _k\left [\epsilon_1(k)-\epsilon_{-1}(k),0\right].
% \eeq
\beq
E_{dir} =min _\kk\left [\epsilon_1(\kk)-\epsilon_{-1}(\kk)\right].
\eeq
i.e. the minimum of the difference between the dispersions at each $\kk$
point in the first MBZ. In Fig.~\ref{Fig:DirBandGap}(a), we show a color plot of the
direct band gap of the system as a function of $\Delta$ and the twist
angle $\theta$. We see two distinct lines of zero direct gap in the
$\Delta-\theta$ plane: one starting at a finite $\Delta \sim 7 ~meV$
at $\theta =1^\circ$, increasing to $\Delta \sim 14~meV$ around
$2.1^\circ$, before going down at large angles. The second line starts
from $1.45^\circ$ at $\Delta=0$, rises sharply and almost merges
with the earlier line around $2.8^\circ$. These lines are clearly
visible as black contours in the color plot. We note that for angles
below $1.45^\circ$, the second line continues on the $\theta$ axis, as the
system is gapless in the absence of an electric field. This can also
be seen from the line-cuts along the $\Delta$ axis shown in
Fig.~\ref{Fig:DirBandGap}(b), where the gap at $\theta=1.3^\circ$ is
$0$ at $\Delta=0$. The twist angles can then be divided into three
regimes depending on the behaviour of the gap closing transitions: (i)
the
low angle regime, from $1^\circ$ to $1.45^\circ$, where there is no
gap at zero electric field. The gap increases with $\Delta$ before
coming down and closing at the $K_M$
point at a finite electric field.  A representative line cut at $\theta
=1.3^\circ$ is shown in Fig.~\ref{Fig:DirBandGap}(b) with solid black
line. In this case the gap closes at $\Delta=8.8~meV$. (ii) The intermediate angle range, from $1.45^\circ$ to
$2.8^\circ$, where the direct band gap is finite at $\Delta=0$, and
then goes through two gap closing transitions. A representative line
cut is shown at $\theta=1.75^\circ$ in ~\ref{Fig:DirBandGap}(b) with a
dashed red line. Here the gap closes first at $8.8~meV$, then rises
again before closing at $12.8~meV$. Beyond this point the gap rises once
again, as seen in the figure. (iii) A large angle regime from
$2.8^\circ$ onwards, where
the two transitions have come so close to each other, that for
all practical purposes, we will see it as a single
transition. This is shown in the dash-dotted blue line in
Fig.~\ref{Fig:DirBandGap}(b), corresponding to a line cut at
$5^\circ$, where the common gap closing happens around $8.1~meV$.
We note that the gap closing transitions are absent in the
particle-hole symmetric minimal model, where the valence and
conduction bands remain gapped at all finite electric fields at all angles (Fig.~\ref{abab_minimal}(a)).

We have
two distinct gap-closing transitions in the $\Delta-\theta$
plane. While the upper line of transition matches with the gap closing at the $K_M$
point, the lower line requires further investigations to understand
its origin. Clearly the gap does not close at one of the high
symmetry points, since the band dispersions of
Fig.~\ref{Fig:BandDispAB-AB} do not hint at a second gap closing
transition at finite electric fields. To understand this transition, we note that the Dirac
point in a standard bilayer graphene in an electric field splits into
a central Dirac cone and three satellite Dirac cones due to the
presence of the trigonal warping term
$\gamma_3$~\cite{RozhkovSboychakovRakhmanovNori2016} in the dispersion of a single
bilayer graphene. The associated Lifshitz
transitions and the resultant quantum Hall degeneracy
breaking~\cite{VMKBSTWFIE2015,ZTGHMZCSW2009} have been observed experimentally. A
similar situation is arising here. At intermediate angles, along the lower dark line in Fig.~\ref{Fig:DirBandGap}(a), the gap is
closing at the three satellite Dirac points around the $K_M$ point. The gap then opens up
again and closes at the central Dirac point ($K_M$ point) at a
larger value of $\Delta$, corresponding to the second
transition. Since the satellite Dirac points do not lie
on the high symmetry axes, this does not show up in
Fig.~\ref{Fig:BandDispAB-AB}.

To
see this clearly, in Fig.~\ref{Fig:Satellite_Dirac} (a) -(e), we show
a $3$ dimensional plot of the band dispersion around the $K_M$
point for a system with twist angle
$\theta =1.75^\circ$ for AB-AB stacked TDBLG, where the system has two clear gap-closing
transitions at $\Delta =8.8~meV$ and $\Delta=12.8~meV$ respectively. In
Fig.~\ref{Fig:Satellite_Dirac} (a), we show the situation at
$\Delta =5.0~meV$, before crossing the first transition point. The
three satellite Dirac like points in the conduction(valence) band have
a lower (higher)
dispersion, and consequently a lower momentum dependent direct gap, $E_{dir}(\kk) =
\epsilon_1(\kk)-\epsilon_{-1}(\kk)$, than the $K_M$ point. The
corresponding color plot of the  the momentum dependent gap $E(\kk)$ is
shown in  Fig.~\ref{Fig:Satellite_Dirac} (f), where it is clear that
the lowest gap is not at the $K_M$ point.
Fig.~\ref{Fig:Satellite_Dirac} (b) depicts the situation at
$\Delta =8.8~meV$ at the first transition point. The satellite Dirac
points in the conduction and valence bands touch each other driving
the direct band gap to zero, as seen in Fig.~\ref{Fig:Satellite_Dirac}
(g). As $\Delta$ is increased to $10.8~meV$ in Fig.~\ref{Fig:Satellite_Dirac}
(c), the satellite Dirac points
in the valence and conduction bands move away from each other, leading
to a finite gap in the system (see Fig.~\ref{Fig:Satellite_Dirac}
(h)). At the same time, the dispersions of the valence and the conduction band at the central Dirac point $K_M$ are moving towards each other in this regime of $\Delta$.
 In Fig.~\ref{Fig:Satellite_Dirac}
(d), at $\Delta=12.8~meV$, the two bands touch each other through a single Dirac cone at the $K_M$ point, leading to a second gap closing, as shown in
Fig.~\ref{Fig:Satellite_Dirac}(i). The satellite Dirac points are
gapped out at this value of $\Delta$. On increasing $\Delta$ further
to $15~meV$, both the satellite and the central Dirac point becomes
gapped, as seen in Fig.~\ref{Fig:Satellite_Dirac} (e) and (j).

The clear understanding of the mechanism of multiple transitions is a key new result in this paper. While the satellite Dirac points are also formed in a BLG in perpendicular electric field, they do not drive any transitions in that case, since the bands get further and further away from each other with increasing electric fields. This description of the mechanism of the transitions will also help us understand the topological changes that are occuring at these transitions in the following sections. We note that the trigonal warping plays a key role in forming the four Dirac points resulting in these transitions. This assertion will be strengthened later, when we systematically study the effects of different band parameters on the phase diagram.

  \subsection{AB-BA stacked TDBLG}

  We now turn our attention to the electric field dependent band
  structure of the AB-BA stacked twisted double bilayer graphene. The
  picture that emerges here is similar to the case of AB-AB stacked materials, with
  important differences at low twist angles.

  The band dispersion of the system at a low twist angle of
  $1.3^\circ$ is plotted in Fig ~\ref{Fig:BandDispAB-BA}(a)-(d) for $\Delta =0~meV$, $\Delta =7~meV$, $\Delta =10.6~meV$ and $\Delta =15~meV$
  respectively. The direct bandgap is finite at zero electric field
  in this case, in contrast to the situation for the AB-AB stacked
  TDBLG. The band gap at the $K_M$ point gradually goes down with
  increasing $\Delta$ till it closes around $\Delta \sim
  10.6~meV$. Beyond this, the gap opens up again as seen at
  $\Delta=15~meV$ in Fig ~\ref{Fig:BandDispAB-BA} (d). Fig
  ~\ref{Fig:BandDispAB-BA} (e) -(h) shows the band structure with
  increasing $\Delta$ for $\theta=1.75^\circ$, while Fig
  ~\ref{Fig:BandDispAB-BA} (i) -(l) shows this for
  $\theta=3.0^\circ$. The trends are similar with the $K_M$ gap
  closing around $\Delta=13.2~meV$ for $1.75^\circ$ and around
  $\Delta=12~meV$ for $3^\circ$.

  The direct bandgap for the AB-BA stacking is shown as a color plot
  in the $\Delta-\theta$ plane in Fig.~\ref{Fig:DirBandGap}(c). The
  trends are very similar to that of the AB-AB stacking, with the
  exception that the band gap is finite at low twist angles in zero
  electric field. The twist angles can once again be divided into
  three regions based on the qualitative dependence of the direct band
  gap with $\Delta$, similar to the case of AB-AB stacking. This is
  shown in Fig.~\ref{Fig:DirBandGap}(d) with three representative
  line-cuts  at $\theta=1.3^\circ$ (solid black line), $1.75^\circ$
  (red dashed line)  and $5.0^\circ$ (blue dash-dotted line). 
  
  While the general behaviour of the gap in the $\Delta-\theta$ plane is similar for AB-AB and AB-BA stacking, there is a crucial difference at low twist angles. While the AB-AB stacked TDBLG has zero direct band gap at zero electric field, the AB-BA stacking shows a finite band gap even at zero electric field in this regime.

  \begin{figure}[t]
    \includegraphics[width=\columnwidth]{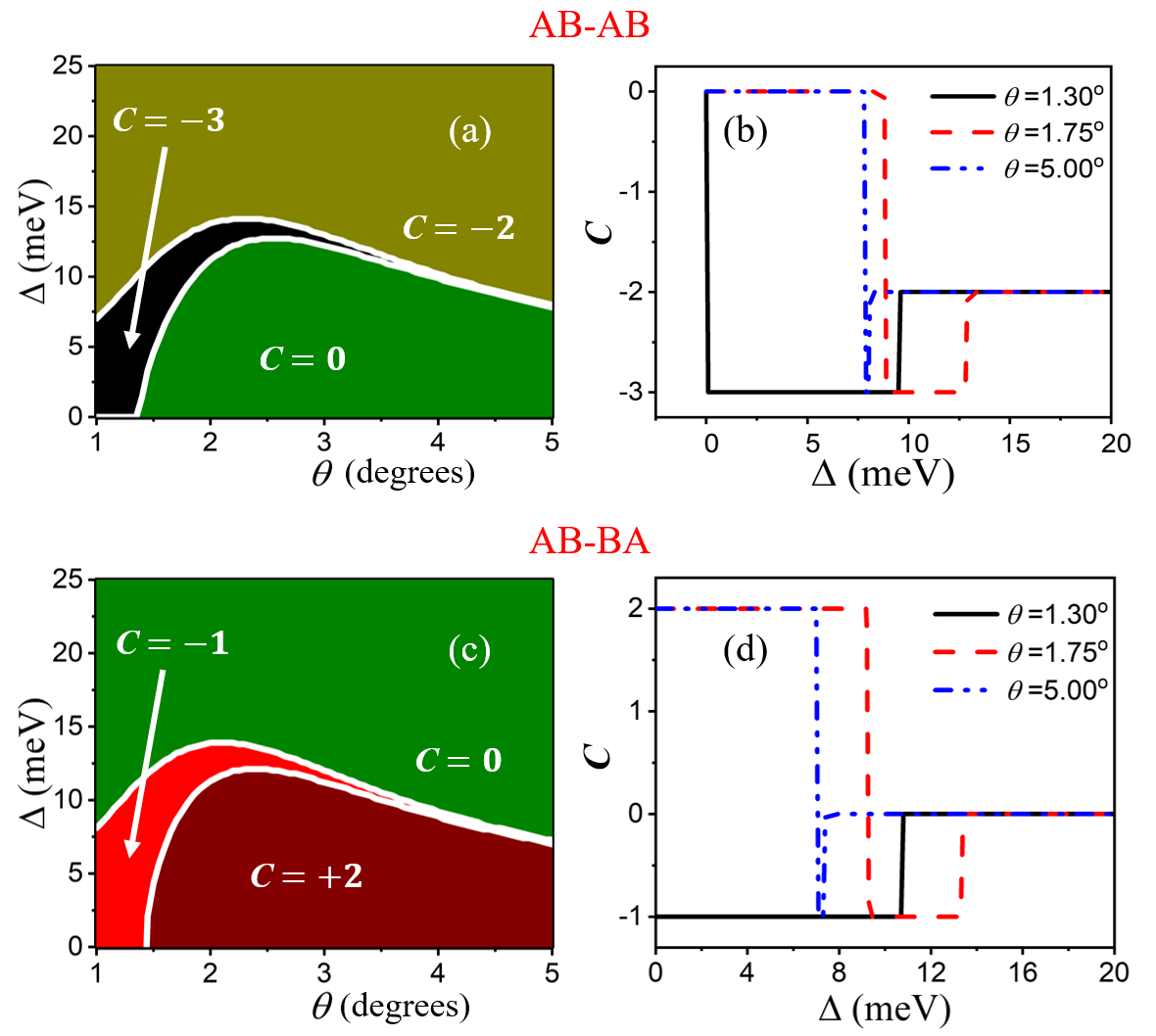}
   \caption{(a) and (b): Topological phase diagram for AB-AB stacked
     TDBLG. (a) Color plot showing the Chern number $C$ of the valence band in
     the $\Delta-\theta$ plane. Note the two transitions corresponding
     to a jump in $C$ of $-3$ and $+1$. (b) Line cuts at fixed
     $\theta=1.3^\circ$(solid black line), $1.75^\circ$ (dashed red
     line) and $5^\circ$ (dot-dashed blue line), showing the different
     behaviour of the variation of the Chern number with
     $\Delta$. These three angles respectively correspond to the low, intermediate
     and high twist angle regimes mentioned in the text. (c) Color
     plot of the valence band Chern number and (d) line cuts
     corresponding to three different regimes of twist angle for the
     AB-BA stacked TDBLG. Note that the Chern number jumps from $0$ to
     $-3$ to $-2$ in (a) and (b), while it jumps from $2$ to $-1$ to
     $0$ in (c) and (d) for intermediate angle regime.}
   \label{Fig:Chernphasediag}
  \end{figure}

\section{Topological transitions and phase diagram}
\label{chern}  

The closing of band-gaps without any associated change in symmetry of
the system in twisted double bilayer graphene points to the
possibility that these gap closings are associated with transitions in
topological character of the bands. We now focus on the topological phase diagram of the system by studying the Chern numbers of the bands in the $\Delta-\theta$ plane

In TDBLG, due to the possible band 
crossings, the total Berry phase has to be constructed by including
all bands upto a certain band, with the highest band index being
$n$. The matrix-valued Berry connection $\vec{\mathcal{A}}^n({\bf k})$ and
the corresponding Berry curvature $\vec{\Omega}^n({\bf k})$ are given
by
\bqa
\nonumber \vec{\mathcal{A}}^n_{\alpha\beta}(\kk)&=& i \langle \psi_\alpha({\bf
  k})|\nabla_{{\bf k}}\psi_\beta({\bf k})\rangle\\
\vec{\Omega}^n_{\alpha\beta}({\bf k}) &=& \nabla_{{\bf k}} \times
\vec{\mathcal{A}}^n_{\alpha\beta}(\kk),
\label{Eq:Berry}
\eqa
where $|\psi_\alpha({\bf k})\rangle $ is the Bloch wavefunction of the
corresponding band, and $\alpha$, $\beta$ run over band indices upto
$n$. The total Chern number of the $n$ bands~\cite{Hatsugai2005} is then defined by 
\beq
C_n=  \frac{1}{2\pi}\int_{ MBZ} d^2 {\bf k}
\mathrm{Tr}~\left[\vec{\Omega}^n(\kk) \cdot \hat{z}\right].
\label{Eq:Chern}
\eeq
% \beq
% \Omega^{n}_{\alpha\beta}(\kk)= i \langle n_\alpha|\nabla_k|n_\beta\rangle,
% \label{Eq:Berry}
% \eeq
% where $n$ denotes the number of bands being considered to calculate the Berry connection, starting from the lowest one.
% The indices $\alpha$ and $\beta$
% run over all bands which are lower in energy than the $n$th band, including the
% band with index $n$.  
This is well defined when the next higher band dispersion is at a
finite direct gap from the band being considered. Thus, in our phase
diagram, this can be uniquely defined for the valence/conduction band at all points other than at the points where the direct gap closes. 
% The determinant of this matrix, integrated over
% the \moire Brillouin zone, gives the total Chern number of all the bands. 
% The Chern number of the band,  can be
% calculated for both the valence and conduction band, except along the direct
% gap closing lines.
Note that this is the Chern number for the mini-band
around the $K$ valley of the original graphene Brillouin zone. One can
do a similar construction for the bands around the $K'$ valley of the
graphene Brillouin zone.

\subsection{AB-AB stacked twisted double bilayer graphene}
 \begin{figure}[t]
    \includegraphics[width=\columnwidth]{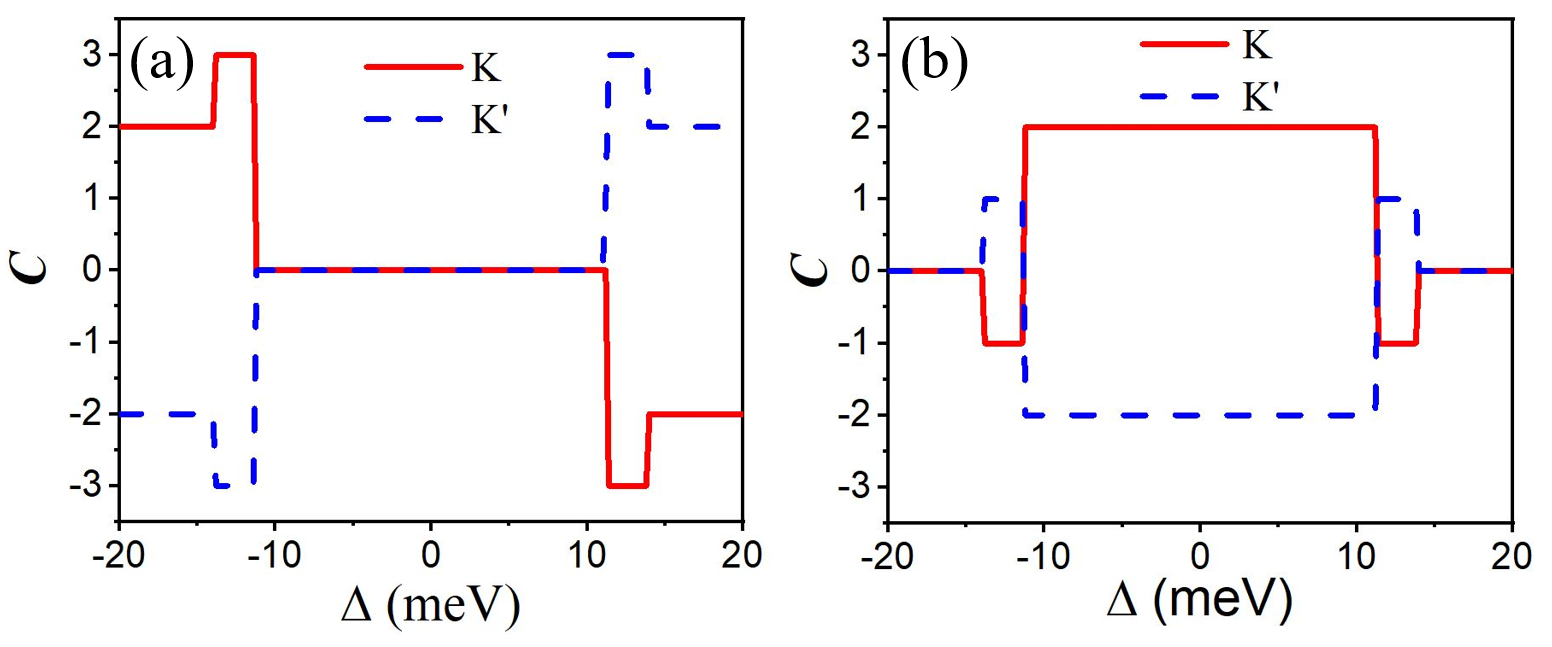}
   \caption{Chern number of the valence band in
     $K$ and $K'$ valleys as a function of $\Delta$ for
     $\theta=1.75^\circ$ in (a) AB-AB stacked and (b) AB-BA stacked
     TDBLG. Note that the Chern number in $K$ and $K'$ valleys are
     negative of each other. Also note that $C$ is antisymmetric under
     $\Delta \rightarrow -\Delta$ in (a), while it is symmetric under
     the same transformation in (b).}
   \label{Fig:ChernSym}
  \end{figure}

Let us first consider the Chern number of the valence band of an AB-AB
stacked TDBLG around the $K$ valley of the original graphene Brillouin
zone. The topological phase diagram (or the Chern number as a function
of $\Delta$ and $\theta$) is plotted as a color plot in
Fig.~\ref{Fig:Chernphasediag}(a). At low twist angle, the band is
trivial exactly at $\Delta=0$. At an infinitesimal $\Delta$, it has a Chern
number of $-3$. As $\Delta$ is increased, the Chern number jumps to
$-2$ as soon as one crosses the gap closing point. This is clearly
seen in the line-cut along the $\Delta$ axis at a fixed
$\theta=1.3^\circ$ in Fig.~\ref{Fig:Chernphasediag}(b) ( black solid
line). At intermediate angles, where the system undergoes two gap
closing transitions, the band is trivial as $\Delta$ is increased,
until one hits the first transition, when the Chern number jumps to
$-3$, and then it jumps back to $-2$, when the second transition
point is crossed. This is clearly seen in the   line-cut along the $\Delta$ axis at a fixed
$\theta=1.75^\circ$ in Fig.~\ref{Fig:Chernphasediag}(b) (red dashed
line). Finally, at  large twist angle these two transitions are so
close that for all practical purposes, the Chern number jumps from $0$
to $-2$. However, even at an angle of $5^\circ$, there are actually
two jumps, one from $0$ to $-3$ and the other from $-3$ to $-2$, as
seen in the line cut in Fig.~\ref{Fig:Chernphasediag}(b) ( blue dot-dashed
line), although the small regime of $\Delta$ between the two jumps would make it impossible to see this in experiments.
 We note that the Chern number for the conduction band is the
negative of the Chern number of the valence band for AB-AB stacked
TDBLG{\cite{ChebroluChittariJung2019}}, and hence a similar looking phase diagram can be constructed by considering the Chern number of the conduction band.

\begin{figure}[t]%
 \includegraphics[width=\columnwidth]{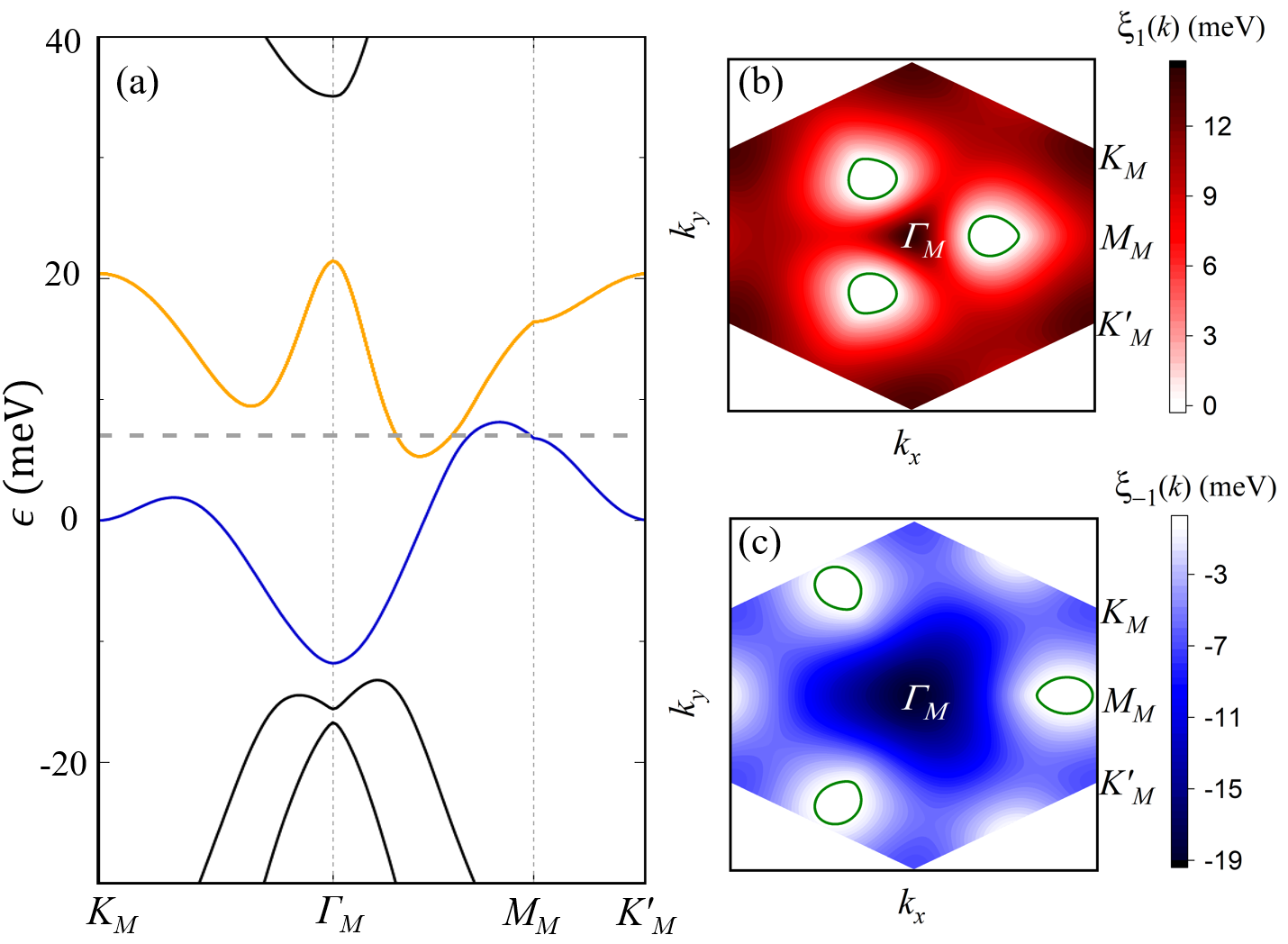}
 \caption{(a) The band structure of AB-BA stacked TDBLG for $\theta=1.33^\circ$ for potential difference $\Delta=0$. The bands with dispersion
 $\epsilon_1(\kk)$ (conduction) and $\epsilon_{-1}(\kk)$ (valence) plotted in orange and blue colors respectively, while $\epsilon_F$
 is denoted by a the grey dashed line. Note that the Fermi level cuts
 both conduction and valence bands at multiple places. (b) The color
 plot of the conduction band dispersion measured from the Fermi level,
 $\xi_{1}(\kk)=\epsilon_1(\kk)-\epsilon_F$. The three white pockets denote filled portions of the band where $\epsilon_1<\epsilon_F$. (c) 
 The color plot of the valence band dispersion measured from the Fermi
 level $\xi_{-1}(\kk)=\epsilon_{-1}(\kk)-\epsilon_F$. The three white pockets denote areas of the band with $\epsilon_{-1}>\epsilon_F$.
As shown here, the partial filling of bands and resultant metallic nature can occur even when there is a finite direct band gap 
 between $\epsilon_1(\kk)$ and $\epsilon_{-1}(\kk)$ at all points in the \moire Brillouin zone. }
 \label{Fig:Bandoverlap} 
\end{figure}  

The change in the Chern number of the bands can be understood from the
picture of $4$ Dirac cones ( a central one at $K_M$ point and three
satellite cones around it) in TDBLG, which drove the two gap closing transitions in the system. The satellite Dirac cones each carry a chirality index of
$-1$, while the central Dirac cone carries an opposite chirality index
of $+1$ (for the valence band). As the electric field is increased at intermediate and high
twist angles, the first gap closing transition occurs when the
bands touch at the satellite Dirac points. The presence of three gap
closing points with a chirality index of $-1$ (i.e. a Berry monopole
of strength $-3$ at the gap closing point) explains the jump in the
Chern number by $-3$. The second transition, which is mediated by the
band touching at the central Dirac cone at $K_M$ point, corresponds to
a Berry monopole of $+1$, and hence the Chern number jumps by $+1$ at
this transition. At low twist angles, as we have said before, the
first transition is shifted to infinitesimally small electric fields,
and hence the band has a Chern number of $-3$ at low electric fields,
before hitting the transition mediated by band touching at the central
Dirac point, with the Chern number jumping to $-2$.

We would like to note that the above Chern numbers are calculated for
bands around the $K$ valley. Similarly, bands around the $K'$ valley have Chern numbers, which are negative of the $K$ valley
Chern numbers. This is shown in Fig.~\ref{Fig:ChernSym}(a), where the
Chern number is plotted as a function of $\Delta$ for $\theta=1.75^\circ$. Since
these bands are degenerate, there is no net Chern number of the
system. However, the system will show anomalous valley Hall effect if the
valley degeneracy is broken. 
%%%%%%%%%%%%%%repeaated later%%%%%%%%%%%%%%%%%%%%%%%%%%%%%%%%
% The valley Chern number can be measured
% in measurements of non-local resistance~\cite{Roth294}or in the degeneracy pattern
% of the corresponding Landau levels in presence of magnetic
% fields~\cite{LLExpt} or by considering how the band gap closes in
% presence of external magnetic fields~\cite{Bernevig}.
%%%%%%%%%%%%%%%%%%%%%%%%%%%%%%%%%%%%%%%%%%%%%%%%%%%%%%%%%%%%%
Fig.~\ref{Fig:ChernSym}(a) also shows that for AB-AB stacked TDBLG, the  Chern number of the bands
in each valley change sign if the direction of the electric field is reversed. 

We thus obtain a complete understanding of the Chern number jumps in
the AB-AB stacked TDBLG in terms of the splitting of the bilayer Dirac
point into a central Dirac point and satellites due to trigonal warping.

\subsection{AB-BA stacked twisted double bilayer graphene}

While the basic picture of gap closing transitions is very similar in
AB-AB and AB-BA stacked TDBLG, they differ quite a bit when it comes
to the topological phase diagram. For the AB-BA stacking, we once
again focus on the Chern
number of the valence band around the $K$ point in the graphene
Brillouin zone, and construct a phase diagram in the $\Delta-\theta$
plane based on the Chern number of the valence band. In this case,
contrary to the AB-AB stacking, the Chern number of the valence and
the conduction bands are same~\cite{ChebroluChittariJung2019}, and hence a similar
phase diagram would be obtained if the Chern number of the conduction
band is studied.

The Chern number of the AB-BA stacked TDBLG is shown in a color plot
in the $\Delta-\theta$ plane in Fig.~\ref{Fig:Chernphasediag} (c). At intermediate twist angles, the
valence band has a Chern number $2$ at zero electric field. This is
different from the AB-AB stacking, where the band is trivial in this regime. As
$\Delta$ is increased and hits the first gap-closing transition, the
Chern number jumps to $-1$. Beyond the second gap-closing transition,
the Chern number jumps to $0$, so that the band is trivial at large
electric fields. This is clearly seen in the representative line cut
at $\theta=1.75^\circ$ in Fig.~\ref{Fig:Chernphasediag} (d). At large
angles, the regime where the Chern number is $-1$ shrinks to zero, and
it seems that the Chern number jumps from $2$ to $0$ directly, as seen
in the line cut at $\theta=5^\circ$ in Fig.~\ref{Fig:Chernphasediag}
(d). Finally, at low angles below $1.45^\circ$, the valence band has a
Chern number of $-1$ at zero electric field, which jumps to $0$ at the
gap closing transition. Contrary to the case of AB-AB stacked model, here
the direct gap does not close at $\Delta=0$ for small angles.

Although the individual Chern numbers and the
topological phase diagram for AB-BA stacked TDBLG is quite different
from that of the AB-AB stacked TDBLG, it is important to note that the
change of Chern numbers of $-3$ and $+1$ across the two transitions is
same for these two stacking. This can once again be explained in
terms of $3$ satellite Dirac points and a central Dirac point with
opposite chirality indices. The first transition takes place at the
satellite Dirac points, with a Chern number change of $-3$, while the
central Dirac point causes a jump of $+1$ at the second transition.

Once again, we note that the the Chern numbers of the band in the
$K'$ valley is negative of the Chern number of the same band in the
$K$ valley, as seen from Fig.~\ref{Fig:ChernSym}(b). Unlike the case of
AB-AB stacking, in this case we find that the Chern numbers are
invariant under the change of direction of the electric field. This is
because the AB-BA stacking looks the same whether one looks from the
top or from the bottom and hence the Chern numbers are invariant under
the reversal of the electric field.

\section{The Observable Phase Diagram\label{observable}}
\begin{figure}[t]%
 \includegraphics[width=\columnwidth]{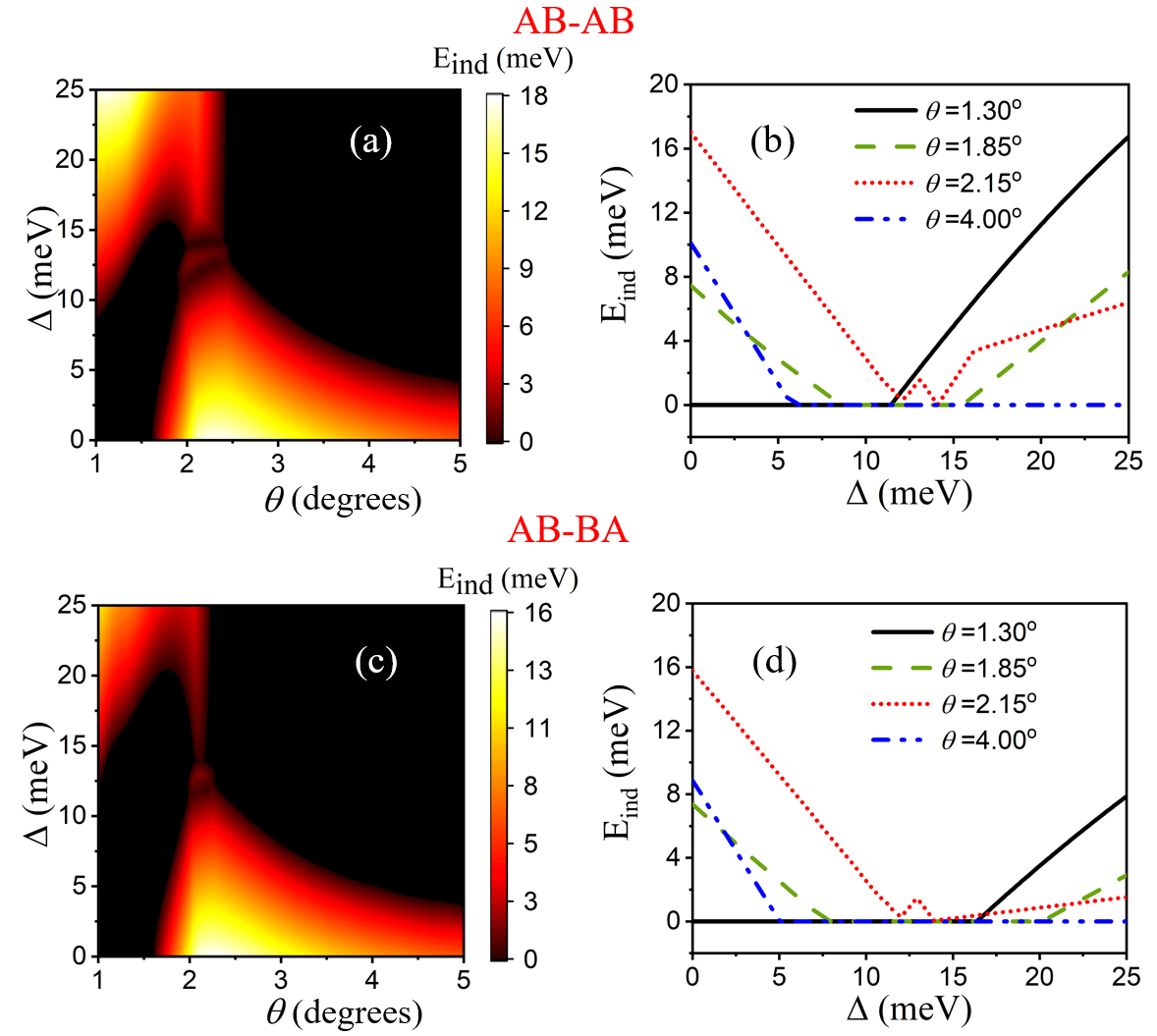}
 \caption{(a): A color plot of the indirect band gap $E_{ind}$ between the
   conduction and valence bands in the twist angle
   ($\theta$)-potential difference ($\Delta$) plane for the AB-AB
   stacked twisted double bilayer graphene. The black regions, where
   $E_{ind}$ vanishes, corresponds to metallic states with Fermi surfaces. (b) Representative line cuts showing the variation
of the indirect band gap $E_{ind}$ with $\Delta$ at fixed $\theta$ for
AB-AB stacking. The values of $\theta=1.3^\circ$ (solid black line),
$1.85^\circ$ (green dashed line), $2.15^\circ$ (red dotted line) and
$4.0^\circ$ (blue dot-dashed line) correspond respectively to the four
regimes defined in the text with distinct variation of $E_{ind}$ with
$\Delta$. (c) Color plot and (d) line cuts of $E_{ind}$ for the AB-BA
stacked TDBLG. It shows similar trends as the AB-AB stacked TDBLG.}
 \label{Fig:IndBandGap} 
\end{figure}
The theoretically calculated band gaps and the Chern numbers provide
an useful starting point to understand the phenomenology of TDBLG;
however they alone are not sufficient to even qualitatively describe
the experimental observations seen in this system. For example, one would expect
from the calculated phase diagram that the AB-AB stacked TDBLG around
the low twist angle of $1.1^\circ$ 
would be gapped at low electric fields with the gap closing and
reopening around the first transition point. However, it has now been
established beyond doubt by several experiments ~\cite{BZTWMT2019,AdakSinhaGhoraiSanganiVarmaWatanabeTaniguchiSensarmaDeshmukh2020},
that at the charge neutrality point, this system remains a metal till
a reasonably large electric field is applied.

The key reason for the mismatch between the experiments and the
calculations of the previous sections is the following: While the
dispersions of the valence and conduction bands are separated from
each other at each $\kk$ point, there are parts of the phase diagram
where the bands as a whole overlap in energy; i.e. the dispersion of
the conduction band at some momentum can be lower than the dispersion
of the valence band at some other momentum. This situation is shown in
Fig.~\ref{Fig:Bandoverlap} (a) for $\theta=1.33^\circ$ and $\Delta
=0~meV$ for AB-BA stacked TDBLG. At the charge neutrality point, the Fermi energy then
passes through both bands (shown as a dashed line in
Fig.~\ref{Fig:Bandoverlap} (a)), leading to creation of Fermi surfaces
and resultant metallic behaviour of the system.  Note that the AB-BA stacked TDBLG 
has a direct band band gap and therefore a well-defined Chern number $(C=-1)$ even at $\Delta=0~meV$. To get a complete picture of the Fermi surfaces in the \moire Brillouin zone, we consider a color plot of the dispersion of the conduction band, as measured from the
Fermi energy, $\xi_1(\kk)=\epsilon_1(\kk)-\epsilon_F$ in
Fig.~\ref{Fig:Bandoverlap} (b). The red regions are regions with
$\xi_1(\kk) >0$, which remain unoccupied at the charge neutrality point,
while the three white blobs surrounded by the thick lines have
$\xi_1(\kk) <0$, i.e. they are occupied by electrons, giving rise to the
three small electron pockets in the conduction band. The corresponding
Fermi surfaces are shown by the thick green lines. Similarly, in
Fig.~\ref{Fig:Bandoverlap} (c), we plot the dispersion of the valence
band, as measured from the Fermi energy,
$\xi_{-1}(\kk)=\epsilon_{-1}(\kk)-\epsilon_F$. Here the blue regions have
$\xi_{-1}(\kk) <0$, i.e. they are occupied by electrons. The three white
regions surrounded by thick lines have $\xi_{-1}(\kk) >0$, leading to
these three hole pockets in the valence band. The Fermi surfaces are
shown by the thick green lines.

In this paper, we will solely focus on the situation at the charge
neutrality point. The extension to finite carrier densities will be
taken up in a later work. In this case, given the energy overlap of the bands in some parameter regimes, it
useful to define an indirect band gap

\beq
E_{ind}=max[  min_\kk [\epsilon_{1}(\kk)]-max_k[\epsilon_{-1}(\kk)], 0].
\eeq
The indirect band gap correctly predicts a gapless system when there are Fermi surfaces due to band overlaps. Further, due to the small size of the \moire Brillouin zone in the angle range we are considering, one can expect disorder potentials to scatter electrons across the \moire Brillouin zones. Hence, the system would be insulating only when the indirect band gap is finite, although the value of the transport gap can be lower than the single particle gap, as has been previously seen in bilayer graphene in the presence of electric field~\cite{MinAbergelLHwangSarma2011,Oostinga_2007,Ohta951}.

\begin{figure}[t]%
 \includegraphics[width=\columnwidth]{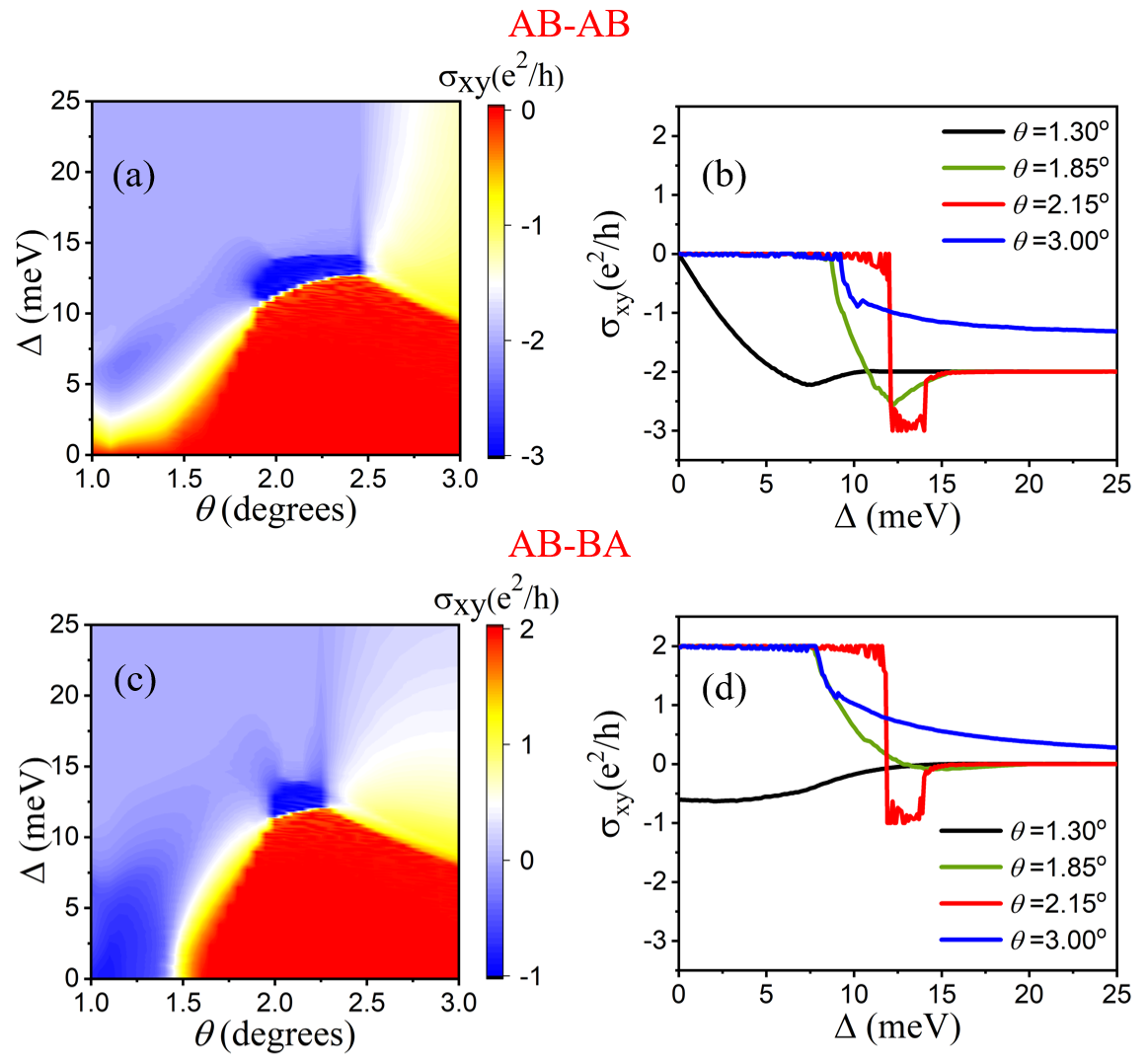}
 \caption{(a) A color plot of the anomalous valley-Hall conductivity in the twist angle
   ($\theta$)-potential difference ($\Delta$) plane for the AB-AB
   stacked TDBLG. The yellow and white regions corresponds to metallic regions
   while the the red, light and dark blue regions mark insulator phases with quantized values 
   $\sigma_{xy}$ equal to $0$, $-2$ and $-3$ respectively. 
(b) Line cuts of $\sigma_{xy}$ as function of $\Delta$ for different values of twist angles $\theta$ for
AB-AB stacking. The values of $\theta=1.3^\circ$ (black line),
$1.85^\circ$ (green line), $2.15^\circ$ (red line) and
$3.0^\circ$ (blue line) correspond respectively to the four
regimes where $\sigma_{xy}$ vs $\Delta$ has dissimilar behaviors. (c)
Color plot of $\sigma_{xy}$ in the 
$\theta$-$\Delta$ plane for AB-BA stacked TDBLG. Here too, the yellow and white regions are metallic phases with non-quantized values while 
red, light and dark blue regions have $\sigma_{xy}$ equal to $2$, $0$ and $-1$ respectively. (d) Line cuts of $\sigma_{xy}$
as a function of $\Delta$ for the AB-BA stacked TDBLG for twist angle $\theta$ values equal to $1.3^\circ$, $1.85^\circ$, $2.15^\circ$ and $3.0^\circ$.}
 \label{Fig:AQHE} 
\end{figure}

The indirect band gap for the AB-AB stacked TDBLG is shown in a color
plot in the $\Delta-\theta$ plane in Fig.~\ref{Fig:IndBandGap} (a). The twist angles can now be divided into $4$ distinct regions based on
the behaviour of $E_{ind}$ with $\Delta$. At low twist angles ( $1^\circ$ to $1.7^\circ$), the system is gapless at low electric fields. Beyond a threshold value of $\Delta$, the indirect gap is finite and the system will show insulating behaviour. This matches with experimental data on these systems near magic angles ~\cite{LHKLRYNWTVK2020,Cao_2020,BZTWMT2019, AdakSinhaGhoraiSanganiVarmaWatanabeTaniguchiSensarmaDeshmukh2020}.
 A representative plot of $E_{ind}$ as
function of $\Delta$ in this regime for a fixed
$\theta=1.3^\circ$ is shown in Fig.~\ref{Fig:IndBandGap} (b) with solid
black line. The metallic regime continues till $\Delta \sim 12~meV$,
beyond which the gap rises steadily with $\Delta$ in the insulating
phase. Beyond a critical twist angle of $\sim
1.7^\circ$, the gap is finite at zero electric field.  In this second
regime, the indirect gap closes
to form a metallic region as $\Delta$ is increased, before opening up
again in the high field insulating regime. This behaviour has been
seen in some experiments on samples with a twist angle around
$2^\circ$ ~\cite{VZPZMKTWMEIR2020,LHKLRYNWTVK2020}. A representative plot of $E_{ind}$ as
function of $\Delta$ in this regime for a fixed
$\theta=1.85^\circ$ is shown in Fig.~\ref{Fig:IndBandGap} (b) with dashed
green line. The indirect gap closes at $\Delta \sim 8~meV$, and the
metallic regime continues till $\Delta \sim 16~meV$, beyond which the
gap increases with increasing $\Delta$. In the third regime, there is
a small range of twist angles between $2^\circ$ and $2.7^\circ$, where
the two gap closing transitions can be clearly seen in
Fig.~\ref{Fig:IndBandGap} (a). A representative plot of $E_{ind}$ as
function of $\Delta$ in this regime for a fixed
$\theta=2.15^\circ$ is shown in Fig.~\ref{Fig:IndBandGap} (b) with dotted
red line. The gap closes for the first time around $\Delta =12~meV$,
opens up again and then closes at the second transition at around
$14~meV$. Beyond this, we enter the high field insulating
state. Finally, there is the fourth regime at high twist angles beyond
$\sim 2.7^\circ$, where there is a finite gap at low $\Delta$, which closes to form a metallic
state beyond a threshold $\Delta$ value. There is no signature of an
insulating state at these angles at high electric fields (upto a $\Delta$ of $25~meV$). A
representative plot of $E_{ind}$ as a
function of $\Delta$ in this regime for a fixed
$\theta=4.0^\circ$ is shown in Fig.~\ref{Fig:IndBandGap} (b) with dash-dotted
blue line.
\begin{figure}[t]
\includegraphics[width=\columnwidth]{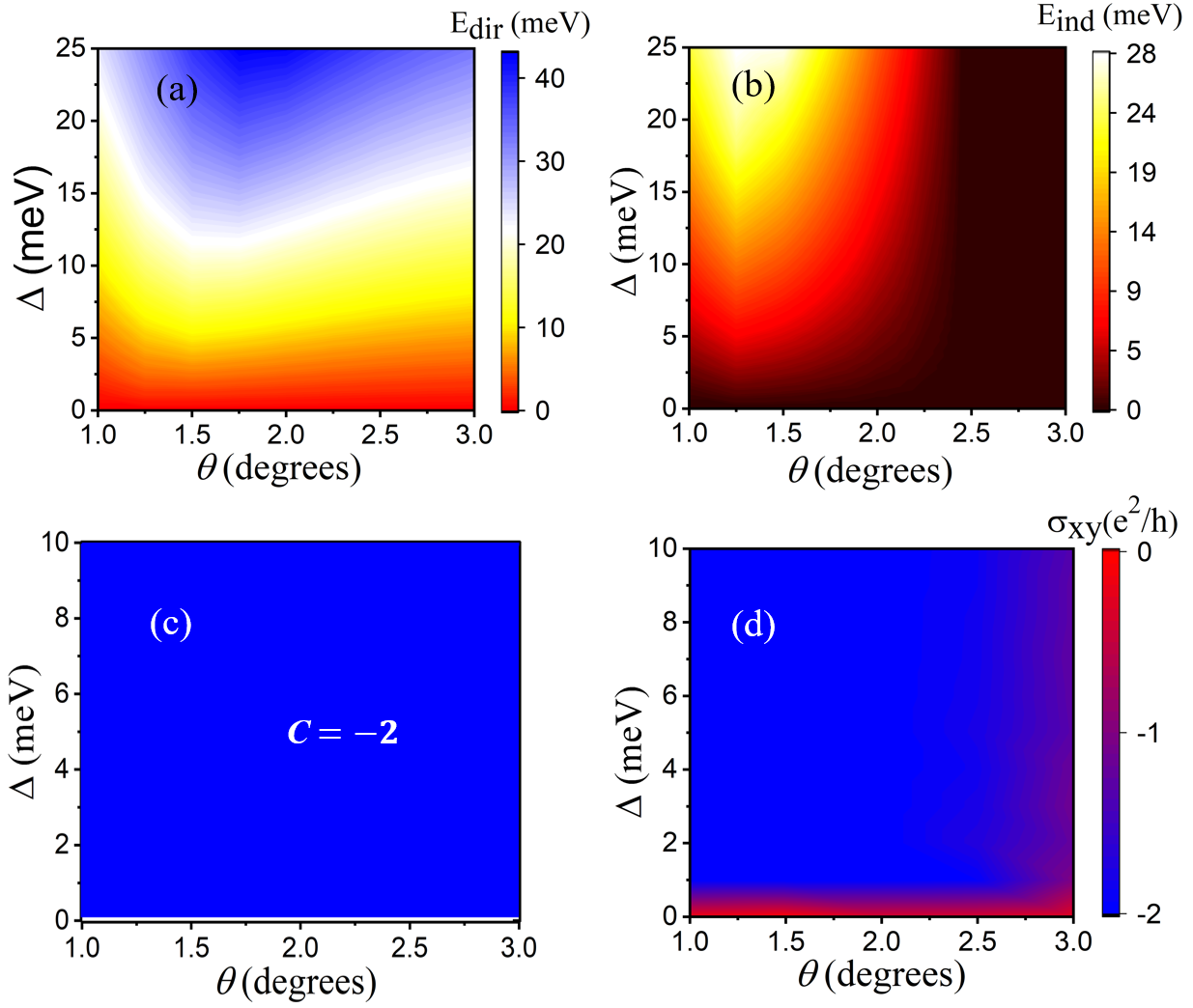}
\caption{ (a)The direct ($\mathrm{E}_\mathrm{dir}$) band gaps for AB-AB stacked TDBLG minimal
model. The band gap increase with increasing $\Delta$ for any fixed value of twist angle $\theta$. (b)  indirect ($\mathrm{E}_\mathrm{ind}$) band gap
is non-zero for $\theta<2.5^\circ$. (c) The Chern number $C=-2$ for all non-zero values of $\Delta$. (d) The valley-Hall conductivity $\sigma_{xy}=-2$
 for all the regions with a nonzero $\mathrm{E}_\mathrm{ind}$. This
 shows that the minimal model only supports a single topological
 insulating phase and there are no transitions as a function of
 electric field at fixed twist angles in this model.}
\label{abab_minimal}
\end{figure}

The indirect bandgap for the AB-BA stacked TDBLG is shown in a color
plot in the $\Delta-\theta$ plane in Fig.~\ref{Fig:IndBandGap} (c). The
phase diagram is qualitatively similar to the phase diagram for the
AB-AB stacking, although the exact numerical values of the critical
twist angle beyond which the system is insulating at zero electric
field, or the numerical value where it reaches the high field
insulating state is slightly different. One notable feature is that
the third regime, where the two transitions are clearly visible has a
shorter spread in this case than the AB-AB stacked TDBLG. The
representative line cuts showing the variation of $E_{ind}$ with
$\Delta$ for the four different regimes are shown in
Fig.~\ref{Fig:IndBandGap} (d) for $\theta=1.30^\circ$, $1.85^\circ$,
$2.15^\circ$ and $4.0^\circ$ respectively. They show behaviour similar
to that of the AB-AB stacked TDBLG.

\begin{figure}[t]
\includegraphics[width=\columnwidth]{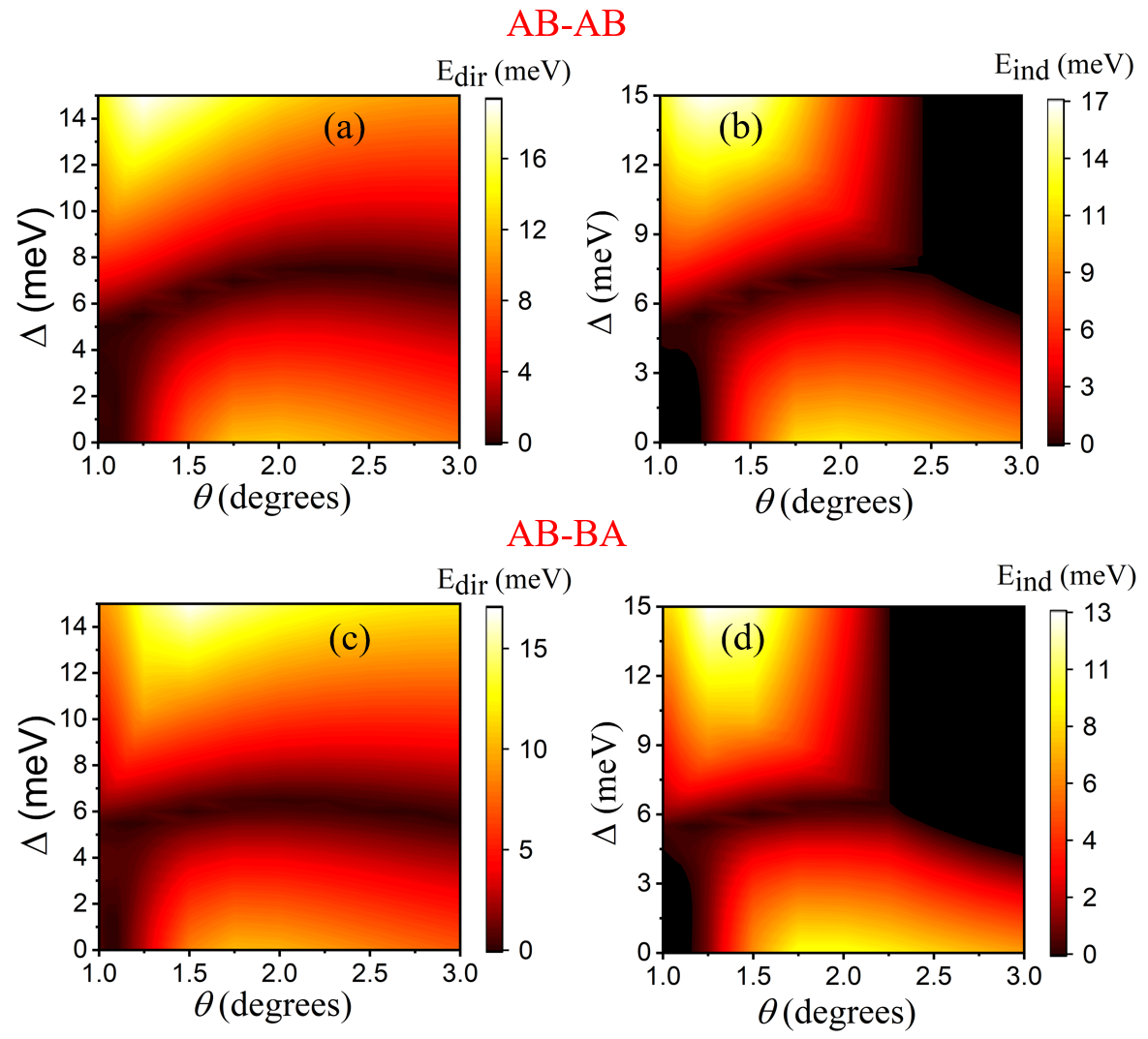}
\caption{(a) The direct band gap ($\mathrm{E}_\mathrm{dir}$) and (b) the indirect band gap ($\mathrm{E}_\mathrm{ind}$) for AB-AB
  stacked TDBLG are plotted in the $\Delta-\theta$ plane. Here we have used the full model but setting only the trigonal warping $\gamma_3$ to zero. In (a), the two dark gap closing lines of
FIG. \ref{Fig:DirBandGap} (a) have collapsed to one. (b) The indirect
band gap ($\mathrm{E}_\mathrm{ind}$) differs 
from FIG. \ref{Fig:IndBandGap}(a) by the absence of the intermediate
gapped phase in the twist angle range
$\sim 2.0^\circ$ to $\sim 2.7^\circ$. The direct ($\mathrm{E}_\mathrm{dir}$) and indirect ($\mathrm{E}_\mathrm{ind}$) band gaps
for AB-BA stacked TDBLG with $\gamma_3=0$ are plotted in (c) and (d) respectively. In (c) and (d), the band gap behaviors are 
similar to that of in AB-AB stacked full model without trigonal warping.}
\label{gap_t30}
\end{figure}
The indirect bandgap provides us with a measure that will
qualitatively if not quantitatively track the transport gap across the
phase space. One similarly needs a measure for observable effects of
the Berry curvature of the bands. To this end, note that the Berry
curvature for each $\kk$ defined in Eq.~\ref{Eq:Berry} 
is a well defined quantity
except at direct gap closing points. When the chemical potential runs through
the bands due to their overlap in energy space, the index of the
highest occupied
band $n(\kk)$ varies from one momentum point to another. We can then calculate
the curvature matrix for different $\kk$ points by taking this into
consideration, and the trace of this matrix, integrated over the
\moire Brillouin zone, gives the intrinsic valley-Hall conductivity of
the system at $T=0$,
%
% \beq
% \sigma_{xy} =\frac{e^2}{h} \sum\limits_{k \in MBZ} \textrm{Det} ~\hat{\Omega}^{n(\kk)}(\kk)\Theta(\epsilon_F-\epsilon_{n}(\kk)), 
% \label{Eq:sigmaxy}
% \eeq
\beq
\sigma_{xy} =\frac{e^2}{2\pi h}  \int_{ MBZ} d^2 {\bf k}
\mathrm{Tr}~\left[{\vec{\Omega}^{n(\kk)}(\kk)} \cdot \hat{z}\right] .
\label{Eq:sigmaxy}
\eeq
Once
again we focus on the charge neutrality point in the system. This
valley Hall conductivity can be measured through non-local resistance
measurements~\cite{Roth294,sinha2020bulk} or by looking at the pattern of Landau fan diagram on
applying a magnetic field~\cite{BurgLianTaniguchiWatanabeBernevigTutuc2020,WuLiuYazyev2020}. In
the insulating phases, where the indirect band gap is non-zero and the
chemical potential is in the band gap, $\sigma_{xy}$ (in units of
$e^2/h$) is quantized and is given by the Chern number of the
band. In
the metallic regimes, the answer depends on the details of the Fermi
surfaces and $\sigma_{xy}$ is not quantized in general.

The intrinsic valley Hall conductivity $\sigma_{xy}$ is shown in the
$\Delta-\theta$ plane as a color plot in Fig.~\ref{Fig:AQHE}(a) for
the AB-AB stacked TDBLG. The insulator at high fields and relatively
lower twist angles, $1^\circ$ to $\sim 1.7^\circ$(first regime),  
has a quantized value of $\sigma_{xy}= -2$, shown in light blue. This steadily rises
to zero (red) as the electric field is decreased through the metallic
phase. For larger values of twist angle ($\theta > 1.7^\circ$), the insulator at low electric fields is
trivial. In the second regime ($\sim 1.7^\circ$ to $\sim 2.0^\circ$), as we increase $\Delta$, $\sigma_{xy}$ crosses
 the metallic region to reach $-2$ in the high field insulator
phase. In the third regime, $\sim 2.0^\circ$ to $\sim 2.7^\circ$, we see that $\sigma_{xy}$ jumps from $0$ (red)
to $-3$ (dark blue) to $-2$ (light blue), as we cross the two transitions, while in the fourth
regime ($\sim 2.7^\circ$ onwards), $\sigma_{xy}$ decreases as we enter the metallic phase, but
does not reach a quantized value. Note that the streaks of yellow color,
corresponding to $\sigma_{xy}=-1$ do not indicate a quantized phase, this is
simply part of $\sigma_{xy}$ steadily decreasing in the metallic phase
as $\Delta$ is increased. To illustrate this, in Fig.~\ref{Fig:AQHE}(b), we show the line-cuts of
$\sigma_{xy}$ as function of $\Delta$ for: (i) $\theta=1.3^\circ$ (black
line), where it smoothly changes from a metallic phase at $\Delta=0$ to a quantized value of $-2$ for $\Delta \gtrsim 10~meV$. 
(ii) $\theta=1.85^\circ$ (green
line), where it remains quantized with $\sigma_{xy}=0$ upto a finite value of $\Delta\sim
8.0~meV$, beyond which the system enters the metallic phase.  This phase extends upto $\Delta\sim15.5~meV$, after 
which $\sigma_{xy}$ again reach a quantized value of $-2$.
(iii) $\theta=2.15^\circ$ (red
line), where it remains $0$ till $\Delta \sim 12~meV$, then abruptly jumps to
$-3$ for a small range of $\Delta$ before jumping back to $-2$ for
$\Delta >15~meV$. Note that the metallic phases in this regime are localised in $\Delta-\theta$ plane, hence causing the sudden changes in $\sigma_{xy}$. (iv) Finally, at $\theta=3.0^\circ$ (blue
line), it remains $0$ till $\Delta \sim 9~meV$, then keeps decreasing
smoothly without reaching a quantized value.
\begin{figure}[t]
\includegraphics[width=\columnwidth]{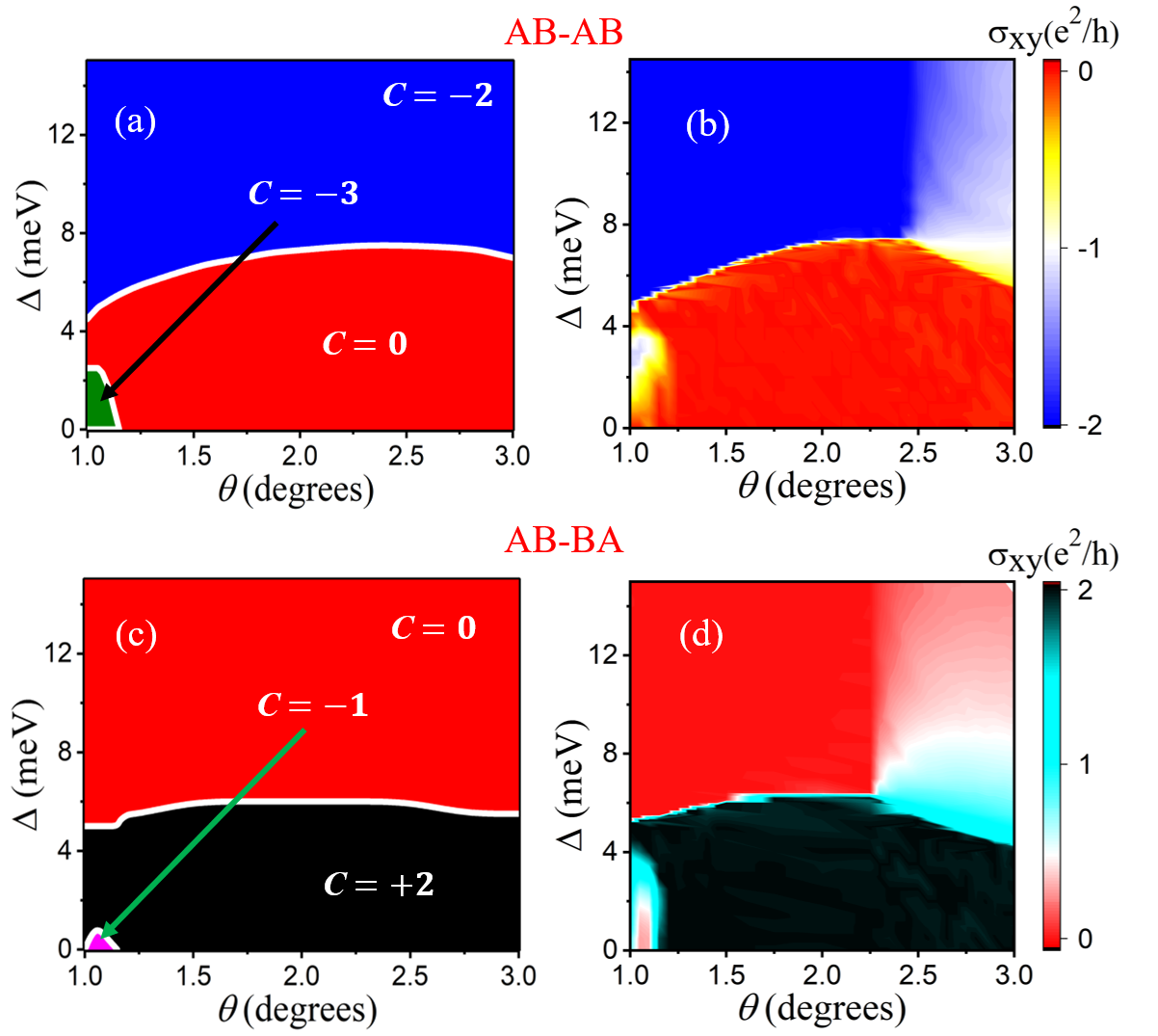}
\caption{(a) The topological phase diagram for the AB-AB stacked TDBLG with only $\gamma_3$ set to zero. 
The key differences from the FIG.\ref{Fig:Chernphasediag}(a) are the following: The Chern number 
jumps from $0$ to $-2$ without passing through a $-3$ phase. The $-3$
phase is present for very low twist angles $\theta < 1.2^\circ$
and very low $\Delta \lesssim 2.5~meV$. b) The valley-Hall conductivity $\sigma_{xy}$ for $\gamma_3=0$ in the AB-AB stacked TDBLG full model. 
The quantized values of $\sigma_{xy}$ are $0$ and $-2$ while the white and yellow
regions corresponds to a non-quantized $\sigma_{xy}$ in the metallic phase. The topological phases (c) 
and $\sigma_{xy}$ (d) for AB-BA TDBLG shows similar behaviour.}
\label{fullmodel_t30_chern}
\end{figure}
 
The valley Hall conductivity of AB-BA stacked TDBLG is shown as a
color plot in Fig.~\ref{Fig:AQHE}(c). The key difference between this
and the AB-AB stacking is that the insulator at low twist angles and
high field is trivial in this case, while the same insulator has a
$\sigma_{xy}$ of $-2$ for AB-AB stacking. Further the insulator at
large twist angle and low field has a $\sigma_{xy}$ of $+2$ for AB-BA
stacking, while it was trivial for the AB-AB stacking. Finally the
intermediate insulator, seen over a small region of twist angles $(\sim 2.0^\circ < \theta < 2.7^\circ)$ and
electric field $(11~meV \mathrm{to} 14~meV)$ has $\sigma_{xy}=-1$ in this case instead of $-3$ for
the case of AB-AB stacking. The representative line cuts for the four
different regimes discussed earlier is shown in
Fig.~\ref{Fig:AQHE}(d), which clearly shows these transitions and the
associated metallic phases with non-quantized $\sigma_{xy}$.

\section{Variation of the Phase Diagram with Model parameters\label{parameter}}
\begin{figure}[t]
\includegraphics[width=\columnwidth]{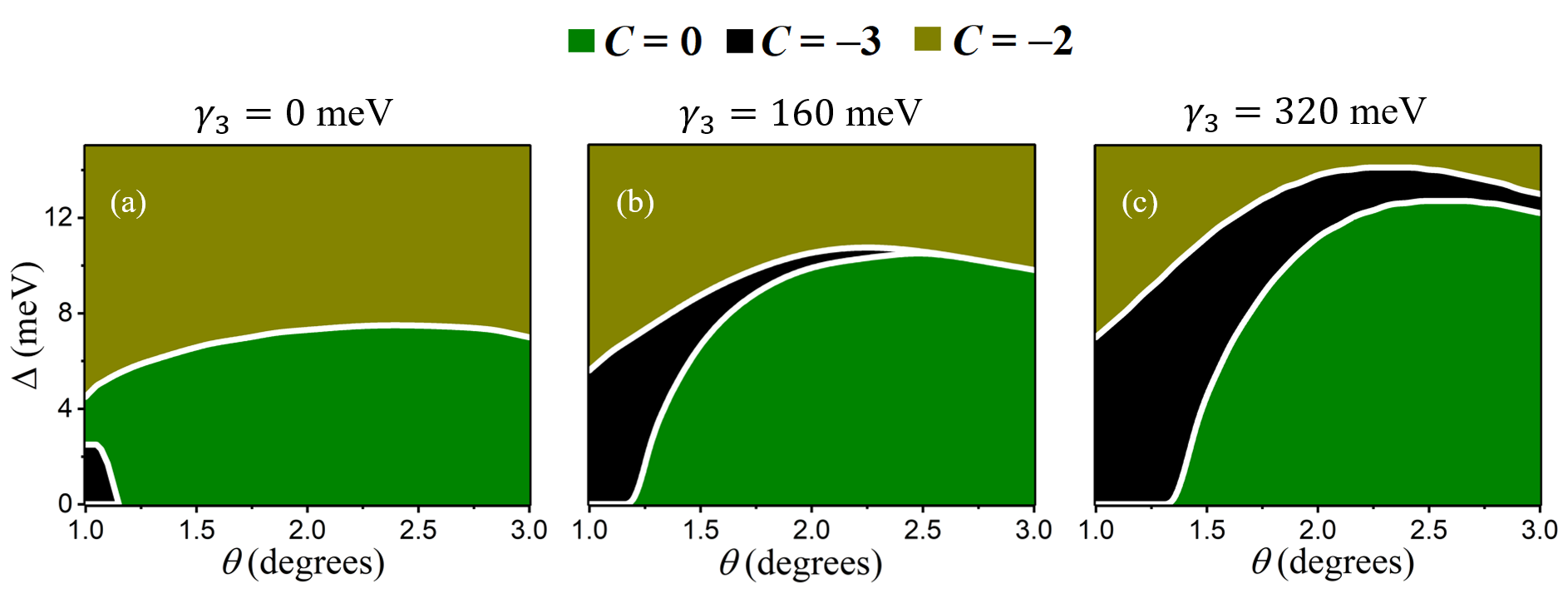}
\caption{ The evolution of the Chern number phase diagram with
  increasing $\gamma_3 =0$ (a), $\gamma_3=~160~meV$ (b) and $320~meV$
  (c) for AB-AB stacked TDBLG. The shrinking of the $C=-3$ phase with decreasing $\gamma_3$ is
  demonstrated in the three panels above. This shows that the trigonal
warping is the crucial term driving the multiple transitions in TDBLG}
\label{fullmodel_phase_t3}
\end{figure}

In all the previous sections, we have worked with a fixed value of
the Hamiltonian parameters of TDBLG. We have used the parameters from
Ref.~{\onlinecite{koshino2019}}, which has been widely used in the
literature. While the key band parameters like the graphene Fermi
velocity and the strongest interlayer tunneling for the bilayer
graphene $\gamma_1$ are fairly well known from ab-initio theory~\cite{CharlierMichenaudGonze1992} as
well as experimental fits to dispersions~\cite{MalardNilssonEliasBrantPlentzAlvesNetoPimenta2007,ZhangLiBasovFoglerHaoMartin2008}, estimates for the other
parameters like $\gamma_3$, $\gamma_4$ and $\Delta^{'}$ vary in the
literature.{\cite{MalardNilssonEliasBrantPlentzAlvesNetoPimenta2007,
LeeKhalafLiuLiuHaoKimVishwanath2019, ZhangLiBasovFoglerHaoMartin2008,KuzmenkoCrasseeMarelBlakeNovoselov2009,RozhkovSboychakovRakhmanovNori2016,
NetoGuineaPeresNovoselovGeim2009,MunozColladoOjedaUsajSofoBalseiro2016,CharlierMichenaudGonze1992,mccannkoshino2013,mccannkoshino2013}} Several estimates and
models also exist in the literature for the interlayer
tunneling across the twisted layers, $u$ and $u'$, with their ratio
ranging from $0$~\cite{LeeKhalafLiuLiuHaoKimVishwanath2019} to $0.8$~\cite{koshino2019}.

\begin{figure*}[t]
\includegraphics[width=\textwidth]{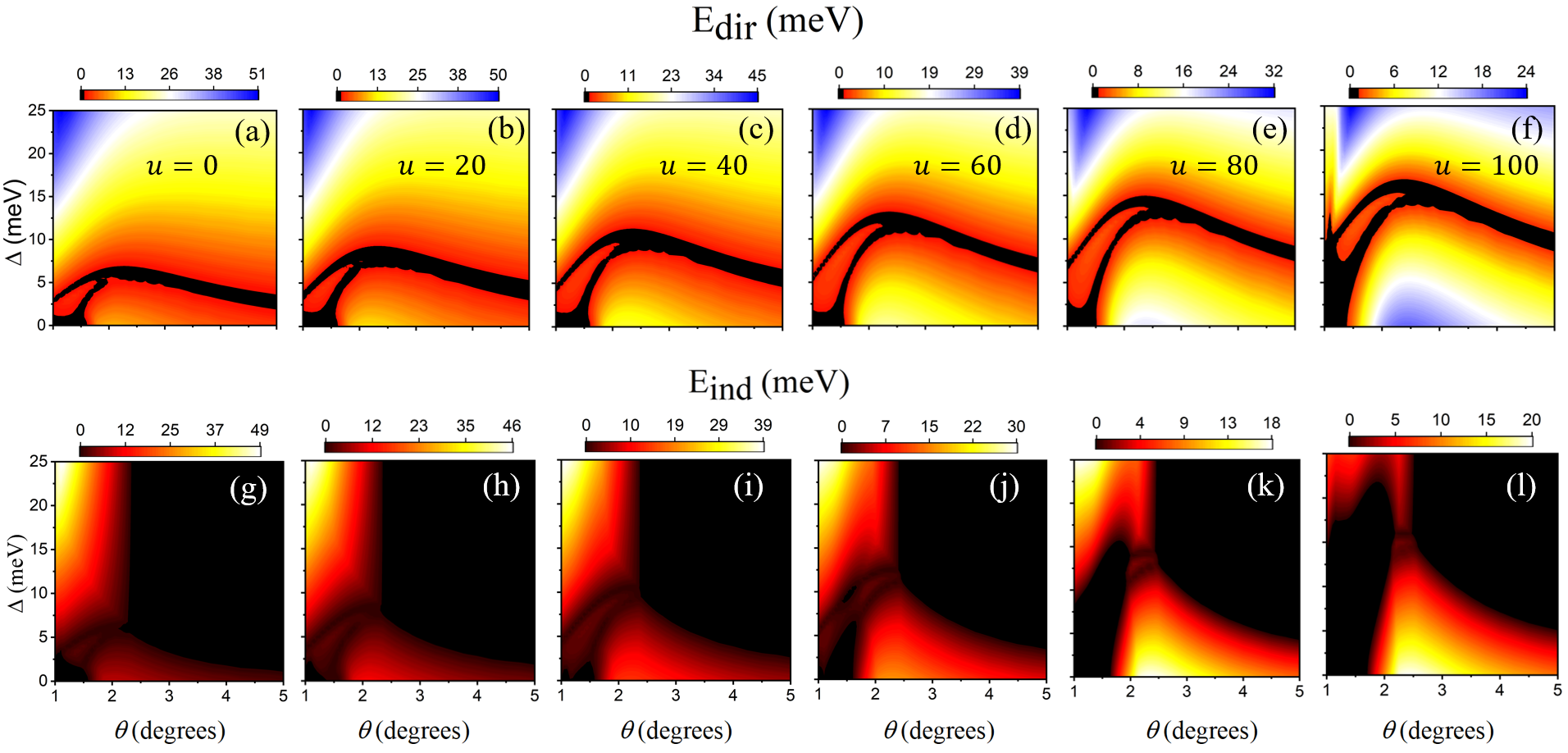}
\caption{(a)-(f) The direct bandgap ($\mathrm{E}_\mathrm{dir}$) of
  AB-AB stacked TDBLG for $u'=100~meV$ and values of $u$ ranging from
  $0$ to $100~meV$ in steps of $20$~ meV. 
At a given twist angle, both the transitions shift to larger $\Delta$
with increasing $u$. (g)-(l) The indirect bandgap ($\mathrm{E}_\mathrm{ind}$) for 
different values of $u$ with a fixed $u'=100~meV$ is plotted. The intermediate gapped region becomes larger with decreasing $u$.}
\label{gap_udep_AB-AB}
\end{figure*}
In this section, we will take a theorist's viewpoint and look at the
phase diagram as a function of some of the model parameters. While it is known that some of the parameters can be tuned by pressure~\cite{ChebroluChittariJung2019}, our main
motivation here is to have a theoretical understanding of the effect
of the different terms in the Hamiltonian on the phase diagram, which
can lead to a better understanding of how to manipulate the electronic
structure of this system. We assume that the graphene in-plane hopping
$\gamma_0$ is fixed to $2.1354~eV$, while $\gamma_1$ is fixed to
$0.4~eV$. We vary the other system parameters to see how the phase
diagram varies in this system.

We first consider the minimal model of TDBLG, which has been used
extensively for its simplicity in the literature. In this case
$\gamma_3$, $\gamma_4$ and $\Delta^{'}$ are set to $0$. This
fine-tuning results in a particle-hole symmetry in the Hamiltonian for
each valley in the system. We take $u=79.7~meV$ and $u'= 97.5~meV$ as before. In this case, for all values of $\theta$,  the dispersion has two quadratic
band touching ``Dirac points'' for each valley when the electric field
is set to zero.  With increasing values of $|\Delta|$, the low energy
bands separate from each other resulting in a direct band gap which is
monotonically increasing with $\Delta$. This is clearly seen in
Fig.~\ref{abab_minimal}(a), which shows a color-plot of the direct band gap of the system
in the $\Delta-\theta$ plane. Note that there are no gap closings at
finite $\Delta$ within this model. A more observable picture, which
takes into account overlap of bands in the energy space, is seen in
Fig.~\ref{abab_minimal}(b), where the indirect bandgap is plotted for the
minimal model. In this case, for twist angles less than $\sim 2.5^\circ$, the indirect band gap opens up at
inifinitisimal $\Delta$ and keeps increasing with $\Delta$. However for $\Delta>2.5^\circ$, 
the indirect band gap is zero as the dispersion of the conduction band at the $K'_M$ point lies below the dispersion of the valence band 
at the $K_M$ point. This leads to metallic behavior at all $\Delta$ for these twist angles. 
This contrasts
with the full model, where at intermediate and high angles, we find a gapped state at
low electric field which makes transition into another gapped state or a
metallic phase through a gap closing at finite electric field. The
Chern number (shown in Fig.~\ref{abab_minimal}(c)) remains fixed to $-2$ in the
whole phase diagram. The anomalous Hall conductivity (shown in Fig.~\ref{abab_minimal}(d)) follows the Chern number 
in the region where indirect band gap is non-zero, and shows non-quantized metallic behavior when the indirect
bandgap vanishes at large twist angles. Note that AB-AB and AB-BA stacked TDBLG
behaves in similar fashion within the minimal model.

\begin{figure*}[t]
\includegraphics[width=\textwidth]{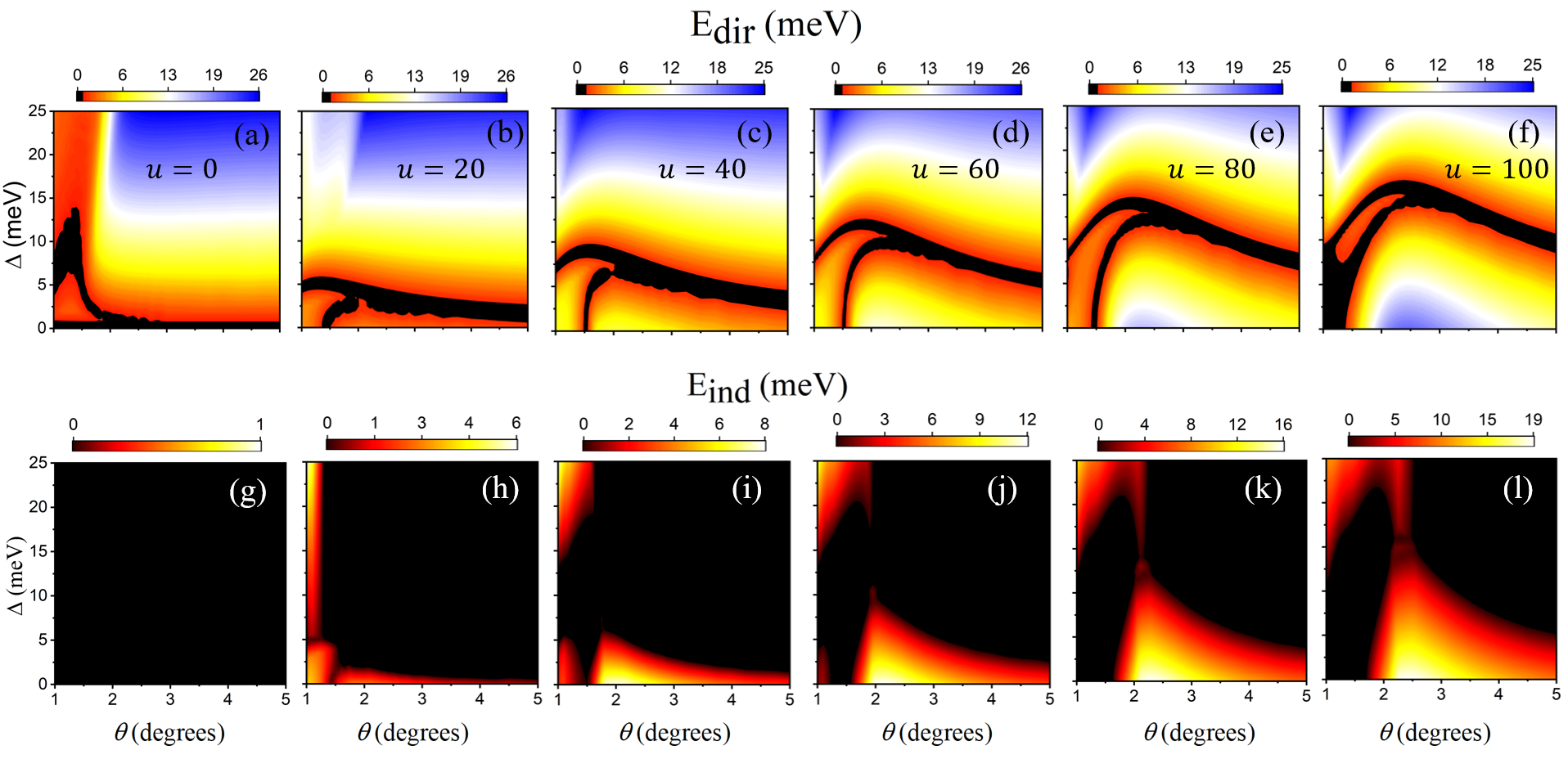}
\caption{(a)-(f) The direct bandgap ($\mathrm{E}_\mathrm{dir}$) of AB-BA stacked TDBLG for $u'=100~meV$ and values of $u$ ranging from
  $0$ to $100~meV$ in steps of $20~meV$.
The nature of the gap changes below $u=20~meV$ where large portions are gapless  (g)-(l) The indirect bandgap ($\mathrm{E}_\mathrm{ind}$) for 
different values of $u$ with a fixed $u'=100~meV$ is plotted. As the value of $u$ is decreased, the metallic region is extended.  }
\label{gap_udep_AB-BA}
\end{figure*}
As we have said before, the presence of a nonzero $\gamma_3$ splits
the Dirac point of the bilayer graphene into four Dirac cones, and
this is crucial for understanding both the multiple transitions and the
Chern number changes associated with it. To see this, we now add to
the minimal model a non-zero $\gamma_4 = 44~meV$ and $\Delta^{'}=
50~meV$, but still set the trigonal warping $\gamma_3=0$. This breaks the particle-hole symmetry in the model and
leads to a single gap-closing as a function of the electric field. This is
seen in Fig.~\ref{gap_t30}  which shows the color plots for the direct and the indirect bandgap in the
$\Delta-\theta$ plane for a AB-AB stacked TDBLG in Fig.~\ref{gap_t30}(a) and (b)
respectively. Fig.~\ref{gap_t30}(c) and (d) plots the same
quantities for a AB-BA stacked TDBLG. We study the Chern number and
the anomalous valley Hall conductivity of the AB-AB stacked TDBLG in
Fig.~\ref{fullmodel_t30_chern}(a) and (b). The Chern number changes
from $0$ at low electric field to $-2$ at the transition in this case. On the other hand,
the Chern number for AB-BA stacking changes from $+2$ at low fields to
$0$ at high electric fields, as seen in
Fig.~\ref{fullmodel_t30_chern}(c). Fig.~\ref{fullmodel_t30_chern}(d)
shows the corresponding valley Hall conductivity, which follows the
Chern number in the insulating phases. We thus see that while breaking
of the particle-hole symmetry is required to obtain the gapped phase
at low electric field at intermediate and large twist angles, it only
leads to a single transition with a Chern number change of $-2$ across
the transition. In this case, there is a qudratic band touching with an associated chirality 
index of $-2$, similar to what happens in a standard bilayer graphene. The two transitions and the Chern number change of
$\pm 3$ and $\pm 1$ is thus clearly driven by the presence of
$\gamma_3$. We would like to note that unlike the case of bilayer
graphene, the azimuthal symmetry in TDBLG is already broken due to the
twist and resultant \moire Brillouin zones; however the splitting of
the single Dirac point into four Dirac points is driven by a finite
$\gamma_3$. To emphasize the effect of $\gamma_3$ on the phase
diagram, in Fig.~\ref{fullmodel_phase_t3}, we plot together the Chern
phase diagram of the AB-AB stacked system for $\gamma_3=0$,
$\gamma_3=160~meV$ and $\gamma_3=320~meV$. We clearly see that the
intermediate phase shrinks with decreasing values of $\gamma_3$.

Finally, we consider the dependence of the phase diagram on the
interlayer couplings between the twisted layers. We work with the
parameters in the earlier sections of this paper. We keep $u'$
fixed at $100~meV$ and vary $u$ from $0$ to $100~meV$ to see how the
phase diagram changes. The direct and indirect bandgap of the AB-AB
stacked TDBLG is shown in Fig.~\ref{gap_udep_AB-AB}. The basic pattern
of two transitions remain intact. From Fig.~\ref{gap_udep_AB-AB} (a)
-(f), we see that as $u$
is increased (I) the low electric field direct bandgap at intermediate
and large twist angles increase and (II) the transitions are pushed
towards higher values of the electric field. If we focus on the
indirect bandgap (Fig.~\ref{gap_udep_AB-AB} (g)
-(l) ), we see that the intermediate phase is more visible at low
$u$. As $u$ is increased, the intermediate gapped phase at low twist
angles is gradually replaced by a metallic state, with a tiny sliver
visible in a small range of twist angles at $u=80~meV$. The general
trend of increased low field gap and higher critical fields with
increasing $u$ hold for AB-BA stacking (Fig.~\ref{gap_udep_AB-BA} (a)
-(f)) as well. However at $u=0$, the system remains gapless (or has a
very small gap) upto very large $\Delta \sim 25 ~meV$ at low twist angles and for all twist angles (upto $\sim 5^{\circ}$ in our study) at very low $\Delta$. Further, if we
look at the indirect bandgap (Fig.~\ref{gap_udep_AB-BA} (g)
-(l)), we find that the intermediate gapped phase vanishes as $u$ is
lowered and the phase diagram is filled up with metallic states for
low values of $u$.

\section{Conclusions}

We have studied the topological phase diagram of an undoped TDBLG as a
function of the twist angle and the perpendicular electric field. We
have looked at both theoretical quantities like the direct bandgap and
the Chern number of low energy bands, as well as observable quantities
like the indirect band gap and the anomalous valley Hall
conductivity. Theoretically, we find three gapped states with different
Chern numbers as we increase electric fields at intermediate
twist angles. There are two gap-closing transitions associated with
these. However, at low twist angle, there is only one gap closing transition. The Chern number change of $-3$ and $+1$ at the two transitions
can be understood in terms of whether the gap closes at the central
Dirac point, or at the three satellite Dirac points created by the
trigonal warping term. At higher twist angles, these two transitions
come close to each other and it is hard to distinguish between them,
resulting in a Chern number change of $-2$ at the transition. The
pattern is same for both AB-AB and AB-BA stacked TDBLG. The Chern
number of the states are different for AB-AB and AB-BA stacking, but
the Chern number changes are same in both cases. This independence of the change in
Chern number on the stacking order further shows that the trigonally warped satellite Dirac points together 
with the main Dirac point plays a pivotal role in causing the two transitions in TDBLG. This is strickingly different 
from BLG where perpendicular electric field is unable to drive any such transitions.

Since the bands overlap in energy, one needs to look at indirect
energy gaps between bands to determine whether one has an insulating or
a metallic state. This results in some of the gapped states being
replaced by multi-band Fermi seas. At low twist angle, the
phases with finite direct bandgap at low electric fields are actually
metallic, whereas at high twist angles, the system becomes metallic at
high electric fields. There is an intermediate region where the three
gapped insulating phases are still visible.

We have studied the systematic variation of this phase diagram with
the model parameters. The particle-hole symmetric minimal model does
not allow any finite field gap closing transitions, whereas a
particle-hole asymmetric model without the trigonal warping term leads
to a single transition with a Chern number change of $-2$. The
intermediate phase in general shrinks with decreasing value of the
trigonal warping term. This clearly shows the importance of trigonal
warping in understanding the phase diagram of TDBLG. Finally we have
studied the dependence of the phase diagram on the value of interlayer
hopping $u$. We see that the transitions are pushed towards higher
values of electric field for increasing $u$. For AB-AB stacking, the
intermediate gapped phase is more visible at low $u$, while the
opposite effect works for an AB-BA stacked TDBLG.

We have thus presented a comprehensive observable topological phase diagram for
the AB-AB and AB-BA stacked TDBLG at zero doping. The extension for
finite doping will be taken up in a future work.

 \acknowledgements {The authors acknowledge useful discussions with Pratap Chandra Adak, Subhajit Sinha and Mandar Deshmukh at TIFR. The authors also acknowledge the use of computational facilities at the Department of Theoretical Physics, Tata Institute of Fundamental Research, Mumbai.{

\bibliography{apssamp} 

\providecommand{\noopsort}[1]{}\providecommand{\singleletter}[1]{#1}%
\begin{thebibliography}{50}%
\makeatletter
\providecommand \@ifxundefined [1]{%
 \@ifx{#1\undefined}
}%
\providecommand \@ifnum [1]{%
 \ifnum #1\expandafter \@firstoftwo
 \else \expandafter \@secondoftwo
 \fi
}%
\providecommand \@ifx [1]{%
 \ifx #1\expandafter \@firstoftwo
 \else \expandafter \@secondoftwo
 \fi
}%
\providecommand \natexlab [1]{#1}%
\providecommand \enquote  [1]{``#1''}%
\providecommand \bibnamefont  [1]{#1}%
\providecommand \bibfnamefont [1]{#1}%
\providecommand \citenamefont [1]{#1}%
\providecommand \href@noop [0]{\@secondoftwo}%
\providecommand \href [0]{\begingroup \@sanitize@url \@href}%
\providecommand \@href[1]{\@@startlink{#1}\@@href}%
\providecommand \@@href[1]{\endgroup#1\@@endlink}%
\providecommand \@sanitize@url [0]{\catcode `\\12\catcode `\$12\catcode
  `\&12\catcode `\#12\catcode `\^12\catcode `\_12\catcode `\%12\relax}%
\providecommand \@@startlink[1]{}%
\providecommand \@@endlink[0]{}%
\providecommand \url  [0]{\begingroup\@sanitize@url \@url }%
\providecommand \@url [1]{\endgroup\@href {#1}{\urlprefix }}%
\providecommand \urlprefix  [0]{URL }%
\providecommand \Eprint [0]{\href }%
\providecommand \doibase [0]{http://dx.doi.org/}%
\providecommand \selectlanguage [0]{\@gobble}%
\providecommand \bibinfo  [0]{\@secondoftwo}%
\providecommand \bibfield  [0]{\@secondoftwo}%
\providecommand \translation [1]{[#1]}%
\providecommand \BibitemOpen [0]{}%
\providecommand \bibitemStop [0]{}%
\providecommand \bibitemNoStop [0]{.\EOS\space}%
\providecommand \EOS [0]{\spacefactor3000\relax}%
\providecommand \BibitemShut  [1]{\csname bibitem#1\endcsname}%
\let\auto@bib@innerbib\@empty
%</preamble>
\bibitem [{\citenamefont {Cao}\ \emph {et~al.}(2018{\natexlab{a}})\citenamefont
  {Cao}, \citenamefont {Fatemi}, \citenamefont {Fang}, \citenamefont
  {Watanabe}, \citenamefont {Taniguchi}, \citenamefont {Kaxiras},\ and\
  \citenamefont {Jarillo-Herrero}}]{CFFWTKJ2018}%
  \BibitemOpen
  \bibfield  {author} {\bibinfo {author} {\bibfnamefont {Y.}~\bibnamefont
  {Cao}}, \bibinfo {author} {\bibfnamefont {V.}~\bibnamefont {Fatemi}},
  \bibinfo {author} {\bibfnamefont {S.}~\bibnamefont {Fang}}, \bibinfo {author}
  {\bibfnamefont {K.}~\bibnamefont {Watanabe}}, \bibinfo {author}
  {\bibfnamefont {T.}~\bibnamefont {Taniguchi}}, \bibinfo {author}
  {\bibfnamefont {E.}~\bibnamefont {Kaxiras}}, \ and\ \bibinfo {author}
  {\bibfnamefont {P.}~\bibnamefont {Jarillo-Herrero}},\ }\href
  {https://doi.org/10.1038/nature26160} {\bibfield  {journal} {\bibinfo
  {journal} {Nature.}\ }\textbf {\bibinfo {volume} {556}},\ \bibinfo {pages}
  {43–50} (\bibinfo {year} {2018}{\natexlab{a}})}\BibitemShut {NoStop}%
\bibitem [{\citenamefont {Carr}\ \emph {et~al.}(2017)\citenamefont {Carr},
  \citenamefont {Massatt}, \citenamefont {Fang}, \citenamefont {Cazeaux},
  \citenamefont {Luskin},\ and\ \citenamefont
  {Kaxiras}}]{CarrMassattFangCazeauxLuskinKaxiras2017}%
  \BibitemOpen
  \bibfield  {author} {\bibinfo {author} {\bibfnamefont {S.}~\bibnamefont
  {Carr}}, \bibinfo {author} {\bibfnamefont {D.}~\bibnamefont {Massatt}},
  \bibinfo {author} {\bibfnamefont {S.}~\bibnamefont {Fang}}, \bibinfo {author}
  {\bibfnamefont {P.}~\bibnamefont {Cazeaux}}, \bibinfo {author} {\bibfnamefont
  {M.}~\bibnamefont {Luskin}}, \ and\ \bibinfo {author} {\bibfnamefont
  {E.}~\bibnamefont {Kaxiras}},\ }\href {\doibase 10.1103/PhysRevB.95.075420}
  {\bibfield  {journal} {\bibinfo  {journal} {Phys. Rev. B}\ }\textbf {\bibinfo
  {volume} {95}},\ \bibinfo {pages} {075420} (\bibinfo {year}
  {2017})}\BibitemShut {NoStop}%
\bibitem [{\citenamefont {Bistritzer}\ and\ \citenamefont
  {MacDonald}(2011)}]{BistritzerMacDonald2011}%
  \BibitemOpen
  \bibfield  {author} {\bibinfo {author} {\bibfnamefont {R.}~\bibnamefont
  {Bistritzer}}\ and\ \bibinfo {author} {\bibfnamefont {A.~H.}\ \bibnamefont
  {MacDonald}},\ }\href@noop {} {\bibfield  {journal} {\bibinfo  {journal}
  {Proceedings of the National Academy of Sciences}\ }\textbf {\bibinfo
  {volume} {108}},\ \bibinfo {pages} {12233} (\bibinfo {year}
  {2011})}\BibitemShut {NoStop}%
\bibitem [{\citenamefont {Shen}\ \emph {et~al.}(2020)\citenamefont {Shen},
  \citenamefont {Chu}, \citenamefont {Wu}, \citenamefont {Li}, \citenamefont
  {Wang}, \citenamefont {Zhao}, \citenamefont {Tang}, \citenamefont {Liu},
  \citenamefont {Tian}, \citenamefont {Watanabe}, \citenamefont {Taniguchi},
  \citenamefont {Yang}, \citenamefont {Meng}, \citenamefont {Shi},
  \citenamefont {Yazyev},\ and\ \citenamefont {Zhang}}]{SCWLWZTLTWTYMSYZ2020}%
  \BibitemOpen
  \bibfield  {author} {\bibinfo {author} {\bibfnamefont {C.}~\bibnamefont
  {Shen}}, \bibinfo {author} {\bibfnamefont {Y.}~\bibnamefont {Chu}}, \bibinfo
  {author} {\bibfnamefont {Q.}~\bibnamefont {Wu}}, \bibinfo {author}
  {\bibfnamefont {N.}~\bibnamefont {Li}}, \bibinfo {author} {\bibfnamefont
  {S.}~\bibnamefont {Wang}}, \bibinfo {author} {\bibfnamefont {Y.}~\bibnamefont
  {Zhao}}, \bibinfo {author} {\bibfnamefont {J.}~\bibnamefont {Tang}}, \bibinfo
  {author} {\bibfnamefont {J.}~\bibnamefont {Liu}}, \bibinfo {author}
  {\bibfnamefont {J.}~\bibnamefont {Tian}}, \bibinfo {author} {\bibfnamefont
  {K.}~\bibnamefont {Watanabe}}, \bibinfo {author} {\bibfnamefont
  {T.}~\bibnamefont {Taniguchi}}, \bibinfo {author} {\bibfnamefont
  {R.}~\bibnamefont {Yang}}, \bibinfo {author} {\bibfnamefont {Z.~Y.}\
  \bibnamefont {Meng}}, \bibinfo {author} {\bibfnamefont {D.}~\bibnamefont
  {Shi}}, \bibinfo {author} {\bibfnamefont {O.~V.}\ \bibnamefont {Yazyev}}, \
  and\ \bibinfo {author} {\bibfnamefont {G.}~\bibnamefont {Zhang}},\ }\href
  {https://doi.org/10.1038/s41567-020-0825-9} {\bibfield  {journal} {\bibinfo
  {journal} {Nat. Phys.}\ }\textbf {\bibinfo {volume} {16}},\ \bibinfo {pages}
  {520–525} (\bibinfo {year} {2020})}\BibitemShut {NoStop}%
\bibitem [{\citenamefont {Su\'arez~Morell}\ \emph {et~al.}(2013)\citenamefont
  {Su\'arez~Morell}, \citenamefont {Pacheco}, \citenamefont {Chico},\ and\
  \citenamefont {Brey}}]{MorellPachecoChicoBrey2013}%
  \BibitemOpen
  \bibfield  {author} {\bibinfo {author} {\bibfnamefont {E.}~\bibnamefont
  {Su\'arez~Morell}}, \bibinfo {author} {\bibfnamefont {M.}~\bibnamefont
  {Pacheco}}, \bibinfo {author} {\bibfnamefont {L.}~\bibnamefont {Chico}}, \
  and\ \bibinfo {author} {\bibfnamefont {L.}~\bibnamefont {Brey}},\ }\href
  {\doibase 10.1103/PhysRevB.87.125414} {\bibfield  {journal} {\bibinfo
  {journal} {Phys. Rev. B}\ }\textbf {\bibinfo {volume} {87}},\ \bibinfo
  {pages} {125414} (\bibinfo {year} {2013})}\BibitemShut {NoStop}%
\bibitem [{\citenamefont {Spanton}\ \emph {et~al.}(2018)\citenamefont
  {Spanton}, \citenamefont {Zibrov}, \citenamefont {Zhou}, \citenamefont
  {Taniguchi}, \citenamefont {Watanabe}, \citenamefont {Zaletel},\ and\
  \citenamefont {Young}}]{Spanton62}%
  \BibitemOpen
  \bibfield  {author} {\bibinfo {author} {\bibfnamefont {E.~M.}\ \bibnamefont
  {Spanton}}, \bibinfo {author} {\bibfnamefont {A.~A.}\ \bibnamefont {Zibrov}},
  \bibinfo {author} {\bibfnamefont {H.}~\bibnamefont {Zhou}}, \bibinfo {author}
  {\bibfnamefont {T.}~\bibnamefont {Taniguchi}}, \bibinfo {author}
  {\bibfnamefont {K.}~\bibnamefont {Watanabe}}, \bibinfo {author}
  {\bibfnamefont {M.~P.}\ \bibnamefont {Zaletel}}, \ and\ \bibinfo {author}
  {\bibfnamefont {A.~F.}\ \bibnamefont {Young}},\ }\href@noop {} {\bibfield
  {journal} {\bibinfo  {journal} {Science}\ }\textbf {\bibinfo {volume}
  {360}},\ \bibinfo {pages} {62} (\bibinfo {year} {2018})}\BibitemShut
  {NoStop}%
\bibitem [{\citenamefont {Zhang}\ \emph {et~al.}(2020)\citenamefont {Zhang},
  \citenamefont {Wang}, \citenamefont {Watanabe}, \citenamefont {Taniguchi},
  \citenamefont {Ueno}, \citenamefont {Tutuc},\ and\ \citenamefont
  {LeRoy}}]{ZhangWangWatanabeTaniguchiUenoTutucLeRoy2020}%
  \BibitemOpen
  \bibfield  {author} {\bibinfo {author} {\bibfnamefont {Z.}~\bibnamefont
  {Zhang}}, \bibinfo {author} {\bibfnamefont {Y.}~\bibnamefont {Wang}},
  \bibinfo {author} {\bibfnamefont {K.}~\bibnamefont {Watanabe}}, \bibinfo
  {author} {\bibfnamefont {T.}~\bibnamefont {Taniguchi}}, \bibinfo {author}
  {\bibfnamefont {K.}~\bibnamefont {Ueno}}, \bibinfo {author} {\bibfnamefont
  {E.}~\bibnamefont {Tutuc}}, \ and\ \bibinfo {author} {\bibfnamefont {B.~J.}\
  \bibnamefont {LeRoy}},\ }\href {\doibase 10.1038/s41567-020-0958-x}
  {\bibfield  {journal} {\bibinfo  {journal} {Nature Physics}\ } (\bibinfo
  {year} {2020}),\ 10.1038/s41567-020-0958-x}\BibitemShut {NoStop}%
\bibitem [{\citenamefont {Novoselov}\ \emph {et~al.}(2016)\citenamefont
  {Novoselov}, \citenamefont {Mishchenko}, \citenamefont {Carvalho},\ and\
  \citenamefont {Castro~Neto}}]{Novoselovaac9439}%
  \BibitemOpen
  \bibfield  {author} {\bibinfo {author} {\bibfnamefont {K.~S.}\ \bibnamefont
  {Novoselov}}, \bibinfo {author} {\bibfnamefont {A.}~\bibnamefont
  {Mishchenko}}, \bibinfo {author} {\bibfnamefont {A.}~\bibnamefont
  {Carvalho}}, \ and\ \bibinfo {author} {\bibfnamefont {A.~H.}\ \bibnamefont
  {Castro~Neto}},\ }\href@noop {} {\bibfield  {journal} {\bibinfo  {journal}
  {Science}\ }\textbf {\bibinfo {volume} {353}} (\bibinfo {year}
  {2016})}\BibitemShut {NoStop}%
\bibitem [{\citenamefont {Cao}\ \emph {et~al.}(2018{\natexlab{b}})\citenamefont
  {Cao}, \citenamefont {Fatemi}, \citenamefont {Demir}, \citenamefont {Fang},
  \citenamefont {Tomarken}, \citenamefont {Luo}, \citenamefont
  {Sanchez-Yamagishi}, \citenamefont {Watanabe}, \citenamefont {Taniguchi},
  \citenamefont {Kaxiras}, \citenamefont {Ashoori},\ and\ \citenamefont
  {Jarillo-Herrero}}]{CFDFTLSWTKAJ2018}%
  \BibitemOpen
  \bibfield  {author} {\bibinfo {author} {\bibfnamefont {Y.}~\bibnamefont
  {Cao}}, \bibinfo {author} {\bibfnamefont {V.}~\bibnamefont {Fatemi}},
  \bibinfo {author} {\bibfnamefont {A.}~\bibnamefont {Demir}}, \bibinfo
  {author} {\bibfnamefont {S.}~\bibnamefont {Fang}}, \bibinfo {author}
  {\bibfnamefont {S.~L.}\ \bibnamefont {Tomarken}}, \bibinfo {author}
  {\bibfnamefont {J.~Y.}\ \bibnamefont {Luo}}, \bibinfo {author} {\bibfnamefont
  {J.~D.}\ \bibnamefont {Sanchez-Yamagishi}}, \bibinfo {author} {\bibfnamefont
  {K.}~\bibnamefont {Watanabe}}, \bibinfo {author} {\bibfnamefont
  {T.}~\bibnamefont {Taniguchi}}, \bibinfo {author} {\bibfnamefont
  {E.}~\bibnamefont {Kaxiras}}, \bibinfo {author} {\bibfnamefont {R.~C.}\
  \bibnamefont {Ashoori}}, \ and\ \bibinfo {author} {\bibfnamefont
  {P.}~\bibnamefont {Jarillo-Herrero}},\ }\href
  {https://doi.org/10.1038/nature26154} {\bibfield  {journal} {\bibinfo
  {journal} {Nature.}\ }\textbf {\bibinfo {volume} {556}},\ \bibinfo {pages}
  {80} (\bibinfo {year} {2018}{\natexlab{b}})}\BibitemShut {NoStop}%
\bibitem [{\citenamefont {Yankowitz}\ \emph {et~al.}(2019)\citenamefont
  {Yankowitz}, \citenamefont {Chen}, \citenamefont {Polshyn}, \citenamefont
  {Zhang}, \citenamefont {Watanabe}, \citenamefont {Taniguchi}, \citenamefont
  {Graf}, \citenamefont {Young},\ and\ \citenamefont {Dean}}]{Yankowitz1059}%
  \BibitemOpen
  \bibfield  {author} {\bibinfo {author} {\bibfnamefont {M.}~\bibnamefont
  {Yankowitz}}, \bibinfo {author} {\bibfnamefont {S.}~\bibnamefont {Chen}},
  \bibinfo {author} {\bibfnamefont {H.}~\bibnamefont {Polshyn}}, \bibinfo
  {author} {\bibfnamefont {Y.}~\bibnamefont {Zhang}}, \bibinfo {author}
  {\bibfnamefont {K.}~\bibnamefont {Watanabe}}, \bibinfo {author}
  {\bibfnamefont {T.}~\bibnamefont {Taniguchi}}, \bibinfo {author}
  {\bibfnamefont {D.}~\bibnamefont {Graf}}, \bibinfo {author} {\bibfnamefont
  {A.~F.}\ \bibnamefont {Young}}, \ and\ \bibinfo {author} {\bibfnamefont
  {C.~R.}\ \bibnamefont {Dean}},\ }\href {\doibase 10.1126/science.aav1910}
  {\bibfield  {journal} {\bibinfo  {journal} {Science}\ }\textbf {\bibinfo
  {volume} {363}},\ \bibinfo {pages} {1059} (\bibinfo {year}
  {2019})}\BibitemShut {NoStop}%
\bibitem [{\citenamefont {Sharpe}\ \emph {et~al.}(2019)\citenamefont {Sharpe},
  \citenamefont {Fox}, \citenamefont {Barnard}, \citenamefont {Finney},
  \citenamefont {Watanabe}, \citenamefont {Taniguchi}, \citenamefont
  {Kastner},\ and\ \citenamefont {Goldhaber-Gordon}}]{Sharpe605}%
  \BibitemOpen
  \bibfield  {author} {\bibinfo {author} {\bibfnamefont {A.~L.}\ \bibnamefont
  {Sharpe}}, \bibinfo {author} {\bibfnamefont {E.~J.}\ \bibnamefont {Fox}},
  \bibinfo {author} {\bibfnamefont {A.~W.}\ \bibnamefont {Barnard}}, \bibinfo
  {author} {\bibfnamefont {J.}~\bibnamefont {Finney}}, \bibinfo {author}
  {\bibfnamefont {K.}~\bibnamefont {Watanabe}}, \bibinfo {author}
  {\bibfnamefont {T.}~\bibnamefont {Taniguchi}}, \bibinfo {author}
  {\bibfnamefont {M.~A.}\ \bibnamefont {Kastner}}, \ and\ \bibinfo {author}
  {\bibfnamefont {D.}~\bibnamefont {Goldhaber-Gordon}},\ }\href@noop {}
  {\bibfield  {journal} {\bibinfo  {journal} {Science}\ }\textbf {\bibinfo
  {volume} {365}},\ \bibinfo {pages} {605} (\bibinfo {year}
  {2019})}\BibitemShut {NoStop}%
\bibitem [{\citenamefont {Po}\ \emph {et~al.}(2018)\citenamefont {Po},
  \citenamefont {Zou}, \citenamefont {Vishwanath},\ and\ \citenamefont
  {Senthil}}]{PoZouVishwanathSenthil2018}%
  \BibitemOpen
  \bibfield  {author} {\bibinfo {author} {\bibfnamefont {H.~C.}\ \bibnamefont
  {Po}}, \bibinfo {author} {\bibfnamefont {L.}~\bibnamefont {Zou}}, \bibinfo
  {author} {\bibfnamefont {A.}~\bibnamefont {Vishwanath}}, \ and\ \bibinfo
  {author} {\bibfnamefont {T.}~\bibnamefont {Senthil}},\ }\href {\doibase
  10.1103/PhysRevX.8.031089} {\bibfield  {journal} {\bibinfo  {journal} {Phys.
  Rev. X}\ }\textbf {\bibinfo {volume} {8}},\ \bibinfo {pages} {031089}
  (\bibinfo {year} {2018})}\BibitemShut {NoStop}%
\bibitem [{\citenamefont {Koshino}(2019)}]{koshino2019}%
  \BibitemOpen
  \bibfield  {author} {\bibinfo {author} {\bibfnamefont {M.}~\bibnamefont
  {Koshino}},\ }\href@noop {} {\bibfield  {journal} {\bibinfo  {journal} {Phys.
  Rev. B}\ }\textbf {\bibinfo {volume} {99}},\ \bibinfo {pages} {235406}
  (\bibinfo {year} {2019})}\BibitemShut {NoStop}%
\bibitem [{\citenamefont {Chebrolu}\ \emph {et~al.}(2019)\citenamefont
  {Chebrolu}, \citenamefont {Chittari},\ and\ \citenamefont
  {Jung}}]{ChebroluChittariJung2019}%
  \BibitemOpen
  \bibfield  {author} {\bibinfo {author} {\bibfnamefont {N.~R.}\ \bibnamefont
  {Chebrolu}}, \bibinfo {author} {\bibfnamefont {B.~L.}\ \bibnamefont
  {Chittari}}, \ and\ \bibinfo {author} {\bibfnamefont {J.}~\bibnamefont
  {Jung}},\ }\href {\doibase 10.1103/PhysRevB.99.235417} {\bibfield  {journal}
  {\bibinfo  {journal} {Phys. Rev. B}\ }\textbf {\bibinfo {volume} {99}},\
  \bibinfo {pages} {235417} (\bibinfo {year} {2019})}\BibitemShut {NoStop}%
\bibitem [{\citenamefont {Haddadi}\ \emph {et~al.}(2020)\citenamefont
  {Haddadi}, \citenamefont {Wu}, \citenamefont {Kruchkov},\ and\ \citenamefont
  {Yazyev}}]{HaddadiWuKruchkovYazyev2020}%
  \BibitemOpen
  \bibfield  {author} {\bibinfo {author} {\bibfnamefont {F.}~\bibnamefont
  {Haddadi}}, \bibinfo {author} {\bibfnamefont {Q.}~\bibnamefont {Wu}},
  \bibinfo {author} {\bibfnamefont {A.~J.}\ \bibnamefont {Kruchkov}}, \ and\
  \bibinfo {author} {\bibfnamefont {O.~V.}\ \bibnamefont {Yazyev}},\
  }\href@noop {} {\bibfield  {journal} {\bibinfo  {journal} {Nano Lett.}\
  }\textbf {\bibinfo {volume} {20}},\ \bibinfo {pages} {2410} (\bibinfo {year}
  {2020})}\BibitemShut {NoStop}%
\bibitem [{\citenamefont {Lin}\ \emph {et~al.}(2020)\citenamefont {Lin},
  \citenamefont {Zhu},\ and\ \citenamefont {Ni}}]{LinZhuNi2020}%
  \BibitemOpen
  \bibfield  {author} {\bibinfo {author} {\bibfnamefont {X.}~\bibnamefont
  {Lin}}, \bibinfo {author} {\bibfnamefont {H.}~\bibnamefont {Zhu}}, \ and\
  \bibinfo {author} {\bibfnamefont {J.}~\bibnamefont {Ni}},\ }\href {\doibase
  10.1103/PhysRevB.101.155405} {\bibfield  {journal} {\bibinfo  {journal}
  {Phys. Rev. B}\ }\textbf {\bibinfo {volume} {101}},\ \bibinfo {pages}
  {155405} (\bibinfo {year} {2020})}\BibitemShut {NoStop}%
\bibitem [{\citenamefont {Liu}\ \emph {et~al.}(2020)\citenamefont {Liu},
  \citenamefont {Hao}, \citenamefont {Khalaf}, \citenamefont {Lee},
  \citenamefont {Ronen}, \citenamefont {Yoo}, \citenamefont {Najafabadi},
  \citenamefont {Watanabe}, \citenamefont {Taniguchi}, \citenamefont
  {Vishwanath},\ and\ \citenamefont {Kim}}]{LHKLRYNWTVK2020}%
  \BibitemOpen
  \bibfield  {author} {\bibinfo {author} {\bibfnamefont {X.}~\bibnamefont
  {Liu}}, \bibinfo {author} {\bibfnamefont {Z.}~\bibnamefont {Hao}}, \bibinfo
  {author} {\bibfnamefont {E.}~\bibnamefont {Khalaf}}, \bibinfo {author}
  {\bibfnamefont {J.~Y.}\ \bibnamefont {Lee}}, \bibinfo {author} {\bibfnamefont
  {Y.}~\bibnamefont {Ronen}}, \bibinfo {author} {\bibfnamefont
  {H.}~\bibnamefont {Yoo}}, \bibinfo {author} {\bibfnamefont {D.~H.}\
  \bibnamefont {Najafabadi}}, \bibinfo {author} {\bibfnamefont
  {K.}~\bibnamefont {Watanabe}}, \bibinfo {author} {\bibfnamefont
  {T.}~\bibnamefont {Taniguchi}}, \bibinfo {author} {\bibfnamefont
  {A.}~\bibnamefont {Vishwanath}}, \ and\ \bibinfo {author} {\bibfnamefont
  {P.}~\bibnamefont {Kim}},\ }\href@noop {} {\bibfield  {journal} {\bibinfo
  {journal} {Nature.}\ }\textbf {\bibinfo {volume} {583}},\ \bibinfo {pages}
  {221} (\bibinfo {year} {2020})}\BibitemShut {NoStop}%
\bibitem [{\citenamefont {Burg}\ \emph {et~al.}(2019)\citenamefont {Burg},
  \citenamefont {Zhu}, \citenamefont {Taniguchi}, \citenamefont {Watanabe},
  \citenamefont {MacDonald},\ and\ \citenamefont {Tutuc}}]{BZTWMT2019}%
  \BibitemOpen
  \bibfield  {author} {\bibinfo {author} {\bibfnamefont {G.~W.}\ \bibnamefont
  {Burg}}, \bibinfo {author} {\bibfnamefont {J.}~\bibnamefont {Zhu}}, \bibinfo
  {author} {\bibfnamefont {T.}~\bibnamefont {Taniguchi}}, \bibinfo {author}
  {\bibfnamefont {K.}~\bibnamefont {Watanabe}}, \bibinfo {author}
  {\bibfnamefont {A.~H.}\ \bibnamefont {MacDonald}}, \ and\ \bibinfo {author}
  {\bibfnamefont {E.}~\bibnamefont {Tutuc}},\ }\href {\doibase
  10.1103/PhysRevLett.123.197702} {\bibfield  {journal} {\bibinfo  {journal}
  {Phys. Rev. Lett.}\ }\textbf {\bibinfo {volume} {123}},\ \bibinfo {pages}
  {197702} (\bibinfo {year} {2019})}\BibitemShut {NoStop}%
\bibitem [{\citenamefont {Adak}\ \emph {et~al.}(2020)\citenamefont {Adak},
  \citenamefont {Sinha}, \citenamefont {Ghorai}, \citenamefont {Sangani},
  \citenamefont {Watanabe}, \citenamefont {Taniguchi}, \citenamefont
  {Sensarma},\ and\ \citenamefont
  {Deshmukh}}]{AdakSinhaGhoraiSanganiVarmaWatanabeTaniguchiSensarmaDeshmukh2020}%
  \BibitemOpen
  \bibfield  {author} {\bibinfo {author} {\bibfnamefont {P.~C.}\ \bibnamefont
  {Adak}}, \bibinfo {author} {\bibfnamefont {S.}~\bibnamefont {Sinha}},
  \bibinfo {author} {\bibfnamefont {U.}~\bibnamefont {Ghorai}}, \bibinfo
  {author} {\bibfnamefont {L.~D.~V.}\ \bibnamefont {Sangani}}, \bibinfo
  {author} {\bibfnamefont {K.}~\bibnamefont {Watanabe}}, \bibinfo {author}
  {\bibfnamefont {T.}~\bibnamefont {Taniguchi}}, \bibinfo {author}
  {\bibfnamefont {R.}~\bibnamefont {Sensarma}}, \ and\ \bibinfo {author}
  {\bibfnamefont {M.~M.}\ \bibnamefont {Deshmukh}},\ }\href {\doibase
  10.1103/PhysRevB.101.125428} {\bibfield  {journal} {\bibinfo  {journal}
  {Phys. Rev. B}\ }\textbf {\bibinfo {volume} {101}},\ \bibinfo {pages}
  {125428} (\bibinfo {year} {2020})}\BibitemShut {NoStop}%
\bibitem [{\citenamefont {Zhang}\ \emph {et~al.}(2009)\citenamefont {Zhang},
  \citenamefont {Tang}, \citenamefont {Girit}, \citenamefont {Hao},
  \citenamefont {Martin}, \citenamefont {Zettl}, \citenamefont {Crommie},
  \citenamefont {Shen},\ and\ \citenamefont {Wang}}]{ZTGHMZCSW2009}%
  \BibitemOpen
  \bibfield  {author} {\bibinfo {author} {\bibfnamefont {Y.}~\bibnamefont
  {Zhang}}, \bibinfo {author} {\bibfnamefont {T.-T.}\ \bibnamefont {Tang}},
  \bibinfo {author} {\bibfnamefont {C.}~\bibnamefont {Girit}}, \bibinfo
  {author} {\bibfnamefont {Z.}~\bibnamefont {Hao}}, \bibinfo {author}
  {\bibfnamefont {M.~C.}\ \bibnamefont {Martin}}, \bibinfo {author}
  {\bibfnamefont {A.}~\bibnamefont {Zettl}}, \bibinfo {author} {\bibfnamefont
  {M.~F.}\ \bibnamefont {Crommie}}, \bibinfo {author} {\bibfnamefont {Y.~R.}\
  \bibnamefont {Shen}}, \ and\ \bibinfo {author} {\bibfnamefont
  {F.}~\bibnamefont {Wang}},\ }\href
  {https://link.aps.org/doi/10.1103/PhysRevB.90.155451} {\bibfield  {journal}
  {\bibinfo  {journal} {Nature.}\ }\textbf {\bibinfo {volume} {459}},\ \bibinfo
  {pages} {820–823} (\bibinfo {year} {2009})}\BibitemShut {NoStop}%
\bibitem [{\citenamefont {Castro}\ \emph {et~al.}(2010)\citenamefont {Castro},
  \citenamefont {Novoselov}, \citenamefont {Morozov}, \citenamefont {Peres},
  \citenamefont {dos Santos}, \citenamefont {Nilsson}, \citenamefont {Guinea},
  \citenamefont {Geim},\ and\ \citenamefont {Neto}}]{CNMPSNGGN2010}%
  \BibitemOpen
  \bibfield  {author} {\bibinfo {author} {\bibfnamefont {E.~V.}\ \bibnamefont
  {Castro}}, \bibinfo {author} {\bibfnamefont {K.~S.}\ \bibnamefont
  {Novoselov}}, \bibinfo {author} {\bibfnamefont {S.~V.}\ \bibnamefont
  {Morozov}}, \bibinfo {author} {\bibfnamefont {N.~M.~R.}\ \bibnamefont
  {Peres}}, \bibinfo {author} {\bibfnamefont {J.~M. B.~L.}\ \bibnamefont {dos
  Santos}}, \bibinfo {author} {\bibfnamefont {J.}~\bibnamefont {Nilsson}},
  \bibinfo {author} {\bibfnamefont {F.}~\bibnamefont {Guinea}}, \bibinfo
  {author} {\bibfnamefont {A.~K.}\ \bibnamefont {Geim}}, \ and\ \bibinfo
  {author} {\bibfnamefont {A.~H.~C.}\ \bibnamefont {Neto}},\ }\href {\doibase
  10.1088/0953-8984/22/17/175503} {\bibfield  {journal} {\bibinfo  {journal}
  {Journal of Physics: Condensed Matter}\ }\textbf {\bibinfo {volume} {22}},\
  \bibinfo {pages} {175503} (\bibinfo {year} {2010})}\BibitemShut {NoStop}%
\bibitem [{\citenamefont {Varlet}\ \emph {et~al.}(2015)\citenamefont {Varlet},
  \citenamefont {Mucha-Kruczynski}, \citenamefont {Bischoff}, \citenamefont
  {Simonet}, \citenamefont {Taniguchi}, \citenamefont {Watanabe}, \citenamefont
  {Falko}, \citenamefont {Ihn},\ and\ \citenamefont
  {Ensslin}}]{VMKBSTWFIE2015}%
  \BibitemOpen
  \bibfield  {author} {\bibinfo {author} {\bibfnamefont {A.}~\bibnamefont
  {Varlet}}, \bibinfo {author} {\bibfnamefont {M.}~\bibnamefont
  {Mucha-Kruczynski}}, \bibinfo {author} {\bibfnamefont {D.}~\bibnamefont
  {Bischoff}}, \bibinfo {author} {\bibfnamefont {P.}~\bibnamefont {Simonet}},
  \bibinfo {author} {\bibfnamefont {T.}~\bibnamefont {Taniguchi}}, \bibinfo
  {author} {\bibfnamefont {K.}~\bibnamefont {Watanabe}}, \bibinfo {author}
  {\bibfnamefont {V.}~\bibnamefont {Falko}}, \bibinfo {author} {\bibfnamefont
  {T.}~\bibnamefont {Ihn}}, \ and\ \bibinfo {author} {\bibfnamefont
  {K.}~\bibnamefont {Ensslin}},\ }\href@noop {} {\bibfield  {journal} {\bibinfo
   {journal} {Synthetic Metals}\ }\textbf {\bibinfo {volume} {210}},\ \bibinfo
  {pages} {19} (\bibinfo {year} {2015})}\BibitemShut {NoStop}%
\bibitem [{\citenamefont {Varlet}\ \emph {et~al.}(2014)\citenamefont {Varlet},
  \citenamefont {Bischoff}, \citenamefont {Simonet}, \citenamefont {Watanabe},
  \citenamefont {Taniguchi}, \citenamefont {Ihn}, \citenamefont {Ensslin},
  \citenamefont {Mucha-Kruczy\ifmmode~\acute{n}\else \'{n}\fi{}ski},\ and\
  \citenamefont {Fal'ko}}]{VBSWTIEMF2014}%
  \BibitemOpen
  \bibfield  {author} {\bibinfo {author} {\bibfnamefont {A.}~\bibnamefont
  {Varlet}}, \bibinfo {author} {\bibfnamefont {D.}~\bibnamefont {Bischoff}},
  \bibinfo {author} {\bibfnamefont {P.}~\bibnamefont {Simonet}}, \bibinfo
  {author} {\bibfnamefont {K.}~\bibnamefont {Watanabe}}, \bibinfo {author}
  {\bibfnamefont {T.}~\bibnamefont {Taniguchi}}, \bibinfo {author}
  {\bibfnamefont {T.}~\bibnamefont {Ihn}}, \bibinfo {author} {\bibfnamefont
  {K.}~\bibnamefont {Ensslin}}, \bibinfo {author} {\bibfnamefont
  {M.}~\bibnamefont {Mucha-Kruczy\ifmmode~\acute{n}\else \'{n}\fi{}ski}}, \
  and\ \bibinfo {author} {\bibfnamefont {V.~I.}\ \bibnamefont {Fal'ko}},\
  }\href {\doibase 10.1103/PhysRevLett.113.116602} {\bibfield  {journal}
  {\bibinfo  {journal} {Phys. Rev. Lett.}\ }\textbf {\bibinfo {volume} {113}},\
  \bibinfo {pages} {116602} (\bibinfo {year} {2014})}\BibitemShut {NoStop}%
\bibitem [{\citenamefont {Lee}\ \emph {et~al.}(2019)\citenamefont {Lee},
  \citenamefont {Khalaf}, \citenamefont {Liu}, \citenamefont {Liu},
  \citenamefont {Hao}, \citenamefont {Kim},\ and\ \citenamefont
  {Vishwanath}}]{LeeKhalafLiuLiuHaoKimVishwanath2019}%
  \BibitemOpen
  \bibfield  {author} {\bibinfo {author} {\bibfnamefont {J.~Y.}\ \bibnamefont
  {Lee}}, \bibinfo {author} {\bibfnamefont {E.}~\bibnamefont {Khalaf}},
  \bibinfo {author} {\bibfnamefont {S.}~\bibnamefont {Liu}}, \bibinfo {author}
  {\bibfnamefont {X.}~\bibnamefont {Liu}}, \bibinfo {author} {\bibfnamefont
  {Z.}~\bibnamefont {Hao}}, \bibinfo {author} {\bibfnamefont {P.}~\bibnamefont
  {Kim}}, \ and\ \bibinfo {author} {\bibfnamefont {A.}~\bibnamefont
  {Vishwanath}},\ }\href@noop {} {\bibfield  {journal} {\bibinfo  {journal}
  {Nat. Commun.}\ }\textbf {\bibinfo {volume} {10}},\ \bibinfo {pages} {5333}
  (\bibinfo {year} {2019})}\BibitemShut {NoStop}%
\bibitem [{\citenamefont {Rozhkov}\ \emph {et~al.}(2016)\citenamefont
  {Rozhkov}, \citenamefont {Sboychakov}, \citenamefont {Rakhmanov},\ and\
  \citenamefont {Nori}}]{RozhkovSboychakovRakhmanovNori2016}%
  \BibitemOpen
  \bibfield  {author} {\bibinfo {author} {\bibfnamefont {A.}~\bibnamefont
  {Rozhkov}}, \bibinfo {author} {\bibfnamefont {A.}~\bibnamefont {Sboychakov}},
  \bibinfo {author} {\bibfnamefont {A.}~\bibnamefont {Rakhmanov}}, \ and\
  \bibinfo {author} {\bibfnamefont {F.}~\bibnamefont {Nori}},\ }\href@noop {}
  {\bibfield  {journal} {\bibinfo  {journal} {Physics Reports}\ }\textbf
  {\bibinfo {volume} {648}},\ \bibinfo {pages} {1} (\bibinfo {year}
  {2016})}\BibitemShut {NoStop}%
\bibitem [{\citenamefont {Geim}\ and\ \citenamefont
  {Novoselov}(2007)}]{GeimNovoselov2007}%
  \BibitemOpen
  \bibfield  {author} {\bibinfo {author} {\bibfnamefont {A.}~\bibnamefont
  {Geim}}\ and\ \bibinfo {author} {\bibfnamefont {K.}~\bibnamefont
  {Novoselov}},\ }\href@noop {} {\bibfield  {journal} {\bibinfo  {journal}
  {Nature Mater}\ }\textbf {\bibinfo {volume} {6}},\ \bibinfo {pages} {183}
  (\bibinfo {year} {2007})}\BibitemShut {NoStop}%
\bibitem [{\citenamefont {Malard}\ \emph {et~al.}(2007)\citenamefont {Malard},
  \citenamefont {Nilsson}, \citenamefont {Elias}, \citenamefont {Brant},
  \citenamefont {Plentz}, \citenamefont {Alves}, \citenamefont {Neto},\ and\
  \citenamefont {Pimenta}}]{MalardNilssonEliasBrantPlentzAlvesNetoPimenta2007}%
  \BibitemOpen
  \bibfield  {author} {\bibinfo {author} {\bibfnamefont {L.}~\bibnamefont
  {Malard}}, \bibinfo {author} {\bibfnamefont {J.}~\bibnamefont {Nilsson}},
  \bibinfo {author} {\bibfnamefont {D.}~\bibnamefont {Elias}}, \bibinfo
  {author} {\bibfnamefont {J.}~\bibnamefont {Brant}}, \bibinfo {author}
  {\bibfnamefont {F.}~\bibnamefont {Plentz}}, \bibinfo {author} {\bibfnamefont
  {E.}~\bibnamefont {Alves}}, \bibinfo {author} {\bibfnamefont {A.~C.}\
  \bibnamefont {Neto}}, \ and\ \bibinfo {author} {\bibfnamefont
  {M.}~\bibnamefont {Pimenta}},\ }\href@noop {} {\bibfield  {journal} {\bibinfo
   {journal} {Phys. Rev. B}\ }\textbf {\bibinfo {volume} {76}},\ \bibinfo
  {pages} {201401} (\bibinfo {year} {2007})}\BibitemShut {NoStop}%
\bibitem [{\citenamefont {Zhang}\ \emph {et~al.}(2008)\citenamefont {Zhang},
  \citenamefont {Li}, \citenamefont {Basov}, \citenamefont {Fogler},
  \citenamefont {Hao},\ and\ \citenamefont
  {Martin}}]{ZhangLiBasovFoglerHaoMartin2008}%
  \BibitemOpen
  \bibfield  {author} {\bibinfo {author} {\bibfnamefont {L.~M.}\ \bibnamefont
  {Zhang}}, \bibinfo {author} {\bibfnamefont {Z.~Q.}\ \bibnamefont {Li}},
  \bibinfo {author} {\bibfnamefont {D.~N.}\ \bibnamefont {Basov}}, \bibinfo
  {author} {\bibfnamefont {M.~M.}\ \bibnamefont {Fogler}}, \bibinfo {author}
  {\bibfnamefont {Z.}~\bibnamefont {Hao}}, \ and\ \bibinfo {author}
  {\bibfnamefont {M.~C.}\ \bibnamefont {Martin}},\ }\href@noop {} {\bibfield
  {journal} {\bibinfo  {journal} {Phys. Rev. B}\ }\textbf {\bibinfo {volume}
  {78}},\ \bibinfo {pages} {235408} (\bibinfo {year} {2008})}\BibitemShut
  {NoStop}%
\bibitem [{\citenamefont {Castro~Neto}\ \emph {et~al.}(2009)\citenamefont
  {Castro~Neto}, \citenamefont {Guinea}, \citenamefont {Peres}, \citenamefont
  {Novoselov},\ and\ \citenamefont {Geim}}]{NetoGuineaPeresNovoselovGeim2009}%
  \BibitemOpen
  \bibfield  {author} {\bibinfo {author} {\bibfnamefont {A.~H.}\ \bibnamefont
  {Castro~Neto}}, \bibinfo {author} {\bibfnamefont {F.}~\bibnamefont {Guinea}},
  \bibinfo {author} {\bibfnamefont {N.~M.~R.}\ \bibnamefont {Peres}}, \bibinfo
  {author} {\bibfnamefont {K.~S.}\ \bibnamefont {Novoselov}}, \ and\ \bibinfo
  {author} {\bibfnamefont {A.~K.}\ \bibnamefont {Geim}},\ }\href {\doibase
  10.1103/RevModPhys.81.109} {\bibfield  {journal} {\bibinfo  {journal} {Rev.
  Mod. Phys.}\ }\textbf {\bibinfo {volume} {81}},\ \bibinfo {pages} {109}
  (\bibinfo {year} {2009})}\BibitemShut {NoStop}%
\bibitem [{\citenamefont {Kuzmenko}\ \emph {et~al.}(2009)\citenamefont
  {Kuzmenko}, \citenamefont {Crassee}, \citenamefont {van~der Marel},
  \citenamefont {Blake},\ and\ \citenamefont
  {Novoselov}}]{KuzmenkoCrasseeMarelBlakeNovoselov2009}%
  \BibitemOpen
  \bibfield  {author} {\bibinfo {author} {\bibfnamefont {A.~B.}\ \bibnamefont
  {Kuzmenko}}, \bibinfo {author} {\bibfnamefont {I.}~\bibnamefont {Crassee}},
  \bibinfo {author} {\bibfnamefont {D.}~\bibnamefont {van~der Marel}}, \bibinfo
  {author} {\bibfnamefont {P.}~\bibnamefont {Blake}}, \ and\ \bibinfo {author}
  {\bibfnamefont {K.~S.}\ \bibnamefont {Novoselov}},\ }\href@noop {} {\bibfield
   {journal} {\bibinfo  {journal} {Phys. Rev. B}\ }\textbf {\bibinfo {volume}
  {80}},\ \bibinfo {pages} {165406} (\bibinfo {year} {2009})}\BibitemShut
  {NoStop}%
\bibitem [{\citenamefont {McCann}\ and\ \citenamefont
  {Koshino}(2013)}]{mccannkoshino2013}%
  \BibitemOpen
  \bibfield  {author} {\bibinfo {author} {\bibfnamefont {E.}~\bibnamefont
  {McCann}}\ and\ \bibinfo {author} {\bibfnamefont {M.}~\bibnamefont
  {Koshino}},\ }\href@noop {} {\bibfield  {journal} {\bibinfo  {journal}
  {Reports on Progress in Physics}\ }\textbf {\bibinfo {volume} {76}},\
  \bibinfo {pages} {056503} (\bibinfo {year} {2013})}\BibitemShut {NoStop}%
\bibitem [{\citenamefont {Koshino}\ \emph {et~al.}(2018)\citenamefont
  {Koshino}, \citenamefont {Yuan}, \citenamefont {Koretsune}, \citenamefont
  {Ochi}, \citenamefont {Kuroki},\ and\ \citenamefont
  {Fu}}]{KoshinoYuanKoretsuneOchiKurokiFu2018}%
  \BibitemOpen
  \bibfield  {author} {\bibinfo {author} {\bibfnamefont {M.}~\bibnamefont
  {Koshino}}, \bibinfo {author} {\bibfnamefont {N.~F.~Q.}\ \bibnamefont
  {Yuan}}, \bibinfo {author} {\bibfnamefont {T.}~\bibnamefont {Koretsune}},
  \bibinfo {author} {\bibfnamefont {M.}~\bibnamefont {Ochi}}, \bibinfo {author}
  {\bibfnamefont {K.}~\bibnamefont {Kuroki}}, \ and\ \bibinfo {author}
  {\bibfnamefont {L.}~\bibnamefont {Fu}},\ }\href {\doibase
  10.1103/PhysRevX.8.031087} {\bibfield  {journal} {\bibinfo  {journal} {Phys.
  Rev. X}\ }\textbf {\bibinfo {volume} {8}},\ \bibinfo {pages} {031087}
  (\bibinfo {year} {2018})}\BibitemShut {NoStop}%
\bibitem [{\citenamefont {Nam}\ and\ \citenamefont
  {Koshino}(2017)}]{NamKoshino2017}%
  \BibitemOpen
  \bibfield  {author} {\bibinfo {author} {\bibfnamefont {N.~N.~T.}\
  \bibnamefont {Nam}}\ and\ \bibinfo {author} {\bibfnamefont {M.}~\bibnamefont
  {Koshino}},\ }\href {\doibase 10.1103/PhysRevB.96.075311} {\bibfield
  {journal} {\bibinfo  {journal} {Phys. Rev. B}\ }\textbf {\bibinfo {volume}
  {96}},\ \bibinfo {pages} {075311} (\bibinfo {year} {2017})}\BibitemShut
  {NoStop}%
\bibitem [{\citenamefont {Uchida}\ \emph {et~al.}(2014)\citenamefont {Uchida},
  \citenamefont {Furuya}, \citenamefont {Iwata},\ and\ \citenamefont
  {Oshiyama}}]{UchidaFuruyaIwataOshiyama2014}%
  \BibitemOpen
  \bibfield  {author} {\bibinfo {author} {\bibfnamefont {K.}~\bibnamefont
  {Uchida}}, \bibinfo {author} {\bibfnamefont {S.}~\bibnamefont {Furuya}},
  \bibinfo {author} {\bibfnamefont {J.-I.}\ \bibnamefont {Iwata}}, \ and\
  \bibinfo {author} {\bibfnamefont {A.}~\bibnamefont {Oshiyama}},\ }\href
  {\doibase 10.1103/PhysRevB.90.155451} {\bibfield  {journal} {\bibinfo
  {journal} {Phys. Rev. B}\ }\textbf {\bibinfo {volume} {90}},\ \bibinfo
  {pages} {155451} (\bibinfo {year} {2014})}\BibitemShut {NoStop}%
\bibitem [{\citenamefont {Moon}\ and\ \citenamefont
  {Koshino}(2013)}]{MoonKoshino2013}%
  \BibitemOpen
  \bibfield  {author} {\bibinfo {author} {\bibfnamefont {P.}~\bibnamefont
  {Moon}}\ and\ \bibinfo {author} {\bibfnamefont {M.}~\bibnamefont {Koshino}},\
  }\href {\doibase 10.1103/PhysRevB.87.205404} {\bibfield  {journal} {\bibinfo
  {journal} {Phys. Rev. B}\ }\textbf {\bibinfo {volume} {87}},\ \bibinfo
  {pages} {205404} (\bibinfo {year} {2013})}\BibitemShut {NoStop}%
\bibitem [{\citenamefont {Dai}\ \emph {et~al.}(2016)\citenamefont {Dai},
  \citenamefont {Xiang},\ and\ \citenamefont
  {Srolovitz}}]{DaiXiangSrolovitz2016}%
  \BibitemOpen
  \bibfield  {author} {\bibinfo {author} {\bibfnamefont {S.}~\bibnamefont
  {Dai}}, \bibinfo {author} {\bibfnamefont {Y.}~\bibnamefont {Xiang}}, \ and\
  \bibinfo {author} {\bibfnamefont {D.~J.}\ \bibnamefont {Srolovitz}},\ }\href
  {https://doi.org/10.1021/acs.nanolett.6b02870} {\bibfield  {journal}
  {\bibinfo  {journal} {Nano Lett.}\ }\textbf {\bibinfo {volume} {16}},\
  \bibinfo {pages} {5923–5927} (\bibinfo {year} {2016})}\BibitemShut
  {NoStop}%
\bibitem [{\citenamefont {Wijk}\ \emph {et~al.}(2015)\citenamefont {Wijk},
  \citenamefont {Schuring}, \citenamefont {Katsnelson},\ and\ \citenamefont
  {Fasolino}}]{WijkSchuringKatsnelsonFasolino2015}%
  \BibitemOpen
  \bibfield  {author} {\bibinfo {author} {\bibfnamefont {M.~M.~v.}\
  \bibnamefont {Wijk}}, \bibinfo {author} {\bibfnamefont {A.}~\bibnamefont
  {Schuring}}, \bibinfo {author} {\bibfnamefont {M.~I.}\ \bibnamefont
  {Katsnelson}}, \ and\ \bibinfo {author} {\bibfnamefont {A.}~\bibnamefont
  {Fasolino}},\ }\href {https://doi.org/10.1088/2053-1583/2/3/034010}
  {\bibfield  {journal} {\bibinfo  {journal} {2D Mater.}\ }\textbf {\bibinfo
  {volume} {2}},\ \bibinfo {pages} {034010} (\bibinfo {year}
  {2015})}\BibitemShut {NoStop}%
\bibitem [{\citenamefont {Song}\ \emph {et~al.}(2019)\citenamefont {Song},
  \citenamefont {Wang}, \citenamefont {Shi}, \citenamefont {Li}, \citenamefont
  {Fang},\ and\ \citenamefont {Bernevig}}]{song2019}%
  \BibitemOpen
  \bibfield  {author} {\bibinfo {author} {\bibfnamefont {Z.}~\bibnamefont
  {Song}}, \bibinfo {author} {\bibfnamefont {Z.}~\bibnamefont {Wang}}, \bibinfo
  {author} {\bibfnamefont {W.}~\bibnamefont {Shi}}, \bibinfo {author}
  {\bibfnamefont {G.}~\bibnamefont {Li}}, \bibinfo {author} {\bibfnamefont
  {C.}~\bibnamefont {Fang}}, \ and\ \bibinfo {author} {\bibfnamefont {B.~A.}\
  \bibnamefont {Bernevig}},\ }\href {\doibase 10.1103/PhysRevLett.123.036401}
  {\bibfield  {journal} {\bibinfo  {journal} {Phys. Rev. Lett.}\ }\textbf
  {\bibinfo {volume} {123}},\ \bibinfo {pages} {036401} (\bibinfo {year}
  {2019})}\BibitemShut {NoStop}%
\bibitem [{\citenamefont {Hatsugai}(2005)}]{Hatsugai2005}%
  \BibitemOpen
  \bibfield  {author} {\bibinfo {author} {\bibfnamefont {Y.}~\bibnamefont
  {Hatsugai}},\ }\href@noop {} {\bibfield  {journal} {\bibinfo  {journal} {J.
  Phys. Soc. Jpn.}\ }\textbf {\bibinfo {volume} {74}},\ \bibinfo {pages} {1374}
  (\bibinfo {year} {2005})}\BibitemShut {NoStop}%
\bibitem [{\citenamefont {Min}\ \emph {et~al.}(2011)\citenamefont {Min},
  \citenamefont {Abergel}, \citenamefont {Hwang},\ and\ \citenamefont
  {Das~Sarma}}]{MinAbergelLHwangSarma2011}%
  \BibitemOpen
  \bibfield  {author} {\bibinfo {author} {\bibfnamefont {H.}~\bibnamefont
  {Min}}, \bibinfo {author} {\bibfnamefont {D.~S.~L.}\ \bibnamefont {Abergel}},
  \bibinfo {author} {\bibfnamefont {E.~H.}\ \bibnamefont {Hwang}}, \ and\
  \bibinfo {author} {\bibfnamefont {S.}~\bibnamefont {Das~Sarma}},\ }\href
  {\doibase 10.1103/PhysRevB.84.041406} {\bibfield  {journal} {\bibinfo
  {journal} {Phys. Rev. B}\ }\textbf {\bibinfo {volume} {84}},\ \bibinfo
  {pages} {041406} (\bibinfo {year} {2011})}\BibitemShut {NoStop}%
\bibitem [{\citenamefont {Oostinga}\ \emph {et~al.}(2007)\citenamefont
  {Oostinga}, \citenamefont {Heersche}, \citenamefont {Liu}, \citenamefont
  {Morpurgo},\ and\ \citenamefont {Vandersypen}}]{Oostinga_2007}%
  \BibitemOpen
  \bibfield  {author} {\bibinfo {author} {\bibfnamefont {J.~B.}\ \bibnamefont
  {Oostinga}}, \bibinfo {author} {\bibfnamefont {H.~B.}\ \bibnamefont
  {Heersche}}, \bibinfo {author} {\bibfnamefont {X.}~\bibnamefont {Liu}},
  \bibinfo {author} {\bibfnamefont {A.~F.}\ \bibnamefont {Morpurgo}}, \ and\
  \bibinfo {author} {\bibfnamefont {L.~M.~K.}\ \bibnamefont {Vandersypen}},\
  }\href {\doibase 10.1038/nmat2082} {\bibfield  {journal} {\bibinfo  {journal}
  {Nature Materials}\ }\textbf {\bibinfo {volume} {7}},\ \bibinfo {pages}
  {151–157} (\bibinfo {year} {2007})}\BibitemShut {NoStop}%
\bibitem [{\citenamefont {Ohta}\ \emph {et~al.}(2006)\citenamefont {Ohta},
  \citenamefont {Bostwick}, \citenamefont {Seyller}, \citenamefont {Horn},\
  and\ \citenamefont {Rotenberg}}]{Ohta951}%
  \BibitemOpen
  \bibfield  {author} {\bibinfo {author} {\bibfnamefont {T.}~\bibnamefont
  {Ohta}}, \bibinfo {author} {\bibfnamefont {A.}~\bibnamefont {Bostwick}},
  \bibinfo {author} {\bibfnamefont {T.}~\bibnamefont {Seyller}}, \bibinfo
  {author} {\bibfnamefont {K.}~\bibnamefont {Horn}}, \ and\ \bibinfo {author}
  {\bibfnamefont {E.}~\bibnamefont {Rotenberg}},\ }\href@noop {} {\bibfield
  {journal} {\bibinfo  {journal} {Science}\ }\textbf {\bibinfo {volume}
  {313}},\ \bibinfo {pages} {951} (\bibinfo {year} {2006})}\BibitemShut
  {NoStop}%
\bibitem [{\citenamefont {Cao}\ \emph {et~al.}(2020)\citenamefont {Cao},
  \citenamefont {Rodan-Legrain}, \citenamefont {Rubies-Bigorda}, \citenamefont
  {Park}, \citenamefont {Watanabe}, \citenamefont {Taniguchi},\ and\
  \citenamefont {Jarillo-Herrero}}]{Cao_2020}%
  \BibitemOpen
  \bibfield  {author} {\bibinfo {author} {\bibfnamefont {Y.}~\bibnamefont
  {Cao}}, \bibinfo {author} {\bibfnamefont {D.}~\bibnamefont {Rodan-Legrain}},
  \bibinfo {author} {\bibfnamefont {O.}~\bibnamefont {Rubies-Bigorda}},
  \bibinfo {author} {\bibfnamefont {J.~M.}\ \bibnamefont {Park}}, \bibinfo
  {author} {\bibfnamefont {K.}~\bibnamefont {Watanabe}}, \bibinfo {author}
  {\bibfnamefont {T.}~\bibnamefont {Taniguchi}}, \ and\ \bibinfo {author}
  {\bibfnamefont {P.}~\bibnamefont {Jarillo-Herrero}},\ }\href {\doibase
  10.1038/s41586-020-2260-6} {\bibfield  {journal} {\bibinfo  {journal}
  {Nature}\ }\textbf {\bibinfo {volume} {583}},\ \bibinfo {pages} {215–220}
  (\bibinfo {year} {2020})}\BibitemShut {NoStop}%
\bibitem [{\citenamefont {de~Vries}\ \emph {et~al.}(2020)\citenamefont
  {de~Vries}, \citenamefont {Zhu}, \citenamefont {Portoles}, \citenamefont
  {Zheng}, \citenamefont {Masseroni}, \citenamefont {Kurzmann}, \citenamefont
  {Taniguchi}, \citenamefont {Watanabe}, \citenamefont {MacDonald},
  \citenamefont {Ensslin}, \citenamefont {Ihn},\ and\ \citenamefont
  {Rickhaus}}]{VZPZMKTWMEIR2020}%
  \BibitemOpen
  \bibfield  {author} {\bibinfo {author} {\bibfnamefont {F.~K.}\ \bibnamefont
  {de~Vries}}, \bibinfo {author} {\bibfnamefont {J.}~\bibnamefont {Zhu}},
  \bibinfo {author} {\bibfnamefont {E.}~\bibnamefont {Portoles}}, \bibinfo
  {author} {\bibfnamefont {G.}~\bibnamefont {Zheng}}, \bibinfo {author}
  {\bibfnamefont {M.}~\bibnamefont {Masseroni}}, \bibinfo {author}
  {\bibfnamefont {A.}~\bibnamefont {Kurzmann}}, \bibinfo {author}
  {\bibfnamefont {T.}~\bibnamefont {Taniguchi}}, \bibinfo {author}
  {\bibfnamefont {K.}~\bibnamefont {Watanabe}}, \bibinfo {author}
  {\bibfnamefont {A.~H.}\ \bibnamefont {MacDonald}}, \bibinfo {author}
  {\bibfnamefont {K.}~\bibnamefont {Ensslin}}, \bibinfo {author} {\bibfnamefont
  {T.}~\bibnamefont {Ihn}}, \ and\ \bibinfo {author} {\bibfnamefont
  {P.}~\bibnamefont {Rickhaus}},\ }\href@noop {} {\enquote {\bibinfo {title}
  {Combined minivalley and layer control in twisted double bilayer graphene},}\
  } (\bibinfo {year} {2020}),\ \Eprint {http://arxiv.org/abs/2002.05267}
  {arXiv:2002.05267 [cond-mat.mes-hall]} \BibitemShut {NoStop}%
\bibitem [{\citenamefont {Roth}\ \emph {et~al.}(2009)\citenamefont {Roth},
  \citenamefont {Br{\"u}ne}, \citenamefont {Buhmann}, \citenamefont
  {Molenkamp}, \citenamefont {Maciejko}, \citenamefont {Qi},\ and\
  \citenamefont {Zhang}}]{Roth294}%
  \BibitemOpen
  \bibfield  {author} {\bibinfo {author} {\bibfnamefont {A.}~\bibnamefont
  {Roth}}, \bibinfo {author} {\bibfnamefont {C.}~\bibnamefont {Br{\"u}ne}},
  \bibinfo {author} {\bibfnamefont {H.}~\bibnamefont {Buhmann}}, \bibinfo
  {author} {\bibfnamefont {L.~W.}\ \bibnamefont {Molenkamp}}, \bibinfo {author}
  {\bibfnamefont {J.}~\bibnamefont {Maciejko}}, \bibinfo {author}
  {\bibfnamefont {X.-L.}\ \bibnamefont {Qi}}, \ and\ \bibinfo {author}
  {\bibfnamefont {S.-C.}\ \bibnamefont {Zhang}},\ }\href@noop {} {\bibfield
  {journal} {\bibinfo  {journal} {Science}\ }\textbf {\bibinfo {volume}
  {325}},\ \bibinfo {pages} {294} (\bibinfo {year} {2009})}\BibitemShut
  {NoStop}%
\bibitem [{\citenamefont {Sinha}\ \emph {et~al.}(2020)\citenamefont {Sinha},
  \citenamefont {Adak}, \citenamefont {Kanthi}, \citenamefont {Chittari},
  \citenamefont {Sangani}, \citenamefont {Watanabe}, \citenamefont {Taniguchi},
  \citenamefont {Jung},\ and\ \citenamefont {Deshmukh}}]{sinha2020bulk}%
  \BibitemOpen
  \bibfield  {author} {\bibinfo {author} {\bibfnamefont {S.}~\bibnamefont
  {Sinha}}, \bibinfo {author} {\bibfnamefont {P.~C.}\ \bibnamefont {Adak}},
  \bibinfo {author} {\bibfnamefont {R.~S.~S.}\ \bibnamefont {Kanthi}}, \bibinfo
  {author} {\bibfnamefont {B.~L.}\ \bibnamefont {Chittari}}, \bibinfo {author}
  {\bibfnamefont {L.~D.~V.}\ \bibnamefont {Sangani}}, \bibinfo {author}
  {\bibfnamefont {K.}~\bibnamefont {Watanabe}}, \bibinfo {author}
  {\bibfnamefont {T.}~\bibnamefont {Taniguchi}}, \bibinfo {author}
  {\bibfnamefont {J.}~\bibnamefont {Jung}}, \ and\ \bibinfo {author}
  {\bibfnamefont {M.~M.}\ \bibnamefont {Deshmukh}},\ }\href@noop {} {\enquote
  {\bibinfo {title} {Bulk valley transport and berry curvature spreading at the
  edge of flat bands},}\ } (\bibinfo {year} {2020}),\ \Eprint
  {http://arxiv.org/abs/2004.14727} {arXiv:2004.14727 [cond-mat.mes-hall]}
  \BibitemShut {NoStop}%
\bibitem [{\citenamefont {Burg}\ \emph {et~al.}(2020)\citenamefont {Burg},
  \citenamefont {Lian}, \citenamefont {Taniguchi}, \citenamefont {Watanabe},
  \citenamefont {Bernevig},\ and\ \citenamefont
  {Tutuc}}]{BurgLianTaniguchiWatanabeBernevigTutuc2020}%
  \BibitemOpen
  \bibfield  {author} {\bibinfo {author} {\bibfnamefont {G.~W.}\ \bibnamefont
  {Burg}}, \bibinfo {author} {\bibfnamefont {B.}~\bibnamefont {Lian}}, \bibinfo
  {author} {\bibfnamefont {T.}~\bibnamefont {Taniguchi}}, \bibinfo {author}
  {\bibfnamefont {K.}~\bibnamefont {Watanabe}}, \bibinfo {author}
  {\bibfnamefont {B.~A.}\ \bibnamefont {Bernevig}}, \ and\ \bibinfo {author}
  {\bibfnamefont {E.}~\bibnamefont {Tutuc}},\ }\href@noop {} {\enquote
  {\bibinfo {title} {Evidence of emergent symmetry and valley chern number in
  twisted double-bilayer graphene},}\ } (\bibinfo {year} {2020}),\ \Eprint
  {http://arxiv.org/abs/2006.14000} {arXiv:2006.14000 [cond-mat.mes-hall]}
  \BibitemShut {NoStop}%
\bibitem [{\citenamefont {Wu}\ \emph {et~al.}(2020)\citenamefont {Wu},
  \citenamefont {Liu},\ and\ \citenamefont {Yazyev}}]{WuLiuYazyev2020}%
  \BibitemOpen
  \bibfield  {author} {\bibinfo {author} {\bibfnamefont {Q.}~\bibnamefont
  {Wu}}, \bibinfo {author} {\bibfnamefont {J.}~\bibnamefont {Liu}}, \ and\
  \bibinfo {author} {\bibfnamefont {O.~V.}\ \bibnamefont {Yazyev}},\
  }\href@noop {} {\enquote {\bibinfo {title} {Landau levels as a probe for band
  topology in graphene moiré superlattices},}\ } (\bibinfo {year} {2020}),\
  \Eprint {http://arxiv.org/abs/2005.10620} {arXiv:2005.10620
  [cond-mat.mes-hall]} \BibitemShut {NoStop}%
\bibitem [{\citenamefont {Charlier}\ \emph {et~al.}(1992)\citenamefont
  {Charlier}, \citenamefont {Michenaud},\ and\ \citenamefont
  {Gonze}}]{CharlierMichenaudGonze1992}%
  \BibitemOpen
  \bibfield  {author} {\bibinfo {author} {\bibfnamefont {J.-C.}\ \bibnamefont
  {Charlier}}, \bibinfo {author} {\bibfnamefont {J.-P.}\ \bibnamefont
  {Michenaud}}, \ and\ \bibinfo {author} {\bibfnamefont {X.}~\bibnamefont
  {Gonze}},\ }\href {\doibase 10.1103/PhysRevB.46.4531} {\bibfield  {journal}
  {\bibinfo  {journal} {Phys. Rev. B}\ }\textbf {\bibinfo {volume} {46}},\
  \bibinfo {pages} {4531} (\bibinfo {year} {1992})}\BibitemShut {NoStop}%
\bibitem [{\citenamefont {Munoz}\ \emph {et~al.}(2016)\citenamefont {Munoz},
  \citenamefont {Collado}, \citenamefont {Usaj}, \citenamefont {Sofo},\ and\
  \citenamefont {Balseiro}}]{MunozColladoOjedaUsajSofoBalseiro2016}%
  \BibitemOpen
  \bibfield  {author} {\bibinfo {author} {\bibfnamefont {F.}~\bibnamefont
  {Munoz}}, \bibinfo {author} {\bibfnamefont {H.~P.~O.}\ \bibnamefont
  {Collado}}, \bibinfo {author} {\bibfnamefont {G.}~\bibnamefont {Usaj}},
  \bibinfo {author} {\bibfnamefont {J.~O.}\ \bibnamefont {Sofo}}, \ and\
  \bibinfo {author} {\bibfnamefont {C.~A.}\ \bibnamefont {Balseiro}},\
  }\href@noop {} {\bibfield  {journal} {\bibinfo  {journal} {Phys. Rev. B}\
  }\textbf {\bibinfo {volume} {93}},\ \bibinfo {pages} {235443} (\bibinfo
  {year} {2016})}\BibitemShut {NoStop}%
\end{thebibliography}%

\end{document}